\PassOptionsToPackage{pdfpagelabels=false}{hyperref}
\documentclass[useAMS,usenatbib]{mnras}
\pdfoutput=1
\usepackage[pdftex]{graphicx}
\usepackage{amsmath}
\usepackage{amssymb}
\usepackage{natbib}
\usepackage{geometry}
\usepackage{rotating,booktabs,multirow}
\usepackage{color}
\usepackage{times}
\usepackage{tabularx}

\newcommand{\bea}{\begin{eqnarray}}
\newcommand{\eea}{\end{eqnarray}}

\newcommand{\Ms}{{\rm M}_\odot}
\def \MS{M_{\rm stars}}

\title[SFRs in IllustrisTNG]{The star-formation activity of IllustrisTNG galaxies: main sequence, UVJ diagram, quenched fractions, and systematics}

\author[Donnari et al.]
{Martina Donnari$^{1}$\thanks{E-mail: donnari@mpia-hd.mpg.de}, 
Annalisa Pillepich$^{1}$,
Dylan Nelson$^{2}$, 
Mark Vogelsberger$^{3}$,
Shy Genel$^{4,5}$,\newauthor
Rainer Weinberger$^{6}$,
Federico Marinacci$^{3}$,
Volker Springel$^{2}$,
and Lars Hernquist$^{6}$\\
$^{1}$ Max-Planck-Institut f\"{u}r Astronomie, K\"{o}nigstuhl 17, 69117 Heidelberg, Germany\\
$^{2}$ Max-Planck-Institut f\"{u}r Astrophysik, Karl-Schwarzschild-Str. 1, D-85741 Garching, Germany\\
$^{3}$ Kavli Institute for Astrophysics and Space Research, Massachusetts Institute of Technology,
Cambridge, MA 02139, USA\\
$^{4}$Center for Computational Astrophysics, Flatiron Institute, 162 Fifth Avenue, New York, NY 10010, USA\\
$^{5}$Columbia Astrophysics Laboratory, Columbia University, 550 West 120th Street, New York, NY 10027, USA\\
$^{6}$ Harvard-Smithsonian Center for Astrophysics, 60 Garden Street, Cambridge, MA 02138, USA}

\begin{document}

\pagerange{\pageref{firstpage}--\pageref{lastpage}} \pubyear{2018}

\maketitle

\label{firstpage}

\maketitle

\begin{abstract}

We select galaxies from the IllustrisTNG hydrodynamical simulations
($\MS > 10^9 \, \Ms$ at $0\le z \le 2$) and characterize the shapes and
evolutions of their UVJ and star-formation rate -- stellar mass
(SFR-$\MS$) diagrams.
We quantify the systematic uncertainties related to different criteria
to classify star-forming vs. quiescent galaxies, different SFR
estimates, and by accounting for the star formation measured within
different physical apertures. 
The TNG model returns the observed features of the UVJ diagram at $z\leq 2$, with a clear
separation between two classes of galaxies. It also returns a tight
star-forming main sequence (MS) for $\MS <10^{10.5}  \, (\Ms)$ with a $\sim 0.3$ dex scatter
at $z\sim0$ in our fiducial choices.
If a UVJ-based cut is adopted, the TNG MS exhibits a downwardly bending at stellar masses
of about $10^{10.5-10.7}~ \Ms$.
Moreover, the model predicts that $\sim80\, (50)$ per cent of $10^{10.5-11}\Ms$ galaxies at $z=0$ $(z=2)$ are quiescent
and the numbers of quenched galaxies at intermediate redshifts and high masses are in better agreement with
observational estimates than previous models.
However, shorter SFR-averaging timescales imply higher normalizations and scatter
of the MS, while smaller apertures lead to underestimating the galaxy
SFRs: overall we estimate the inspected systematic uncertainties to sum up to about $0.2-0.3$ dex in the locus of the MS and to about 15 percentage points in the fraction of quenched galaxies.
While TNG color distributions are clearly bimodal, this is not the case
for the SFR logarithmic distributions in bins of stellar mass (SFR$\gtrsim 10^{-3} ~ \Ms$yr$^{-1}$).
Finally, the slope and $z=0$ normalization of the TNG MS are consistent
with observational findings; however, the locus of the TNG MS remains lower by about $0.2-0.5$ dex at
$0.75\le z < 2$ than the available observational estimates taken at face
value.

\end{abstract}


\section{Introduction}
\label{intro}

The observed distribution of many galaxy properties, such as color, morphology, and star-formation rate, are thought to be bimodal in the nearby Universe and for relatively massive galaxies. 
These properties correlate with galaxy density and stellar mass, suggesting that the processes that regulate the star formation activity of galaxies depend both on a galaxy's stellar mass and the environment in which galaxies evolve \citep{2010Vulcani,2010Peng,2012Vulcani,2016paccagnella,2017Darvish}.

Broadly speaking, galaxies in the local Universe can be mainly divided into two categories: disk-like `star-forming' galaxies that preferentially populate the low-density regions of the Universe, and `passive' or `quenched' galaxies, with older stellar populations and little-to-no star formation, which are mostly observed at high stellar masses and/or populating the high-density environments. The latter type is comprised mainly of spheroidal or elliptical galaxies \citep{2003Kauffmann,2004Baldry}.

As suggested by observations, star-forming galaxies show a tight correlation between their stellar mass and star-formation rate (SFR), this relation being commonly called `star-forming main sequence' \citep[hereafter MS or SFMS; e.g.][]{2007Noeske,2011Rodighiero,2011Wuyts,2012Whitaker,2014Speagle,2015Renzini,2015Schreiber,2015Tasca,2015Shivaei}. The existence of a MS provides an easy and powerful way to discern between star-forming and quenched galaxies, representing, indeed, a suitable upper-limit below which a galaxy of a given stellar mass can be considered quenched.

Although the MS is widely recovered from both observations and theory, two different schools of thought exist over e.g. the shape and locus of the MS. Some observational works point out that at $\MS \gtrsim 10^{10.5} \, \Ms$ there may be a turnover, with the slope of the MS being shallower above this stellar mass value than below \citep{2008Reddy,2011Karim,2012Whitaker,2014Whitaker,2015NLee,2015Gavazzi,2016Tomczak,2016Schreiber,2018BLee}. On the other side, some studies find the MS can be simply described by a single power-law, with the logarithm of the SFR increasing with the logarithm of the stellar mass across the whole studied mass range \citep[see][and references therein for a collection of observational works]{2014Speagle}.

Additionally, observations show that the value of the normalization increases at increasing redshift, with a redshift-independent slope measured in the range ($0.4,1$) and the normalization measured to be in the range of $(-1.0, 0.25)\,\Ms$yr$^{-1}$ at $\sim 10^9 \Ms$ and $z=0-2$, likely depending on sample selection and on the indicator adopted to estimate the SFR of observed galaxies \citep[see e.g.][and references therein]{2014Rodighiero,2014Speagle,2017Santini,2018Jian,2018Pearson}.
In fact, non-negligible systematics potentially affect the observational estimates of the SFRs of galaxies across masses and times, owing to a variety of different assumptions on e.g. the effects of dust attenuation, the choices of the stellar initial mass function, on the  conversions from observed luminosity to SFR and on the levels of contamination \cite[see e.g.][]{2014Speagle,2018Theios,2018Leja}.

Different observational indicators adopted to infer galaxy SFRs, such as H$\alpha$ emission lines, X-ray, UV and IR bands, and the associated definition of star-forming or passive galaxies, can lead to different quantitative results \citep{2017Katsianis,2017Davies,2018Davies}.
Consequently, any comparison between observational and theoretical works should take into account the effects of different observational tracers, since the timescales over which different SFR indicators are thought to be sensitive to can vary from a few -- 5 or 10 megayears, usually associated to H$\alpha$ and UV emission lines -- to hundreds of megayears -- for 24 $\mu$m, IR and FIR bands.

Furthermore, the absence of a standard approach in dealing with SFR in galaxies is outlined by several authors who have highlighted the importance of quantifying the effect of apertures in measuring galaxies' SFRs \citep{2016Richards}, both in the observational data \citep{2005Kewley} and also in comparison with the predictions of theoretical model \citep{2016Guidi}. For example, the Sloan Digital Sky Survey (SDSS) fibre aperture covers only a limited region of a galaxy at low-redshift ($\sim 5$ physical kpc), and thus it is sensitive to only a fractional amount of its H$\alpha$ emission. 

From a theoretical point of view, the whole SFR-M$_{\rm star}$ plane and the MS of star-forming galaxies have been investigated by several authors in analytical galaxy formation models \citep[e.g.][]{2010Bouche,2012Dave,2013Lilly,2015Mitra,2017Mitra}, in semi-analytical models \citep[SAMs: ][]{2008Kang,2012Delucia,2014Hirschmann,2017Henriques} and hydrodynamical simulations \citep[e.g.][]{2014Torrey,2015Bahe,2015Sparre,2017Dave,2017Feldmann}. Although the MS is recovered in both the aforementioned techniques, its evolution with redshift is still a controversial issue, making the comparison among theoretical models and observations still unsatisfactory. 
A general consensus is that the fraction of passive galaxies increases with increasing stellar mass and decreases with increasing redshift. However, an optimal match between theory and observations on how SFR relates to the environment remains elusive, although the use of the state-of-art simulations provide nowadays a powerful tool to investigate this question.

Indeed, the SFR-M$_{\rm star}$ plane is not immune to different interpretations: this is due to the lack of agreement on a universal definition of `quenched' galaxies from both theoretical and observational sides and to the limitations on the minimum SFR values that can be effectively detected in observations and sampled in simulations.
Because of the broadly-recognized galaxy color bimodality that is present at least up to $z\sim 3$, a considerable number of observational studies use a color-color based classification to distinguish between blue/star-forming and red/quenched galaxies \citep[e.g.][]{2011Wuyts,2009Williams,2010Williams,2010Whitaker,2011Whitaker,2012Patel,2012Quadri,2013Muzzin,2013Ilbert,2014tomczak,2016darvish}.
Besides a color bimodality, the data from large-scale surveys like SDSS have been shown to exhibit a double peak in the specific star-formation rate (sSFR = SFR$/ \MS$) distribution of galaxies \citep{2015Ilbert}, with a depression or break that likely depend on the observed galaxy sample but appears independent of galaxy and host halo mass, at least in the $\gtrsim 10^{10}\Ms$ range \citep{2004Brinchmann, 2004Kauffmann, 2012Wetzel}. 
However, recently \cite{2017Feldmann} and \cite{2017Eales,2018Eales} have argued that the observed SFR bimodality and the existence of a `green valley' region in SFR-$\MS$ plane, that separates star-forming from quenched populations, are likely connected to uncertainties and observational biases in the SFR measurements, hence challenging a foundational galaxy-evolution paradigm.

In this paper we aim at providing a thorough characterization of the star formation activity of galaxies produced in the state-of-the-art cosmological magneto-hydrodynamical simulations of \textit{The Next Generation} Illustris project \citep[hereafter IllustrisTNG or TNG:][]{2018Naiman,2018Marinacci,2018Springel,2018Pillepich,2018Nelson}. Particularly, we study galaxies from the TNG100 and TNG300 simulations at redshifts $0\le z\le2$, by extending to higher redshifts and new diagnostics previous TNG analyses of the TNG $z=0$ galaxy colors \citep{2018Nelson}, $z=0$ SFRs \citep{2018Weinberger}, and $z=1-2$ SFRs and black-hole (BH) populations \citep{2018Habouzit}. For instance, in \citealt{2018Nelson} we have shown that the IllustrisTNG model returns a remarkably good agreement in terms of the shape and width of the red sequence and the blue cloud of galaxies in comparison to SDSS data, by improving upon the galaxy optical-color bimodality at the high-mass end ($\sim 10^{10} - 10^{11} ~ \Ms$) of the original Illustris model \citep{2014MNRASVogel, 2014Genel} and thus providing an indirect confirmation of the adopted underlying recipes for stellar and BH feedback. 

The IllustrisTNG simulations therefore provide an interesting starting point to further investigate some of the open-ended questions that mine our observational and theoretical understanding of the evolution of galaxies: how does the MS evolve in time? Is there a bending of the star-forming MS at the high-mass end? How do galaxy selection criteria affect the shape of the MS and the estimates of the fraction of quenched galaxies? What is the quantitative connection between color-color and SFR-based criteria to select for quenched galaxies? And can a color bimodality be in place even without a bimodality in SFR or sSFR?

The structure of our paper is as follows.
In Section \ref{method}, we present the simulations used in this work and we introduce all  of the criteria, definitions, and measurement choices adopted throughout the analysis.
Our IllustrisTNG results are given in Section \ref{results}: we characterize the UVJ diagram of TNG galaxies, quantify the relative number of quenched galaxies as a function of mass and time. We also discuss the morphology of the SFR-$\MS$ plane and quantify the effects that different measurement choices have on the locus and shape of the star-forming MS and its scatter.
A discussion of the main findings, together with an extensive study of the distributions of galaxy colors and SFRs and of the shape of the star-forming MS at the high-mass end, are presented in Section \ref{discussion}. There we attempt a comparison to a selection of observational data taken at face value. Our summary and conclusions are given in Section \ref{summary}.

\section{Methods}
\label{method}

\subsection{The IllustrisTNG simulations}

\begin{table}
\centering
\begin{tabular}{c|c|c|c|c|c}
\hline
Run & L$_{\rm box}$ & N$_{\rm GAS}$ & N$_{\rm DM}$ & m$_{\rm baryon}$ & m$_{\rm DM}$ \\
 & [Mpc h${^{-1}}$] &  &  & [$10^6 \Ms$] &  [$10^6 \Ms$]\\
\hline
\hline
TNG100 & 75  & 1820$^3$ & 1820$^3$ & 1.4 & 7.5\\
TNG300 & 205 & 2500$^3$ & 2500$^3$ & 11  & 59\\
\hline
\end{tabular}
\caption{\label{tab:tng} Properties of the TNG100 and TNG300 simulations. Columns read: 1): simulation name; 2): box-side length; 3) initial number of gas cells; 4) number of dark matter particles; 5) target baryonic mass; 6) dark-matter particle mass.}
\end{table}

The IllustrisTNG simulations\footnote{\url{http://www.tng-project.org}} \citep[][TNG hereafter]{2018Naiman,2018Marinacci,2018Springel,2018Pillepich,2018Nelson} are a series of cosmological magneto-hydrodynamical simulations that model a range of physical processes considered relevant for the formation of galaxies. They build and improve upon the original Illustris project \citep{2013Vogelsberger,2014Torrey,2014vogel,2014Genel,2015Sijacki,2015Nelson} by including, among others, a new black-hole feedback model, magnetohydrodynamics, and a new scheme for galactic winds \citep[see][for details on the TNG model]{2018Weinberger, 2018Pillepich_A}.

The simulations are performed with the \textsc{AREPO} code \citep{2010springel}, which solves Poisson's equation for gravity by employing a tree-particle-mesh algorithm. For magneto-hydrodynamics, the code uses the finite volume method on an unstructured, moving, Voronoi tessellation of the simulation domain.

The initial conditions of the TNG runs have been initialized at redshift $z=127$ assuming a matter density $\Omega_{\rm m} = \Omega_{\rm dm}+\Omega_{\rm b} = 0.3089$, baryonic density $\Omega_{\rm b} =0.0486$, cosmological constant $\Omega_{\Lambda}=0.6911$, Hubble constant $H_0=100 \,h$ km s$^{-1}$ Mpc$^{-1}$ with $h = 0.6774$, normalization $\sigma_8 = 0.8159$ and spectral index $n_s = 0.9667$ \citep{2016planck}.

A number of outcomes from the IllustrisTNG simulations have been shown to be consistent with observations: e.g. the shape and width of the red sequence and the blue cloud of $z=0$ galaxies \citep{2018Nelson}, the distributions of metals in the intra-cluster medium at low redshifts \citep{2018Vogelsberger}, the evolution of the galaxy mass-metallicity relation \citep{2018Torrey} and of the galaxy size-mass relation for star-forming and quiescent galaxies at $0\le z \le 2$ \citep{2018Genel}, the amount and distribution of highly-ionized Oxygen around galaxies \citep{2018NelsonB}, and the dark matter fractions within the bodies of massive galaxies at $z=0$ \citep{2018Lovell}.
All these validations of the model make the IllustrisTNG simulations an ideal laboratory to further investigate galaxy formation and evolution. Of relevance for the analysis at hand, the new physics ingredients included in the TNG model affect the star formation activity of galaxies, thus potentially producing important differences in the locus of the star-forming main sequence and the number of quenched galaxies in comparison to the original Illustris simulation \citep{2015Sparre}.

In this work, we use the simulations called TNG100 and TNG300, that model two cosmological boxes with a side length of 75 $h^{-1}\approx$ 100 Mpc and 205 $h^{-1}\approx$ 300 Mpc, respectively. 
A third simulation, with a side length box of 35 $h^{-1}\approx$ 50 Mpc and increased numerical resolution (TNG50), is still running and will not be discussed in this paper. The main properties of TNG100 and TNG300 are listed in Table~\ref{tab:tng} \citep[see][for more details]{2018Pillepich,2018Nelson}.

\subsection{The galaxy sample}
All of the properties of haloes and subhaloes used in this work are obtained taking advantage of the Friends-of-Friends (FoF) and \texttt{SUBFIND} algorithms \citep{1985Davis,2001springel}, which are used to identify substructures, and hence galaxies, across the simulated volumes. 

For the purpose of this work, we do not distinguish between central and satellite galaxies and select all subhaloes with stellar mass $\MS > 10^9 \, \Ms$. 
This minimum mass corresponds to have only well-resolved galaxies in our sample, namely, with at least one thousand star particles in total in TNG100 and more than one hundred in TNG300 (see the Appendix \ref{appendix} for more details).

The stellar mass is defined, throughout this work, as the sum of all stellar particles within twice the stellar half mass radius, $R_{\rm star,h}$.

With these prescriptions, we gather, at $z=0$, more than 18800 galaxies in TNG100 and more than 253000 galaxies in TNG300. We study galaxy populations between $z=0$ and 2.


\subsection{Star Formation Rates in IllustrisTNG}
\label{SFRinTNG}
In the TNG model, the star formation is treated following \cite{2003Springel}:  the gas is stochastically converted into star particles once its density exceeds $\rm n_H = 0.1 \rm \, cm^{-3}$ and on a time scale determined so that it empirically reproduces the Kennicutt-Schmidt relation \citep{1989Kennicutt}.
Star particles hence represent single-age stellar populations with a Chabrier stellar initial mass function \citep[IMF][]{2003Chabrier}. Gas cells are dubbed {\it star forming} when their density is larger than the above-mentioned threshold and their intrinsic star-formation rate (SFR) is thus non vanishing. The SFR of a gas cell is inversely proportional to a density-dependent time-scale for star formation and is proportional to the subgrid estimate of cold gas mass. Such a SFR is {\it instantaneous}.
Therefore, the SFR of a galaxy can be measured by summing up the `instantaneous' SFRs of any given subset of its star-forming gas cells.

However, from an observational point of view, the SFR of galaxies is generally derived through several diagnostics and adopting the luminosity at different observational bands as an indicator for star formation. As widely discussed in the literature, such star-formation tracers span the full electromagnetic spectrum, from X-ray and ultraviolet (UV) to radio wavelengths, via optical and infrared (IR) bands, and can be based on both continuum and line emissions \citep[see the most recent review on the subject by][]{2012Kennicutt}.
Furthermore, different star-formation indicators are understood to be sensitive to star-formation rates that are averaged over different timescales, ranging from a few to hundreds of million years. Consequently, any comparison between observational and theoretical works, and even among observations, must take into account possible systematic effects due to the different values of SFRs derived via different tracers.

Finally, we note that, even if all galaxies taken into account in this work are resolved in terms of stellar and gas element counts, this is not the case for their SFR values. The finite numerical resolution of the simulation implies a minimum resolvable value for SFR (SFR$_{\rm min}$). Galaxies with Log SFR $<$ Log SFR$_{\rm min}$, have unresolved SFR values and are artificially labeled with SFR$\equiv 0$ in the simulation output. The exact value of SFR$_{\rm min}$ depends on the model, numerical resolution, and averaging timescales over which the SFR is measured (see the Appendix \ref{appendix} for more details).

\subsection{Identification of the star-forming Main Sequence and its scatter}
\label{MS and scatter}
Once a method has been adopted to distinguish between star-forming and quiescent galaxies (see next Section) and irrespective of the definition of a galaxy's SFR, we define the star-forming main sequence (MS) of the galaxy population on the SFR-$M_{\rm stars}$ plane in two ways. 
In the first one, we perform a fit to the median SFR of star-forming galaxies in 0.2-dex bins of galaxy stellar mass in the mass range $\MS =10^9-10^{10.2} \, \Ms$, where the MS can be considered linear, and we therefore linearly extrapolate the locus of the MS to larger galaxy masses. The following fitting equation is adopted throughout:

\begin{equation}
 {\rm Log} \, 
 \left (\frac{ \langle {\rm SFR}\rangle _{\rm sf-ing\,galaxies}}{\Ms ~ \rm yr^{-1}} \right) = \alpha(z) \; {\rm Log} \,\left (\frac{  \MS}{\Ms} \right)+ \beta(z)\; ,
\label{ms}
\end{equation}
where $\alpha$ and $\beta$ denote the logarithmic slope and normalization, respectively, of the MS and are both redshift dependent \citep[see e.g.][]{2014Speagle}. We refer to this choice as ``linearly-extrapolated'' MS.

In the second way, we record the median SFR of star-forming galaxies as a function of galaxy stellar mass (in 0.2-dex logarithmic bins) across the whole range of available stellar masses. In this case, the MS is allowed to deviate from a linear trend, as in fact it does by exhibiting a turnover at about $\MS \gtrsim 10^{10.5} \, \Ms$, as suggested by several observations \citep[][see Section \ref{sec:bending}]{2011Karim,2014Whitaker,2015NLee,2016Tomczak,2018BLee}. We refer to this choice as ``bending'' MS and we fit it with a fourth-order polynomial formula, where appropriate.
Throughout this work, the 1$\sigma$ scatter of the MS is measured by stacking star-forming galaxies in stellar mass bins of 0.2 dex:

\begin{equation}
\label{sigma}
\sigma_{\rm MS} = \sqrt{\frac{\sum_{i=1}^N (\rm SFR- \langle SFR\rangle)^2}{N}} \; ,
\end{equation}
where $N$ is the total number of star-forming galaxies in the considered mass bin and $\langle \rm SFR\rangle$ is their median SFR. This is an intrinsic scatter, i.e. it does not take into account measurement uncertainties.
As we demonstrate in Section \ref{sec:scatter}, although the general features of the 1$\sigma$ scatter of the TNG MS do not depend on the measurement choices, its values changes whenever different timescales, apertures or quenching definitions are considered.

\subsection{Systematics and measurement choices}
\label{systematics}
In this Section we introduce and describe all the measurement choices used throughout this work: for the purposes of measuring the SFRs of simulated galaxies, we introduce SFR estimates based on the mass of stars in a galaxy formed over the last N Myr, with N equal to 10, 50, 200 and 1000 Myr, four physical galaxy apertures, and several selection criteria to separate quenched vs. star-forming galaxies.


\begin{itemize}
\item \textbf{Averaging timescales for SFR}: As already mentioned in the previous Section, the instantaneous SFR measured from the gas is not an observable. Therefore, we additionally derive a galaxy's SFR from the amount of stellar mass formed over the last 10, 50, 200 and 1000 Myr, based on the output formation times of individual  stellar particles, with the goal of connecting to the timescales associated to different SFR indicators. A systematic discussion about the differences across these measures and of how numerical resolution affects their values is given in Appendix \ref{appendix}.
\\
\item \textbf{Apertures}: With the aim of bracketing the typical galaxy apertures used in observational surveys, we measure the galaxy SFRs in different ways: i) by considering all gas cells and/or star particles that are gravitationally bound to the galaxy according to the adopted \texttt{SUBFIND} algorithm; or by additionally restricting the galaxy measurements to the gas cells and/or star particles that are found within the following 3D spherical galactocentric distances: ii) within twice the stellar half-mass radius; iii) within 30 physical kpc (hereafter pkpc), and iv) within 5 pkpc, the latter meant to mock an SDSS fibre aperture. 
\\
\item \textbf{Quenched vs star-forming galaxies}: From both theoretical and observational sides, the separation between ‘quenched’ vs. ‘star-forming’ galaxies is ambiguous and somewhat arbitrary. For this reason, in this work we explore and contrast a number of different classification criteria to separate these two populations of galaxies.
Details about our definitions are provided in the following, together with a schematic summary in Table~ \ref{tab:thresholds}. Throughout this work, we use the adjectives quenched, quiescent, and passive interchangeably and we do not distinguish between star-forming and star-burst galaxies nor between quiescent and green-valley galaxies. All the following classifications can be adopted for any averaging timescales for SFR.
\end{itemize}

\begin{table*}
\large
\centering
\begin{tabular}{l|l|l}
\hline
Classification criteria & Definition of ``Quenched'' galaxies & Notes and comments \\
\hline
\hline
Population-based & $ \rm Log \, sSFR \, \rm(yr^{-1}) < -11 $ & \multicolumn{1}{m{8cm}}{Logarithmic sSFR below $-11 \, \rm yr^{-1}$.} 
\\
& $ \rm Log \, SFR < \rm MS - 2\sigma$ & \multicolumn{1}{m{8cm}}{SFR below $2\sigma$ from the star-forming main sequence.} 
\\
& $ \rm Log \, SFR < \rm MS - 1$ dex & \multicolumn{1}{m{8cm}}{SFR below $1$ dex from the star-forming main sequence.} 
\\
& Color-color diagram & \multicolumn{1}{m{8cm}}{UVJ colors satisfy Eqs.~\ref{TNG_cut}-\ref{TNG_cut2}.} 
\\
Galaxy-based & ${\rm SFR}< \overline{\rm SFR}(t) $& \multicolumn{1}{m{8cm}}{SFR below the lifetime-average SFR.} \\
\hline
\end{tabular}
\caption{\label{tab:thresholds} Description of the different criteria used to separate star-forming vs. quenched galaxies.}
\end{table*}

\subsubsection*{Specific Star Formation Rate threshold}
As often adopted in the literature, we define as ``quenched'' to be those galaxies whose logarithmic sSFR is lower than a given fixed value at any redshift, namely: sSFR $\leq 10^{-11} \, \rm yr^{-1}$. Star-forming galaxies are thus those with sSFR larger than this threshold.
This fixed criterion does not account for the evolution in time of the SFR-$\MS$ plane \citep{2012Whitaker,2014Whitaker,2015NLee,2016Tomczak,2018BLee}, and it well describes the sSFR distributions only in a relatively narrow range of redshift. Yet, such threshold-based separation is often used in observations \citep{2011Mcgee,2013Wetzel,2014Lin,2018Jian}, albeit with somewhat different choices for the threshold value, since it roughly corresponds to the minimum of the bimodal distribution of galaxies' sSFRs \citep[see][and Section \ref{sec:bimodality} of this work for a more detailed discussion]{2004Kauffmann,2012Wetzel}.


\subsubsection*{Distance from the Main Sequence}
A more robust alternative to separate ‘quenched’ vs. ‘star-forming’ galaxies, and thus to identify the MS, involves an iterative algorithm. 
In practice, we stack all galaxies in 0.2-dex bins of stellar mass, measure the median SFR in the bin, and label ‘quenched’ (‘star-forming’) those galaxies whose SFR value falls below (above) a certain relative distance from the median SFR at the corresponding mass. We repeat this procedure by retaining at every step only star-forming galaxies until the median SFR in the bin is converged to a certain accuracy, thus defining the MS of star-forming galaxies. Unless otherwise stated, this is applied to galaxies up $\MS = 10^{10.2} \, \Ms$, otherwise the locus of the MS at every step is obtained by linearly extrapolating Eq.~(\ref{ms}) from lower masses. 
We adopt two relative, mass-dependent distances from the MS to separate between quenched and star-forming galaxies: $2\sigma$ and 1 dex from the median SFR value, following e.g. \cite{2018Fang}. Namely, quenched galaxies are those whose SFR (log SFR) is below $2\sigma$ (1 dex) from the median SFR at the corresponding galaxy stellar mass. 
 
\subsubsection*{Time-averaged SFR threshold}
The star formation histories of galaxies exhibit large variations across their lifetimes. For this reason, similarly to \cite{2016Mistani}, we estimate the time-averaged SFRs of individual galaxies across their merger-tree main branch -- $\overline{\rm SFR(t)}$ -- and define a galaxy to be ``quenched'' if its SFR drops below this value at the time of observation. 
Therefore we use this method to mostly compare to previous theoretical works, without arranging for any comparisons with observations. 


\subsubsection*{Color-color UVJ diagram}
Most of the selection criteria described above are based on the position of galaxies on the SFR-$M_{\rm stars}$ plane.
A powerful tool for selecting quiescent galaxies that is extensively used in observations is the color-color diagram, particularly the rest-frame U-V vs.V-J plane \citep[hereafter UVJ diagram; ][]{2011Wuyts,2009Williams,2010Williams,2010Whitaker,2012Patel,2012Quadri}.

An empirical criterion based on the number density distribution of galaxies on the UVJ diagram is usually adopted to divide the sample into red/quiescent vs. blue/star-forming galaxies: galaxies located approximately on the top-left portion of the UVJ diagram are dubbed ``quenched'', and star-forming otherwise. The slope and the position of the separating diagonal cut in the UVJ diagram are fine-tuned to fall on the break of the bimodal number distribution of galaxies \citep{2009Williams}. Since the rest-frame color distribution could be slightly different from one galaxy sample to another, some authors adjust arbitrarily the cut to better separate the two peaks of the galaxy number density in their sample \citep{2011Whitaker,2013Muzzin,2014tomczak}. We adopt a similar strategy for TNG galaxies: see next Section.

Stellar light of simulated galaxies is determined following \cite{2018Nelson} i.e. adopting the FSPS stellar population synthesis code \citep{2009Conroy, 2010Conroy, 2014ForemanMackey} with the Padova isochrones, MILES stellar library, and assuming a Chabrier initial mass function \citep{2003Chabrier}. The effects of dust attenuation are accounted for by including unresolved and resolved dust (Model C of \citealt{2018Nelson}, with the redshift dependencies outlined there and adopted across the $z=0-2$ range), in one random galaxy projection in the simulated volume. A galaxy's spectrum is the sum of the stellar population spectra of all star particles in a given aperture and is convolved with the rest-frame broad-band filters U, V, and J centered at 365, 551, and 1220 nm, respectively, to obtain U-V and V-J colors.

\subsubsection{Fiducial measurement choices}
\label{sec:fiducial}

Unless otherwise stated, we elect to use as fiducial the following measurement choices: we use SFRs averaged over 200 Myr, obtained from all stars within twice the 3D spherical stellar half-mass radius, and with a separation of quiescent vs. star-forming galaxies based on the TNG UVJ diagram.


\begin{figure*}
\centering
\includegraphics[width=0.49\textwidth]{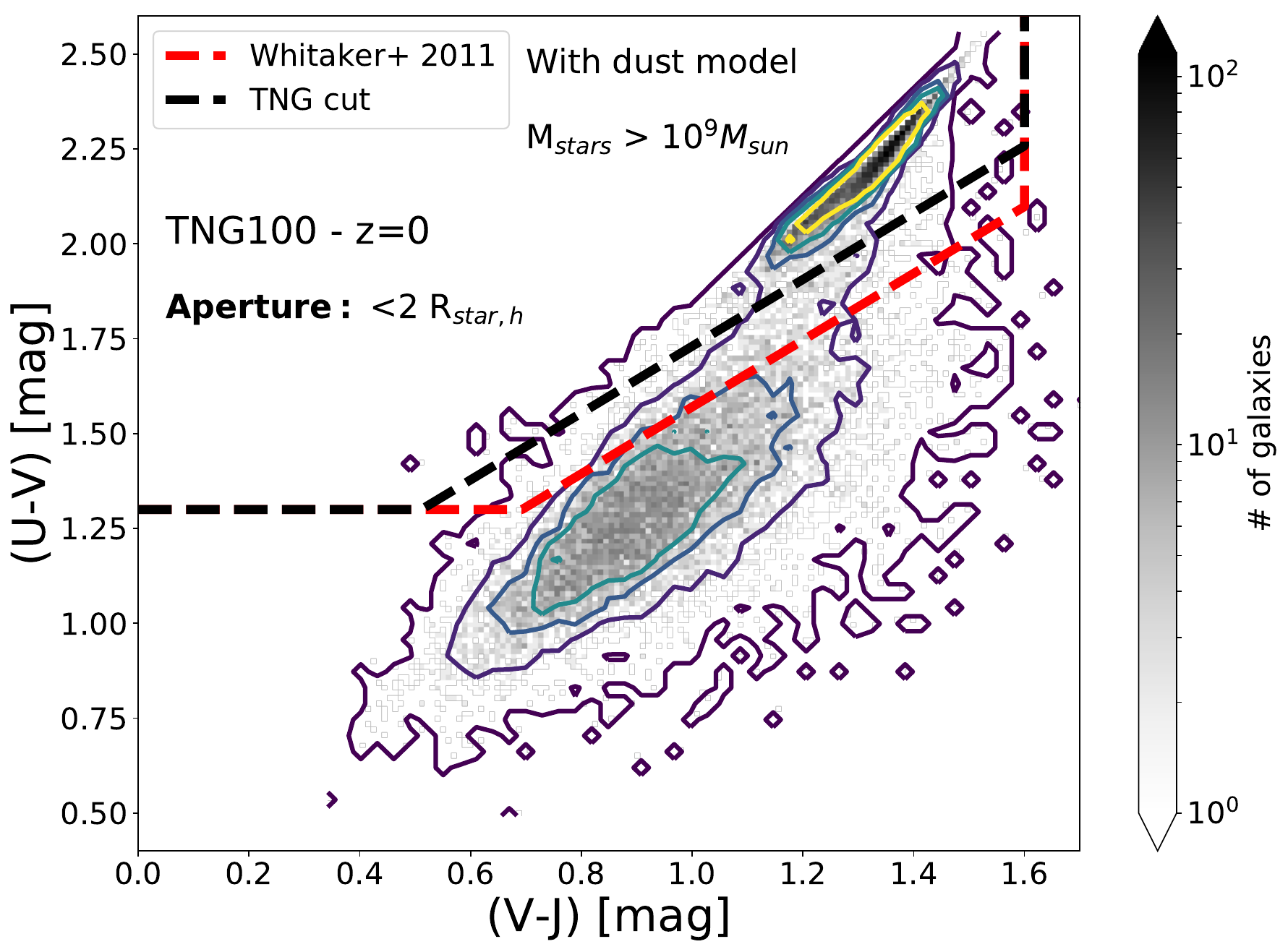}
\includegraphics[width=0.49\textwidth]{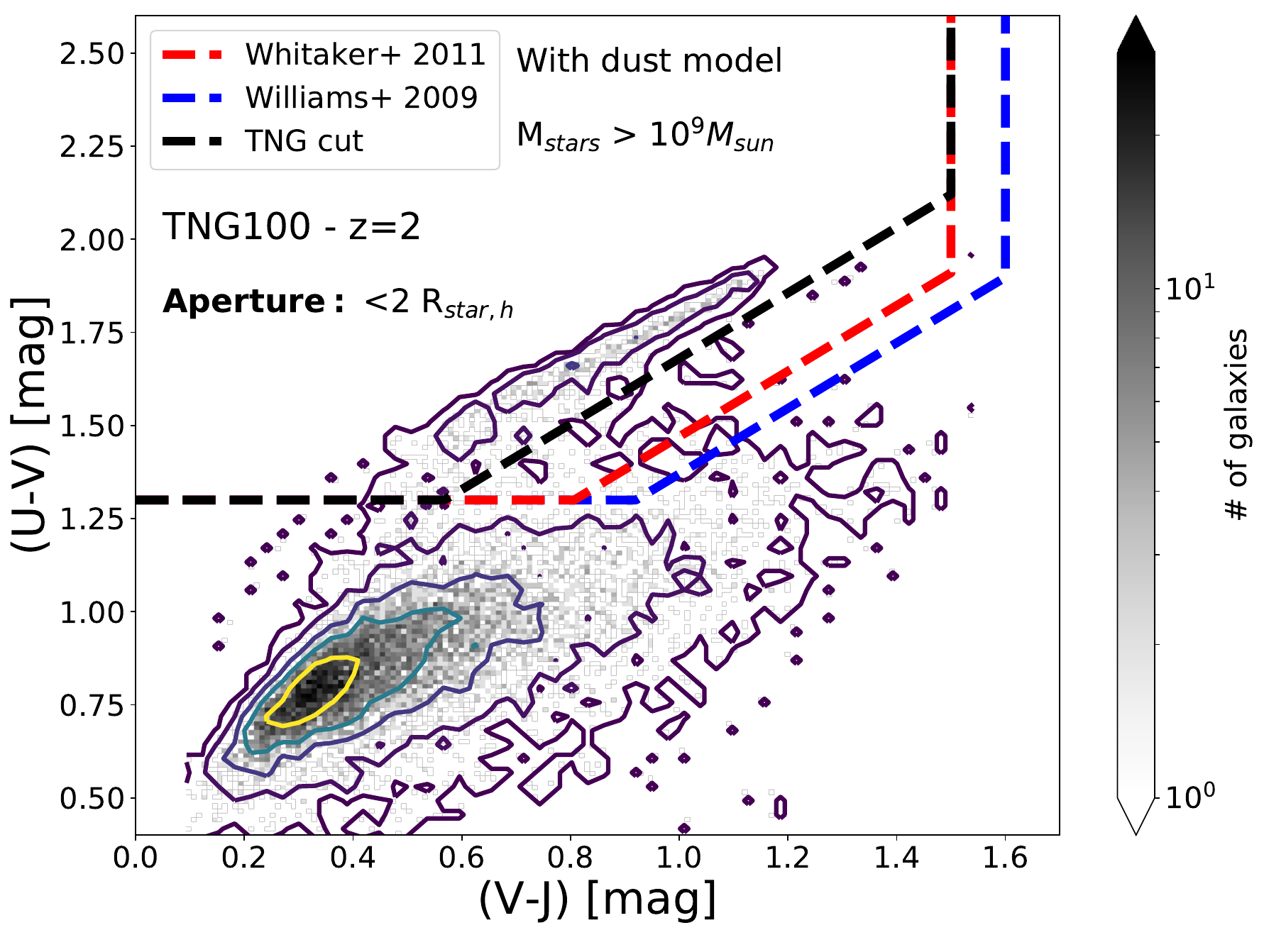}
\caption{\label{fig:UVJ} UVJ diagram for TNG100 galaxies at $z=0$ (left panel) and $z=2$ (right panel). We divide all galaxies in 150x150 bins of the plane (V-J)-(U-V).
The color scale indicates the number of galaxies in each bin and the solid contours encompass 20, 50, 70, 90, and 99 $\%$ of the galaxies. The black dashed line is the cut based on the number density distribution of TNG galaxies. For comparison, two observational thresholds are shown, defined by \citet{2011Whitaker} (red dashed lines) and by \citet{2009Williams} (blue dashed line at $z=2$).}
\end{figure*}


\section{Results}
\label{results}

\subsection{The UVJ diagram of TNG galaxies}
\label{sec:uvj-sfr}

We start the presentation of the TNG results by examining the simulated color-color diagram of TNG galaxies. Fig.~\ref{fig:UVJ} shows the UVJ diagram at $z=0$ (left panel) and $z=2$ (right panel) by modeling the effects of dust extinction as discussed in \cite{2018Nelson}.
We divide all galaxies in 150x150 bins on the (V-J)-(U-V) plane: the color scale indicates the number of galaxies in each bin and the solid contours encompass 20, 50, 70, 90, and 99 per cent of the simulated galaxies with $\MS\ge 10^9\Ms$. 

In both panels, a clear separation between two groups of galaxies is visible. This is the case in terms of galaxy number density and it appears more emphasized at high redshift. The existence of two classes of galaxies in the TNG color-color is a validation of the model in comparison to observational findings. It provides further support to the separation into star-forming vs. quiescent galaxies based on UVJ cuts that we adopt throughout the paper. 

In fact, the red dashed line in Fig.~\ref{fig:UVJ} is the separating threshold to select star-forming vs. quenched galaxies adopted by \cite{2011Whitaker}. The blue dashed line denotes the analog separation adopted by \cite{2009Williams}, which deviates from the red one only at high redshifts. Fig.~\ref{fig:UVJ} demonstrates that, despite the separating red and blue dashed lines are based on observed galaxy samples and observed colors, they can be reasonably well applied also to TNG galaxies on the UVJ diagram and could be adopted to separate quite successfully red and quenched galaxies (in the upper part) from blue and star-forming ones (in the lower portion of the diagram). 

Yet, since the TNG galaxies populate the UVJ diagram in a manner which is broadly consistent with but not identical to the observed ones, to distinguish between star-forming and quenched galaxies, we use the same slope of the separating threshold provided by \cite{2011Whitaker} and we adjust its position to better separate the bimodal number-density distribution of TNG galaxies.

The black dashed line in Fig.~\ref{fig:UVJ} denotes the best estimate to separate TNG galaxies in the UVJ diagram based on the depicted number density distributions and is described by the following equations: 

\begin{eqnarray}
\label{TNG_cut}
(U-V) > 0.88 \, (V-J) + 0.85  &\; [z \leq 1] \\
(U-V) > 0.88 \, (V-J) + 0.80  &\; [1<z<2] \nonumber \, .
\end{eqnarray}

In addition, for the limits on U-V and V-J we arbitrarily use the ones defined in \cite{2011Whitaker}:

\begin{eqnarray}
\label{TNG_cut2}
(U-V) > 1.3 \, ; &\; (V-J) < 1.6 \; & [0 < z \leq 1.5] \nonumber \\
(U-V) > 1.3 \, ; &\; (V-J) < 1.5 \; & [1.5 < z \leq 2] \\
(U-V) > 1.2 \, ; &\; (V-J) < 1.4 \; & [2 < z < 3.5] \nonumber \, .
\end{eqnarray}

Throughout this work, we use the TNG UVJ cut as our fiducial choice and compare our results, whenever necessary, to those obtained with other color cuts widely used in the literature.
We note that, while Fig.~\ref{fig:UVJ} reports the distribution of TNG100 galaxies, a consistent picture emerges from the distribution of TNG300 galaxies, with enhanced statistics but lower numerical resolution. Moreover, it may be useful to point out that the effect of dust in the UVJ plane is to shift galaxies more or less along the diagonal portion of the threshold dashed lines. However we note that, in both TNG100 and TNG300, we do not reproduce the full range of colors that is seen in real galaxies at intermediate redshifts. Specifically, e.g. at $z=2$, the TNG model does not produce the very dusty star-forming galaxies that occupy the upper right corner of the UVJ diagram in the observed samples and for which the vertical boundary is needed. This is similar to what found in the MUFASA simulation \citep{2017Dave} and requires future and further analysis.

\begin{figure*}
\centering

\includegraphics[width=\textwidth]{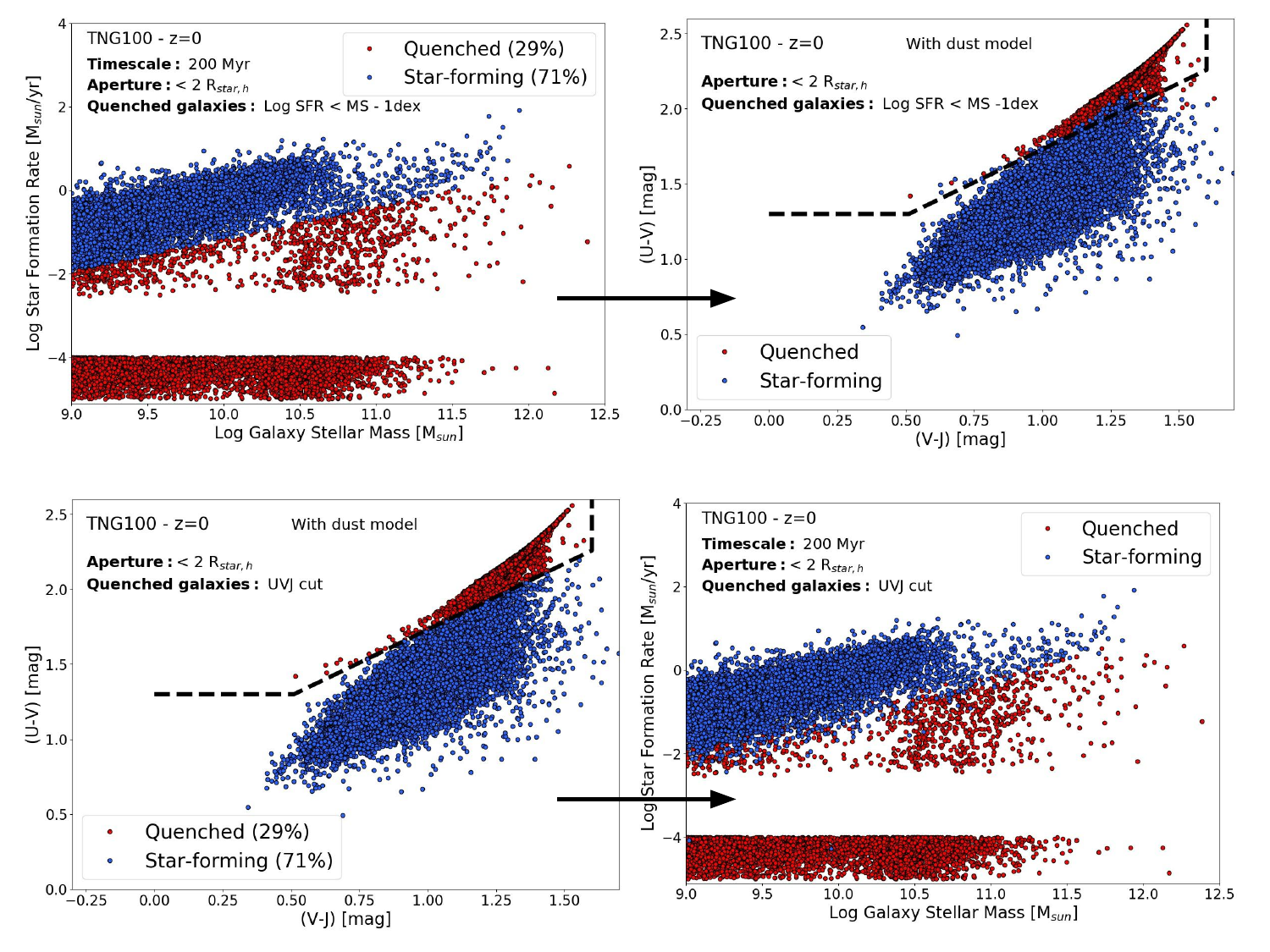}
\caption{\label{fig:SFR-UVJ} Comparison between SFR-based and UVJ color-based criteria to distinguish between quiescent vs. star-forming galaxies. Top: SFR as a function of galaxy stellar mass (left panel) and UVJ diagram (right panel) of TNG100 galaxies at $z=0$. In both panels, we identify quenched galaxies (red circles) if their SFR is below 1 dex from the MS. All the other galaxies are identified as star-forming (blue circles).
Bottom: the same as in the top but for galaxies selected using the TNG UVJ cut indicated with a black dashed line.}
\end{figure*}

\subsection{TNG quenched fractions across times, masses, and operational definitions}
\label{sec:qfracts}

The majority of the star-forming vs. quenched classification criteria adopted in this work can be grouped into two general classes: some are based on the position of each galaxy on the SFR-$\MS$ plane and hence its distance from the star-forming MS; others are based on the position of each galaxy on the color-color diagram. Here, we present a quantitative comparison among these two operational definitions of quenched galaxies. The goal is to understand whether or not SFR-based and UVJ-based classifications lead to quantitatively consistent results in terms of the shape and evolution of the MS and of the amount and evolution of quenched galaxies. 
Specifically, in Fig.~\ref{fig:SFR-UVJ} we examine how our color-color selection for quenched galaxies compares to other SFR-based selections, and vice-versa.

The top left panel of Fig.~\ref{fig:SFR-UVJ} shows the SFR-$\MS$ plane of TNG100 at $z=0$ in which galaxies are color-coded according to their SFR: red/quenched if they are below 1 dex from the MS, blue/star-forming if they are above.

As we discuss more in detail in Appendix~\ref{appendix}, all galaxies with SFR below the minimum resolved values will be assigned to the range $10^{-5} - 10^{-4} \, \Ms \rm \, yr^{-1}$. Even if their actual distribution is  unknown, those galaxies would distribute well below the MS, thus populating the quenched region.

The top right panel shows the UVJ diagrams for the same galaxies, with the same color coding adopted from the definition applied in the left panel. The black dashed line is the separating TNG color-color cut (Eqs. \ref{TNG_cut} and \ref{TNG_cut2}).
A good agreement between the two criteria is manifest: nearly all galaxies identified as blue/star-forming according to their SFRs, in the top left panel, settle below the black line in the top right panel.

This statement holds also in reverse. The bottom left panel of Fig.~\ref{fig:SFR-UVJ} shows the UVJ diagram for TNG100 galaxies, at $z=0$. In this case, the galaxies are color-coded according to their position with respect to the separating threshold (dashed black line). Then, with the same color-code, we show the position of these galaxies in the SFR-$\MS$ plane (bottom right panel). 
As in the previous case, galaxies identified as blue/star-forming accordingly to their color in the bottom left panel, populate the region above Log SFR $>$ MS-1dex in the bottom right panel.

These findings are a further confirmation that the UVJ colors of galaxies correlate well with their SFRs; practically, the simulated UVJ diagram provides a reliable tool to separate galaxies in star-forming and quenched populations by returning results that are overall consistent with SFR-based selections where quiescent galaxies lie 1 dex below the MS. This is in agreement with observational findings by e.g. \cite{2018Fang}. 
In fact, although in the upper panels some star-forming galaxies pass the UVJ color threshold and, vice versa, in the lower panels a fraction of galaxies that are identified as star forming according to their colors turn out to have a quite low SFR. In the whole mass range considered here ($\MS \ge 10^9\Ms$), the fraction of quenched galaxies reads 29 per cent with both criteria. 
Such an excellent quantitative agreement between the two operational definitions of quenched galaxies supports the choice of a UVJ cut tailored to the number density distribution of TNG galaxies.

In fact, even if not shown in Fig.~\ref{fig:SFR-UVJ}, we have repeated the same comparison using the separating UVJ threshold provided by \cite{2011Whitaker} instead of the TNG one. In this case, we find that, at $z=0$, the SFR-based selection leads to a smaller fraction of quenched galaxies in comparison to the UVJ selection, with 7 per cent of galaxies in the studied mass range being dubbed differently by the two different UVJ cuts. While such mismatch is not negligible, all the results obtained with the \cite{2011Whitaker} observational cut are still in qualitative agreement with TNG's, providing a further qualitative validation of the model. 

\begin{figure*}
\centering
\includegraphics[width=0.85\textwidth]{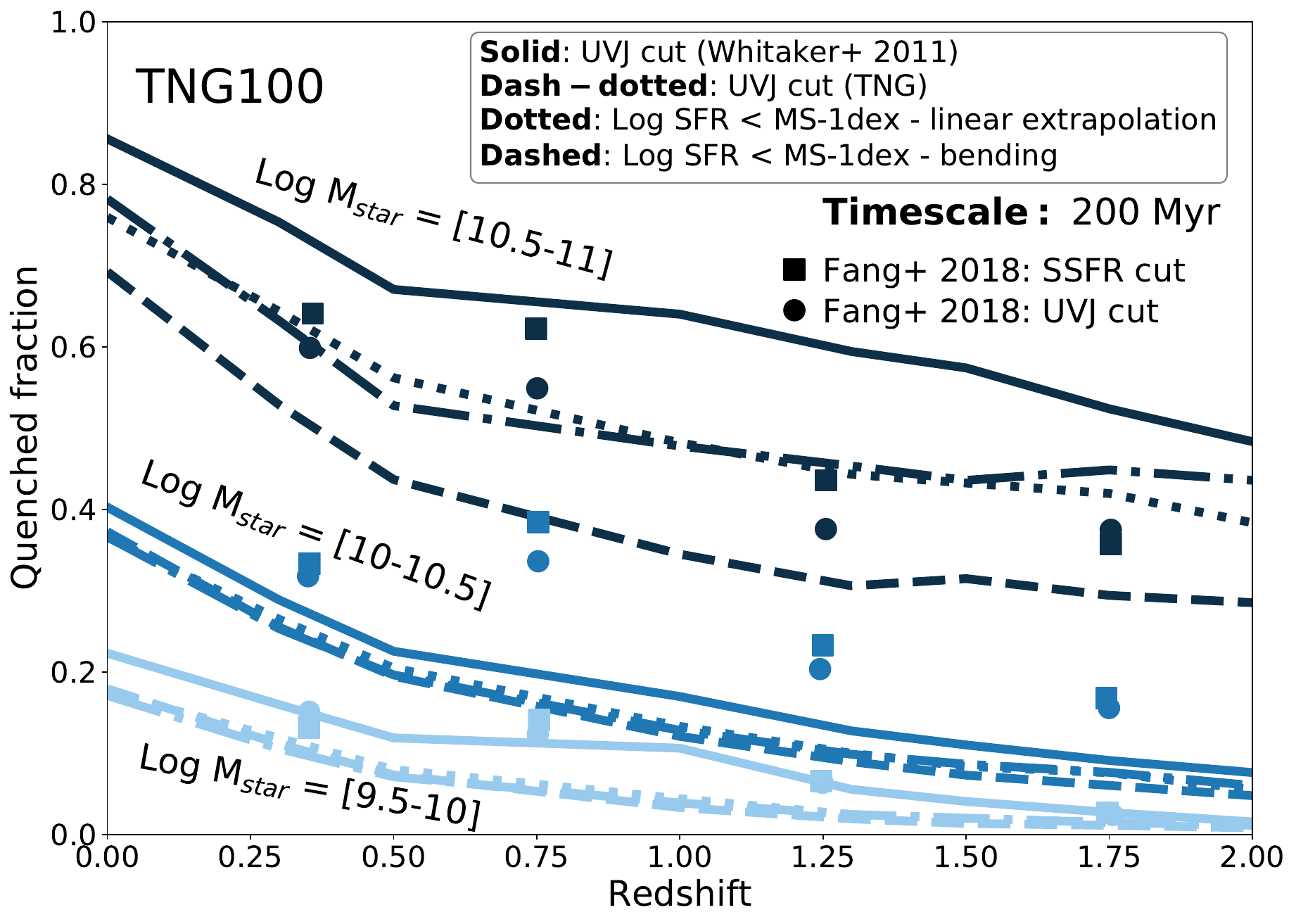}
\includegraphics[width=\textwidth]{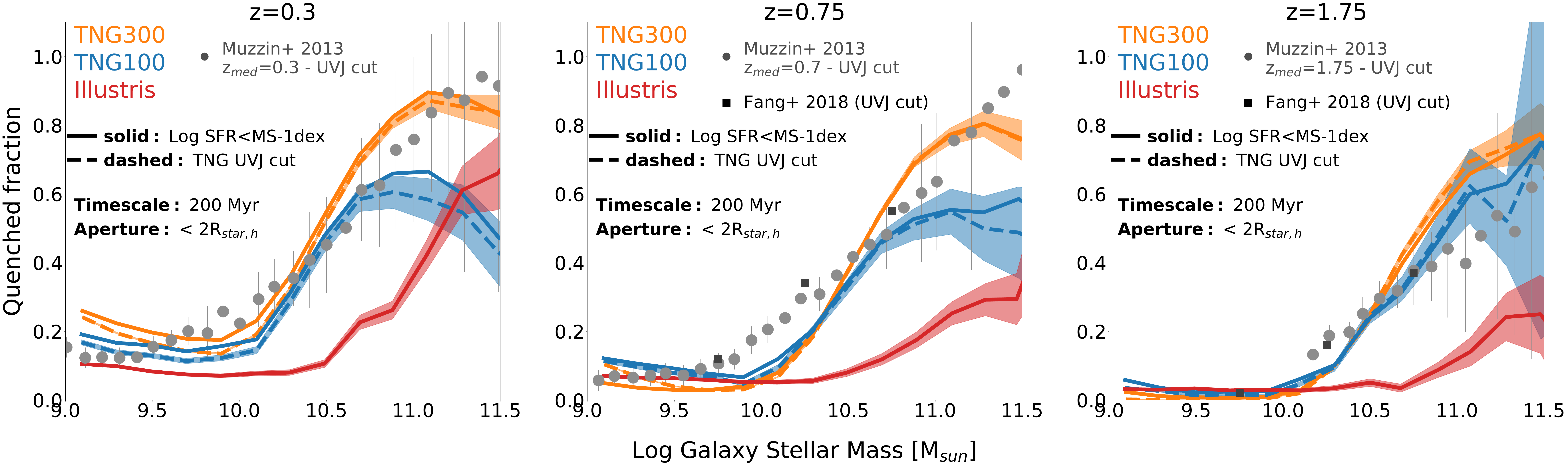}
\caption{\label{fig:Qfrac_methods} Top panel: fractions of quenched galaxies, in three bins of stellar mass (indicated with different colors), as a function of redshift for TNG100. Quenched galaxies are selected in four ways: via the UVJ cut adopted by \citet{2011Whitaker} (solid), via the TNG UVJ cut (dash-dotted) and according to their SFR values with respect to the locus of the star-forming main sequence, i.e. by both extrapolating the linear MS to the high-mass end (dotted) and by using a possibly-bending MS (dashed). Observed quenched fractions of \textit{CANDELS} galaxies \citep{2018Fang} are indicated as filled symbols (squares for the sSFR selection and circles for the UVJ selection). The color-code indicates the stellar mass bins, as in TNG.
At the highest-mass end probed here, different quenched definitions may return quenched fractions that differ by up to 10-20 percentage points.
Bottom panels: fractions of quenched galaxies (centrals and satellites) as a function of stellar mass at $z=0$ (bottom left), $z=0.75$ (bottom central), $z=1.75$ (bottom right), for TNG300 (orange curves), TNG100 (blue curves) and Illustris (red curves). Quenched galaxies are selected in two ways: via the TNG UVJ cut (dashed) and via the linear extrapolation of the MS at the high-mass end (solid). For all curves, we include the effect of 0.2 dex and 0.3 dex uncertainties on $M_{\rm stars}$ and SFRs, respectively. Shaded areas in all panels indicate the Poissonian error associated to the UVJ selection for TNG models and to SFR selection for Illustris.
A selection of observational data are indicated as gray symbols. TNG models return a higher quenched fraction than Illustris, at any redshift.}
\end{figure*}
In the top panel of Fig.~\ref{fig:Qfrac_methods}, we quantify how different selection criteria impact the assessment of the fraction of quenched galaxies across galaxy stellar mass and redshifts. We provide quenched fractions in bins of stellar mass, (indicated with different colors, light blue for low mass galaxies, $10^{9.5}-10^{10}~ \Ms$, blue for intermediate galaxies, $10^{10}-10^{10.5}~ \Ms$ and dark blue for the more massive ones, $10^{10.5}-10^{11}~ \Ms$), in the redshift range $0\le z \le 2$. 
We select quenched galaxies according to their UVJ colors and their SFR values. For the UVJ cut, we assume both the one provided by \cite{2011Whitaker} (solid curves) and the one based on the TNG diagram i.e. Eqs.~\ref{TNG_cut} and \ref{TNG_cut2} (dash-dotted curves). For the SFR-based selection where galaxies are quiescent if their SFR is 1 dex below the star-forming MS, we use both the linear extrapolation of the MS beyond $10^{10.2}\Ms$ (dotted curves) as well as we allow the MS to bend at the high-mass end (dashed curves: see Section~\ref{MS and scatter} for the definitions and Section \ref{sec:bending} for a detailed discussion on the bending of the MS). In all cases, we account for all galaxies that are resolved according to the adopted  stellar mass limit ($\MS\geq10^{9.0}\, \Ms$), hence including in the galaxy counts also those with very low values of SFR.

In general, we find that the fraction of quenched galaxies increases from $z=1.75$ to $z=0.3$, with a larger increment of TNG300 quenched galaxies at larger masses, being about 70-80 per cent at $z=1.75$ and reaching more than 90 per cent at $z=0.3$.
Moreover, at all redshifts, the quenched fractions are higher for galaxies in the $10^{10.5} <\MS \, (\Ms)< 10^{11}$ mass range with respect to the less massive ones, the latter featuring quenched fractions below about 20 per cent at all times.

For galaxies in the stellar mass range $\MS=10^{9.5}-10^{10.5} \, \Ms$, the fractions of quenched galaxies via the two SFR-based criteria are perfectly consistent with one another: dashed vs. dotted curves. As we will see in Section \ref{general properties}, in this mass range the MS is approximately linear. 
Interestingly, in the same stellar mass range, the excellent agreement previously mentioned between the UVJ-based selection -- with the TNG cut -- and the SFR-based one (Fig. \ref{fig:SFR-UVJ}), holds even at high redshifts, as suggested by the overlap among dash-dotted, dotted and dashed curves.
Instead, the UVJ cut of \cite{2011Whitaker} leads to a systematically higher quenched fraction, even though the amplitude of such a discrepancy is less than 0.05 
at all redshifts. 

Let us now focus on more massive galaxies: $10^{10.5} \le \MS \, (\Ms) \le 10^{11}$.
In this mass regime, the TNG UVJ selection leads to fractions of quenched galaxies that are very similar to the ones found with the linear extrapolation of the MS: $\sim 80, 50$, and $40-45$ per cent at $z=0,1,$ and $2$, respectively. A fraction that is systematically lower by 10 percentage points is instead recovered when quiescent galaxies are selected by using a curved MS. As in the case of low-mass galaxies, the UVJ cut provided by \cite{2011Whitaker} turns into higher quenched fractions with respect to the TNG, of about $\sim 10, 20$ and $< 10$ percentage points $z=0, 1,$ and $2$, respectively. 

A similar comparison between UVJ and SFR selections has been made by \cite{2018Fang} (filled symbols in the top panel of Fig.~\ref{fig:Qfrac_methods}), who have examined the fraction of quenched $\textit{CANDELS}$ galaxies, in bins of stellar mass, at $0.2<z<2.5$.
A qualitative and a somewhat quantitative agreement in terms of general redshift and mass trends and results is manifest, barring perhaps the $10^{10-10.5}\Ms$ mass range. However we do not push further for any conclusion, since $\textit{CANDELS}$ SFRs are obtained via UV-optical spectral energy distribution fitting including specific light-to-SFR calibrations we do not mock \footnote{Note that in the \cite{2018Fang} analysis, selection effects are necessarily in place: the CANDELS sample here utilized is selected based on a magnitude limit (H = 24.5) and on the quality of the GALFIT fits. None of such cuts implies rejection of objects based on their SFR values and in fact none of them biases the values of quenched fractions used for comparison in this paper. Similar arguments apply to the samples of \citet{2013Muzzin}.}. 

In the bottom panel of Fig.~\ref{fig:Qfrac_methods} we show the fractions of quenched galaxies (without distinguishing between centrals and satellites) as a function of stellar mass at $z=0.3$ (bottom left), $z=0.75$ (bottom central) and $z=1.75$ (bottom right), for TNG300 (orange), TNG100 (blue) and Illustris (red).
For the TNG models we adopt both the TNG UVJ-based selection (dashed) and the SFR-based selection (linearly-extrapolated MS, solid) to identify quenched galaxies, whereas for Illustris we only use the SFRs-based criterion. In order to take into account observational uncertainties on galaxy stellar masses and SFRs, we add a random Gaussian error to the simulation data, with a width of 0.2 dex for the stellar mass and 0.3 dex for the SFRs.

At any redshift, we find that the fractions of quenched galaxies is systematically higher in TNG than in Illustris: this actually brings the TNG model into better agreement with observational constraints. Observed quenched fractions of COSMOS/UltraVISTA galaxies \citep{2013Muzzin} are indicated as gray filled circles in the bottom panels of Fig. \ref{fig:Qfrac_methods}, where quiescent galaxies are defined via the UVJ cut provided by \cite{2011Whitaker}. For comparison we additionally show the results from \cite{2018Fang} as black filled squares at $z=0.75$ and $z=1.75$ consistently to what is reported on the top panel.
As seen above for the simulations, also for observational results different quenching definitions might cause a variation in the values of the quenched fractions at different redshifts. For example, in \cite{2016darvish} quiescent galaxies are selected via a NUV-r-J cut: although at lower redshift the quenched fractions are in a quite good agreement with other observations \citep[e.g.][]{2013Muzzin,2013moustakas}, at higher redshift -- namely $z \sim 1.75$ -- \cite{2016darvish} provide a fraction of about 15-20 per cent lower than \cite{2013Muzzin}.

Because of the new AGN feedback recipes implemented in the TNG model \footnote{The SMBH model in IllustrisTNG has been significantly updated in comparison to the original Illustris simulation. In particular, in the low-accretion mode regime, the bubble thermal heating feedback implemented in Illustris has been replaced with a kinetic wind. Moreover, SMBHs are seeded at larger masses and a BH mass dependent threshold is used to prevent low-mass SMBHs to enter the low accretion state \cite[see][for more details]{2018Weinberger}. 
The new TNG kinetic feedback at low accretion rates has been demonstrated to be responsible for the quenching of galaxies residing in massive haloes \citep{2018Weinberger} and to improve the galaxy color bimodality in comparison to SDSS \citep{2018Nelson}.}, quenching mechanisms are more efficient in TNG than in Illustris, especially for galaxies more massive than $10^{10} \, \Ms$, for which the fraction of quenched galaxies can be different by up to about 50 percentage points between the two simulations. This is more apparent at higher redshift, $z>0$: the quenched fractions in Illustris reach $\gtrsim$ 20 per cent only for galaxies more massive than $10^{11} \, \Ms$, whereas both TNG100 and TNG300 return fractions in the range 30-80 per cent in the stellar mass range $M_{\rm stars} =10^{10.5}-10^{11.5} \, \Ms$.

However, because of resolution effects, differences between TNG100 and TNG300 are clearly noticeable: at lower resolution the fractions of quenched galaxies are larger. Particularly, at $z=0$ we recover a fraction of about 10-20 per cent higher in TNG300 with respect to TNG100, in the whole mass range, whereas at $z=0.75$, a difference of about 20 percentage points is apparent only for galaxies more massive than $10^{10.5} \, \Ms$.
Besides the mass resolution effects, the volume of the simulated box could play a crucial role. Indeed, TNG300 samples much denser environments with respect to TNG100, and many of the most massive object will be quite different.

In comparison to previous results from other cosmological models, e.g. EAGLE \citep{2015Furlong} and L-Galaxies \citep{2017Henriques}, TNG returns similar quenched fractions at low redshifts, but systematically higher ones at higher redshifts.
As opposed to the TNG model, the fraction of quenched galaxies more massive than $10^{10.5} \, \Ms$ predicted by EAGLE (where passive galaxies are selected via sSFR cut) at $z\sim 1$ and $2$, falls below the observed ones \citep[][and others]{2013Muzzin}, by $\sim$ 10-15 per cent.
Similarly, the passive fractions predicted by L-Galaxies at $z=0.86$ (where passive galaxies are identified with a U-B vs. $M_{\rm stars}$ cut) fall below constraints from COSMOS \citep[compilation by][]{2010Peng} by 15-20 per cent too, suggesting that quenching mechanisms are less effective in those models than what observations imply.


\begin{figure*}
\centering
\includegraphics[width=0.49\textwidth]{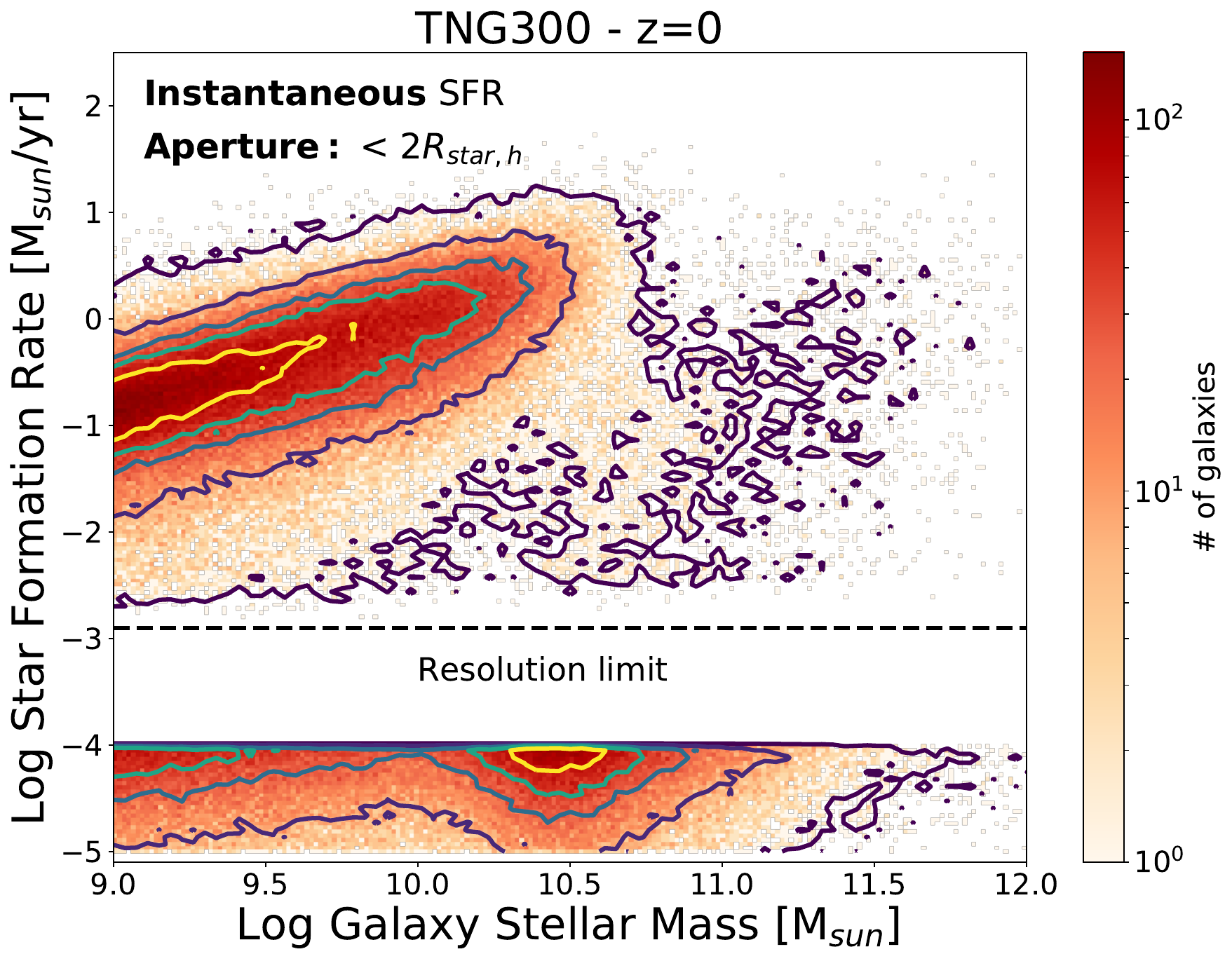}
\includegraphics[width=0.49\textwidth]{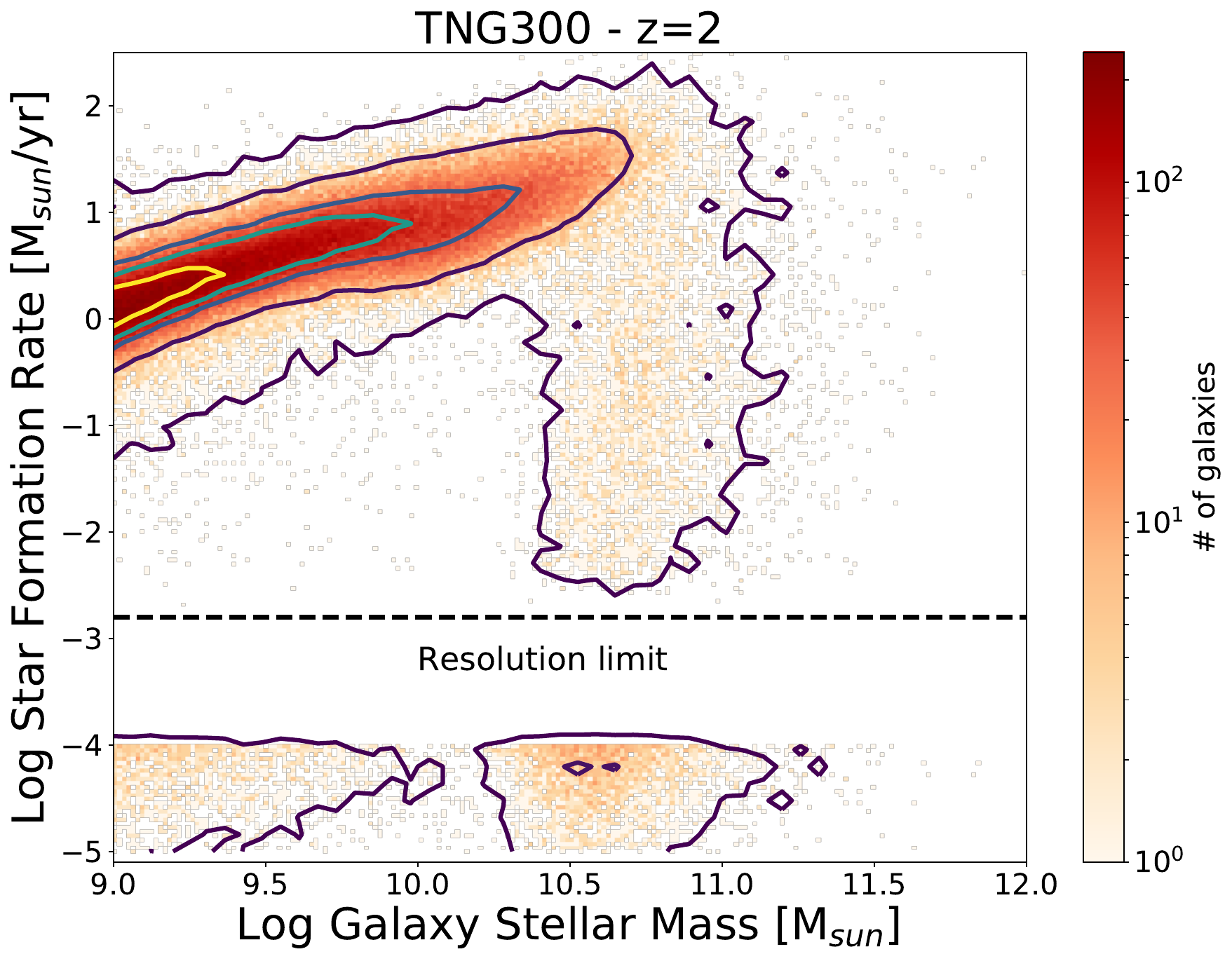}
\includegraphics[width=0.49\textwidth]{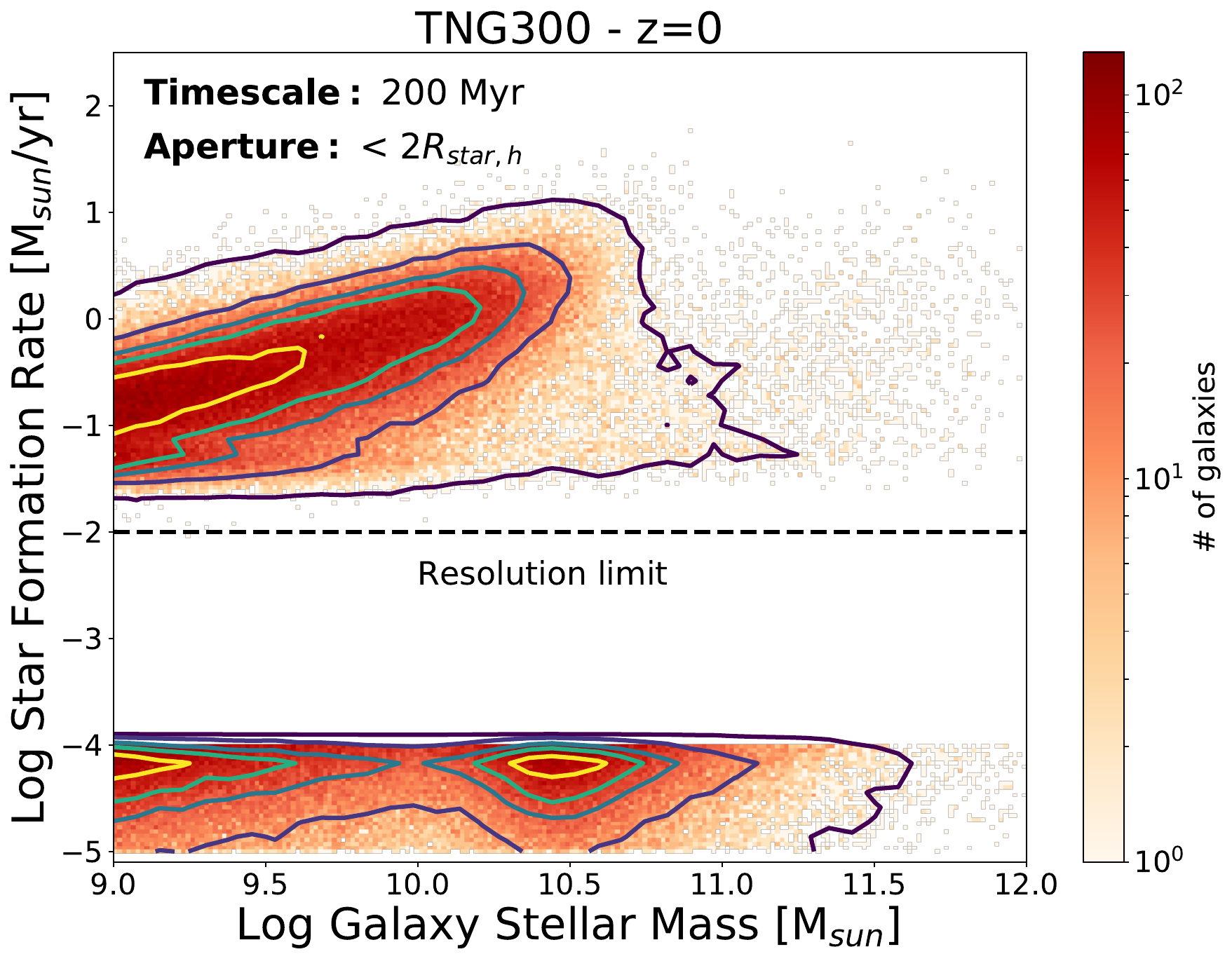}
\includegraphics[width=0.49\textwidth]{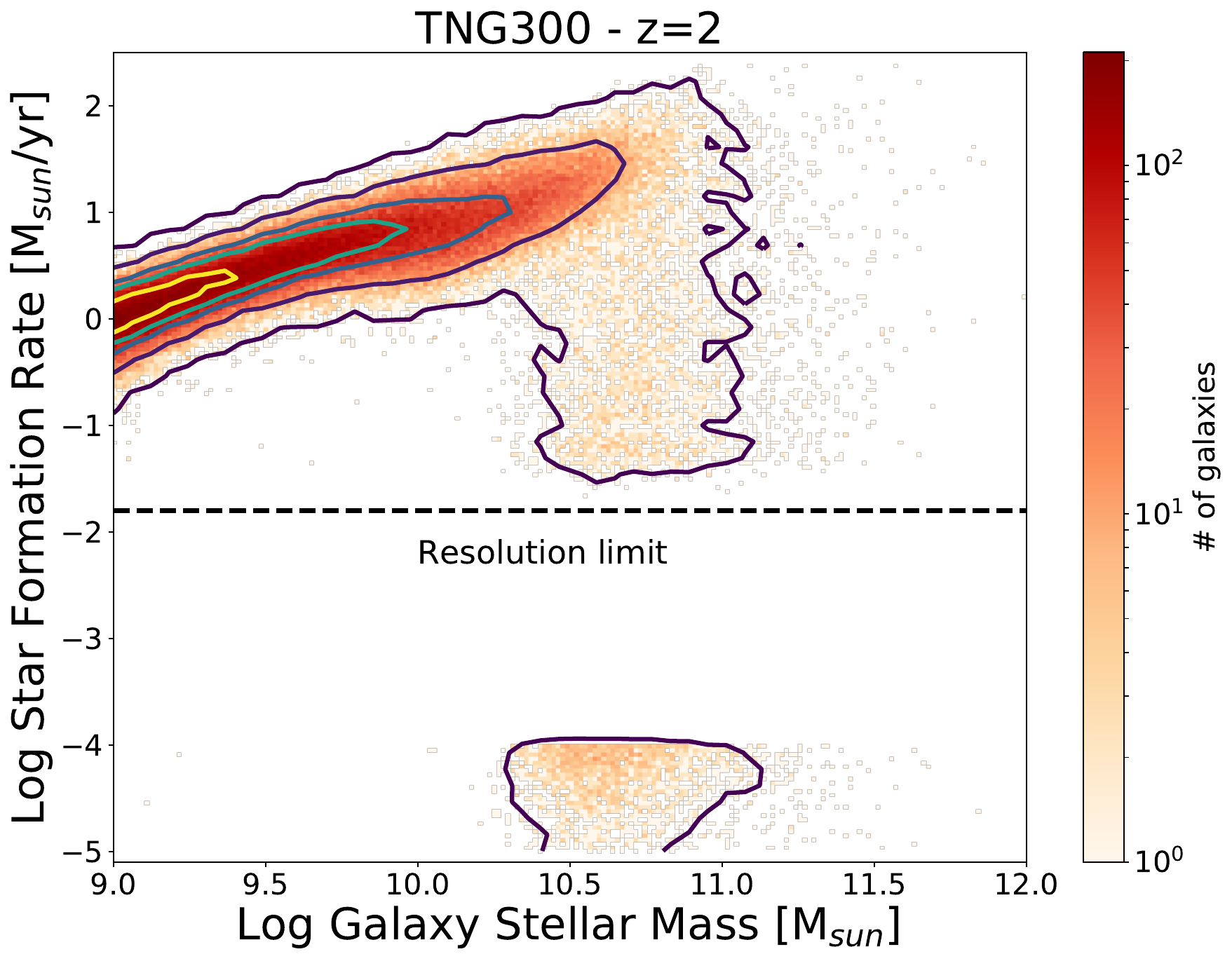}
\caption{\label{fig:SFR-MSTAR} SFR as a function of stellar mass for TNG300 galaxies, at $z=0$ (left panels) and $z=2$ (right panels). The color scale indicates the number of galaxies and the solid contours encompass 20, 50, 70, 90, and 99 $\%$ of the studied galaxies. To all galaxies with unresolved SFR values, we assign a random value in the range $\rm  SFR = 10^{-5} - 10^{-4} \, \Ms \rm \, yr^{-1}$.
Galaxy SFRs are measured within $2\times R_{\rm star,h}$ and are estimated from the instantaneous SFRs of the gas (top panels) and from the stars formed in the last 200 Myr (bottom panels).}
\end{figure*}

\subsection{The SFR-$\MS$ plane of TNG galaxies and their Main Sequence}
\label{general properties}

In this section, we present the general properties of the SFR-$\MS$ plane and characterize the MS of IllustrisTNG galaxies in the redshift range $0\le z \le 2$. 

In Fig.~\ref{fig:SFR-MSTAR}, we show the SFR-$\MS$ plane at $z=0$ (left column) and $z=2$ (right column) for TNG300 galaxies. Here, the SFRs are measured from the instantaneous SFRs of the galaxies' gas cells (top panels) and from the stellar mass formed in the last 200 Myr (bottom panels). In both cases, the physical aperture is $2\times R_{\rm star,h}$.
We divide all galaxies in 200x200 bins on the LogSFR $\times$ Log$\MS$ plane. The color scale in Fig.~\ref{fig:SFR-MSTAR} denotes the number of galaxies in each bin whereas the solid contours encompass 20, 50, 70, 90, and 99 $\%$ of our galaxy sample.

In the stellar mass range $10^9-10^{10.5} \, \Ms$, the SFR-$\MS$ plane exhibits a densely populated region in which galaxies settle: these form the well-known star-forming main sequence that is clearly visible in all four panels, i.e. at low and high redshifts, and for both the instantaneous and averaged SFRs. A smaller number of galaxies appears to fall off from the star-forming main sequence, by exhibiting lower SFRs than MS galaxies at the same stellar mass. Moreover, we find that $\sim 32$ per cent of galaxies have unresolved SFR values, due to the mass resolution limits of the simulation. 
As we discuss more in detail in Appendix~\ref{appendix}, the minimum resolved SFR value is $\rm Log \, SFR ~(\Ms yr^{-1}) \sim -2$, meaning that all galaxies with SFR below this limit, will be assigned to the range $10^{-5} - 10^{-4} \, \Ms \rm \, yr^{-1}$.


Furthermore, we find that the number of these galaxies with very low SFRs decreases to 3 per cent at $z=2$ (right panel), when they are mainly more massive than $>10^{10} \, \Ms$. Indeed, as suggested by observations, at high redshifts, although a quenched population is recovered, the SFR-$\MS$ plane is dominated by star-forming galaxies at all stellar masses \citep[see e.g.][]{2010Whitaker}. 

These general features are confirmed even when the mean SFR is measured over longer timescales (bottom panels). In this case, at $z=0$, the number of galaxies with unresolved SFR values is larger with respect to the top panel ($\sim 37$ per cent). We postpone a comparison among different timescales to Section~\ref{sec:MS_systematics} and a detailed discussion on the effects of numerical resolution to Appendix~\ref{appendix}.

We now focus on the properties of the main sequence, its evolution with time and with galaxy stellar mass.
Henceforth, to identify the MS, we refer to our \textit{fiducial choices} (see Section~\ref{sec:fiducial}): SFRs are averaged over 200 Myr and measured within a physical aperture of $2\times R_{\rm star,h}$; quenched and star-forming galaxies are separated using the TNG UVJ cut, or alternatively using the SFR-based selection ($\rm Log \, SFR < MS - 1 dex$), depending on the cases: throughout the paper we specify whether we adopt one criterion rather than the other.

\begin{figure*}
\centering
\includegraphics[width=0.85\textwidth]{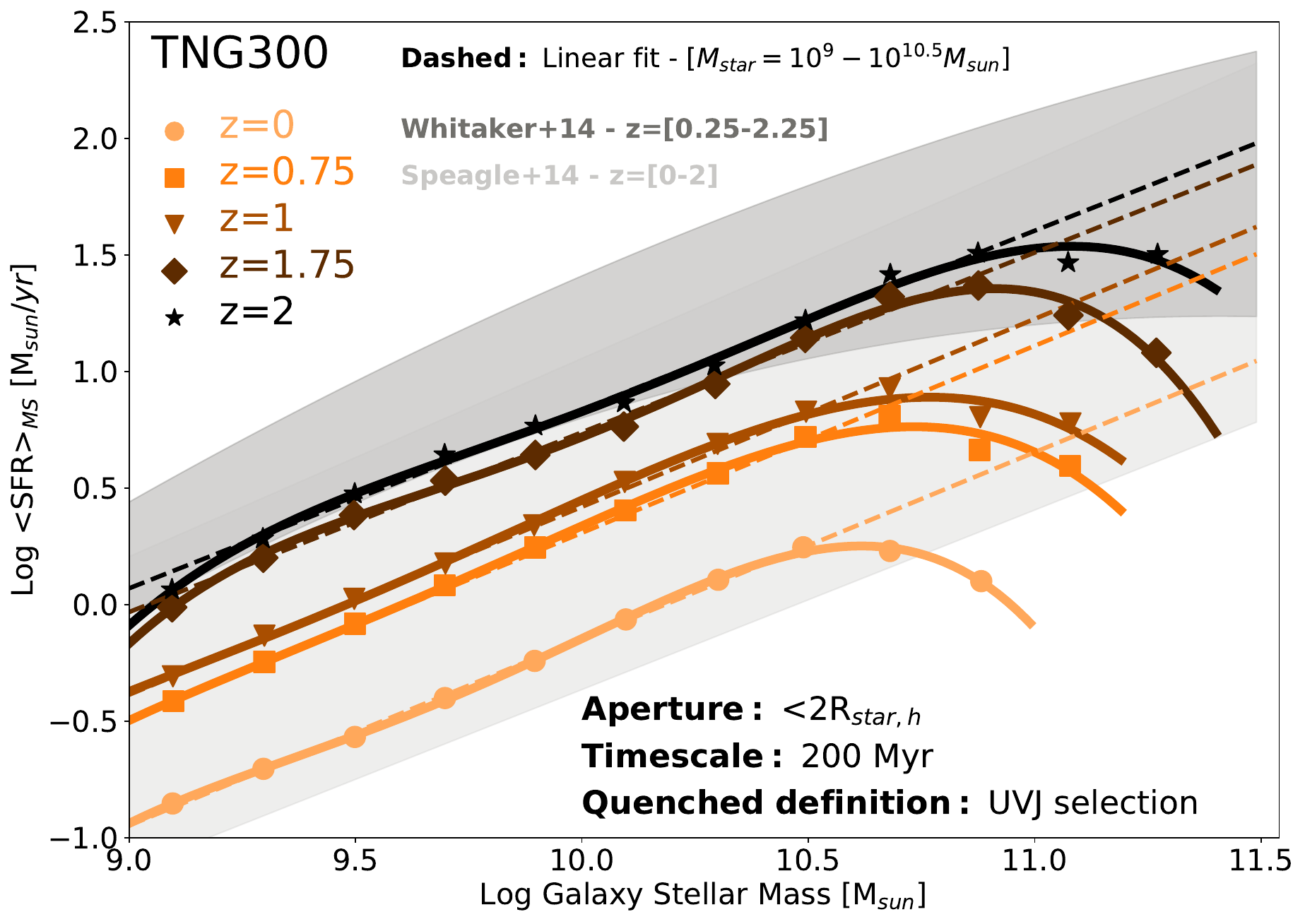}
\includegraphics[width=0.49\textwidth]{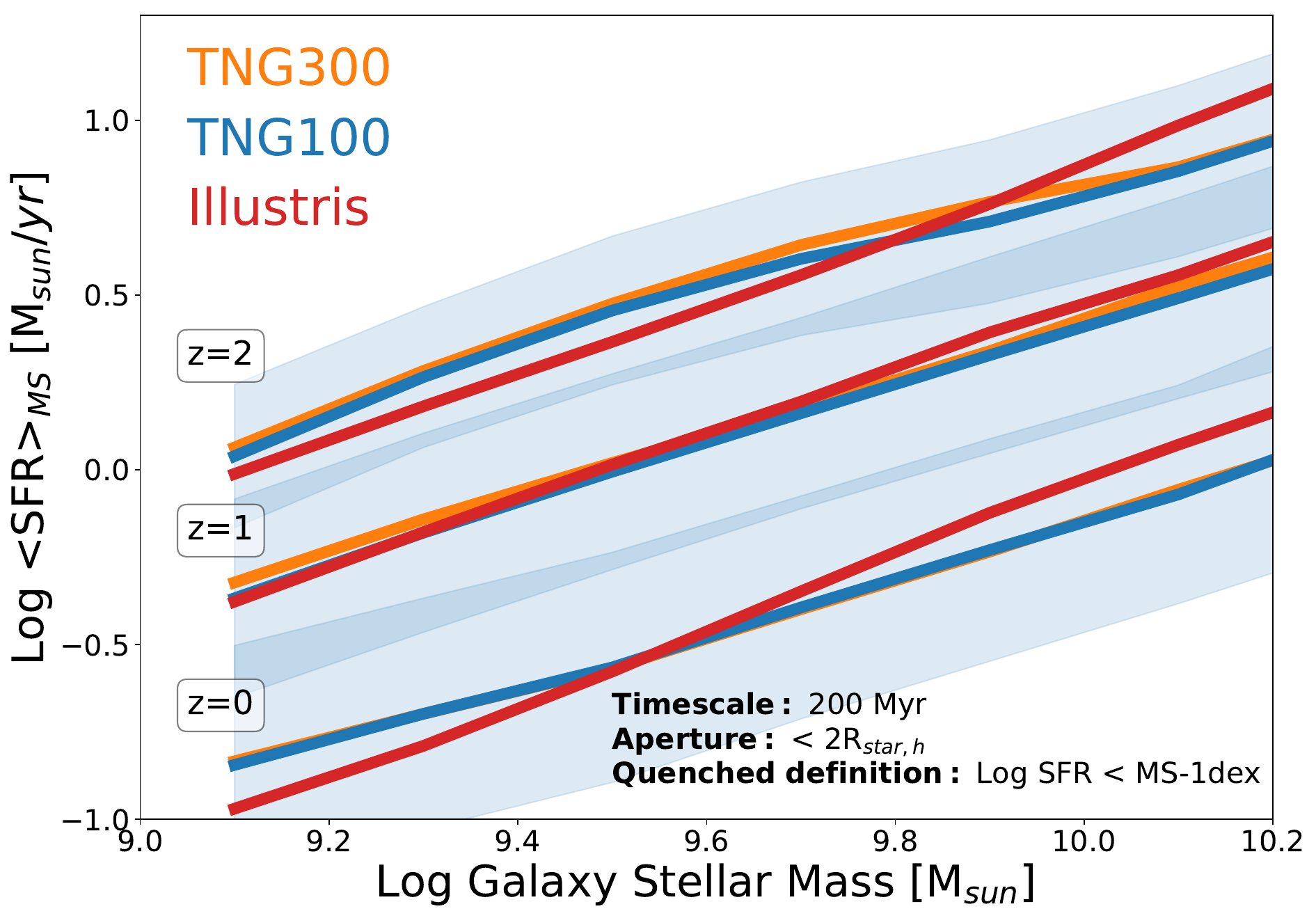}
\includegraphics[width=0.49\textwidth]{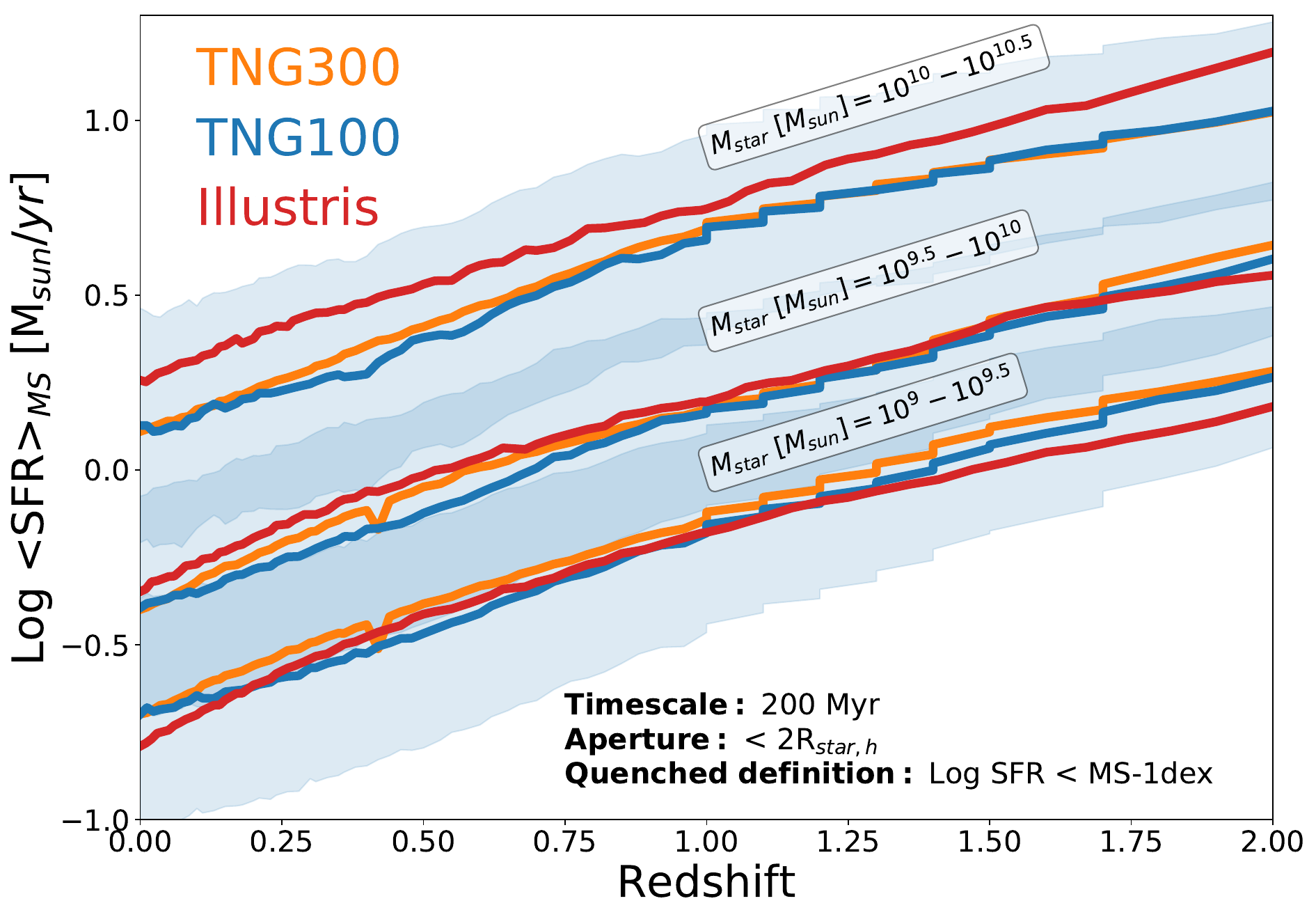}
\caption{\label{fig:SFMS} Main sequence of TNG star-forming galaxies, computed with the following measurement choice: the SFR from stars is averaged over 200 Myr and measured within $2\times R_{\rm star,h}$. Quenched galaxies are neglected in two ways: with a TNG UVJ cut (top panel) and with Log SFR$<$MS-1 dex (bottom panels).
Colors indicate different redshifts: $z=0,0.75,1,1.75,2$ (from light orange to black symbols in the top panel). Dashed lines are the linear fit performed in the mass range $\MS = 10^9-10^{10.5} \, \Ms$ (Eq. \ref{ms}). A clear deviation from a linear trend is evident for galaxies more massive than $10^{10.5} \, \Ms$. In the stellar mass range $\MS = 10^9-10^{10.2} \, \Ms$, the MS of TNG300 galaxies is compared to TNG100 (blue line) and Illustris (red line) at $z = 0, 1, 2$ (bottom left) and by stacking galaxies in three stellar mass bins (bottom right). The shaded areas indicate the 1$\sigma$ scatter of the MS in TNG100. An excellent agreement is noticeable between TNG models, whereas a steeper slope and a higher normalization at the high-mass end is recovered in the MS of Illustris galaxies.}
\end{figure*} 


The top panel of Fig. \ref{fig:SFMS} shows the MS of TNG300 galaxies at $z=0,0.75,1,1.75,2$, computed using the fiducial choices and selecting star-forming galaxies with the UVJ selection.
In addition, we show the results of a fourth order polynomial fit (solid line) and of a linear fit (dashed line) performed in the stellar mass range $\MS = 10^9-10^{10.5} \, \Ms$ (Eq. \ref{ms}), to better highlight the bending of the MS at the high-mass end. 

The relevant observational features of the MS are also recovered for TNG star-forming galaxies. Indeed we find that the normalization of the MS increases toward higher redshifts (from light orange to black line), regardless of stellar mass. 
Moreover, while for galaxies with $\MS < 10^{10.5} \, \Ms$, the MS can be considered linear, at the high-mass end it exhibits a bending, with a shallower slope.
The results of the linear fit as per Eq.~(\ref{ms}) are given in Table~\ref{tab:table_ms}, for reproducibility.

\begin{table}
\centering
\begin{tabular}{c|c|c}
\hline
Redshift & $\alpha$ & $\beta$ \\
\hline
\hline
0     & 0.80 $\pm$ 0.01  & -8.15 $\pm$ 0.11    \\
0.75  & 0.80 $\pm$ 0.01  & -7.72 $\pm$ 0.03  \\
1     & 0.81 $\pm$ 0.02 & -7.64 $\pm$ 0.17    \\ 
1.75  & 0.77 $\pm$ 0.02 & -6.97 $\pm$ 0.24    \\ 
2     & 0.77 $\pm$ 0.03 & -6.83 $\pm$ 0.25  \\
 
\hline
\end{tabular}
\caption{\label{tab:table_ms} Values of the coefficients from the linear fit for TNG300 MS (Fig. \ref{fig:SFMS}), performed in the stellar mass range $M_{\rm stars} =10^9-10^{10.5} \, \Ms$ (Eq.\ref{ms}). The MS is computed with the SFR from stars averaged over 200 Myr, measured within $2 \times R_{\rm star,h}$. Quenched galaxies are neglected if they are below 1 dex from the MS. Columns read: 1) redshift; 2) and 3) coefficients of the linear fit.}
\end{table}

The shaded grey areas in Fig.~\ref{fig:SFMS} represent the MSs of \cite{2014Whitaker} in the redshift range $0.25 < z < 2.25$ (dark gray area) and of a compilation of observational works taken by \cite{2014Speagle} (light gray area) in the redshift range $0 \le z \le 2$. 
The MS of TNG300 galaxies falls within the observational ball park: however,
making a comparison at face-value we find a systematic discrepancy between the MS in our model with respect to the observed values, particularly at high redshifts (see Section \ref{comparison} for more details).

In the bottom panels of Fig.~\ref{fig:SFMS} we show the MS as a function of galaxy stellar mass (left panel) and redshift (right panel), for TNG300 (orange line), TNG100 (blue line) and Illustris (red line) galaxies, in the stellar mass range $\MS =10^9- 10^{10.2} \, \Ms$.
Here, the MS is measured using the fiducial choices and neglecting galaxies with SFR values below 1 dex from the MS.
The shaded areas in both panels represent the 1$\rm \sigma$ scatter of the MS computed for TNG100.
In both panels an excellent agreement between TNG models is manifest: this demonstrates that the different mass resolution and statistical sampling of the two runs affect neither the slope nor the normalization of the MS, at any redshift.

On the other hand, the MS exhibits a steeper slope in Illustris than in the TNG models, particularly at $z=0$ and $z=2$ and at the highest-mass end (left bottom panel). These results expand upon the analysis provided by \cite{2015Sparre} for the Illustris simulation. The inability of Illustris feedback to quench the star formation in massive galaxies ($\MS > 10^{10} \, \Ms$) turns into a $\sim 0.1$ dex higher normalization with respect to TNG at all times (right panel). These results are in line with the findings of \textcolor{blue}{Hayward at al. (in prep)} who study the populations of submm galaxies in both TNG and Illustris at intermediate redshifts. The higher SFRs of Illustris galaxies at the high-mass end are most probably due to the different AGN feedback recipes implemented in Illustris and TNG and subsequently produce different fractions of quenched galaxies as a function of stellar mass, as we demonstrate in Fig. \ref{fig:Qfrac_methods} (see \textcolor{blue}{Donnari et al. in prep.} for a detailed discussion on this topic). 

In summary, the general properties of the star-forming MS of TNG galaxies and its trends with the stellar mass and redshift are in qualitatively agreement with observational findings. We postpone a more quantitative analysis to Section \ref{comparison}, after we quantify the effects of different measurement choices in the next Section.

\subsection{The effects of different measurement choices on the MS}
\label{sec:MS_systematics}


\begin{figure*}
\centering
\includegraphics[width=0.49\textwidth]{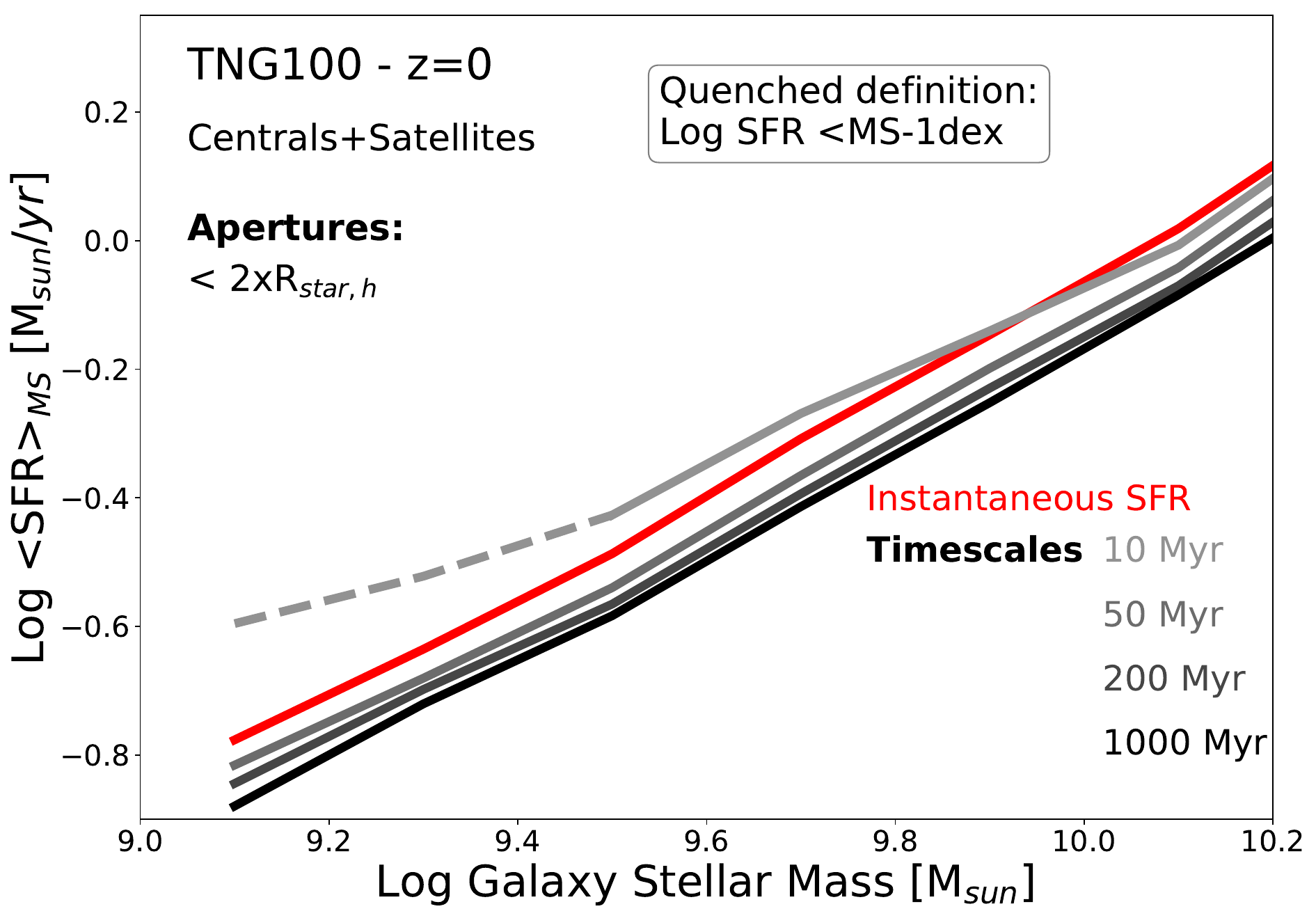}
\includegraphics[width=0.49\textwidth]{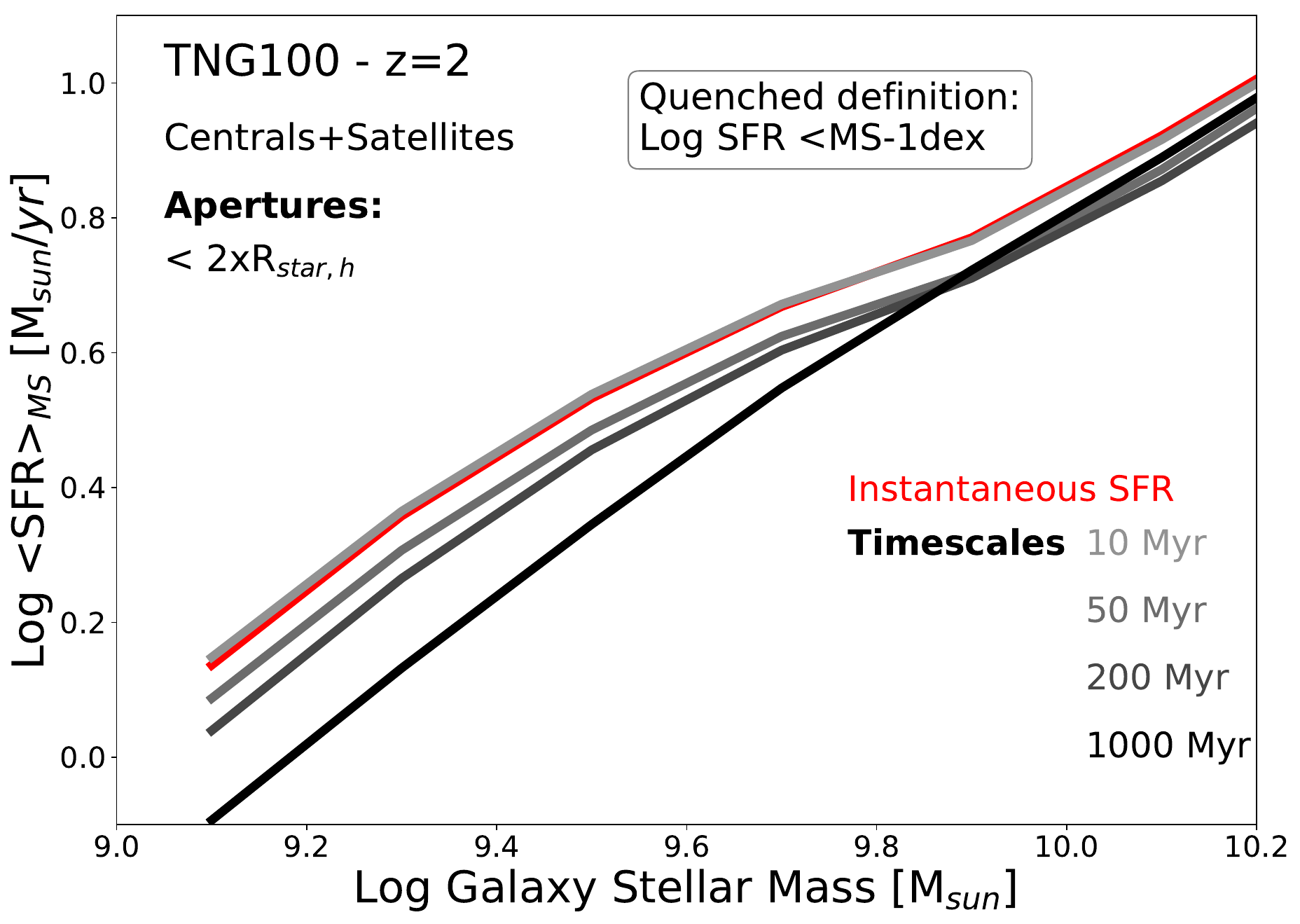}
\includegraphics[width=0.49\textwidth]{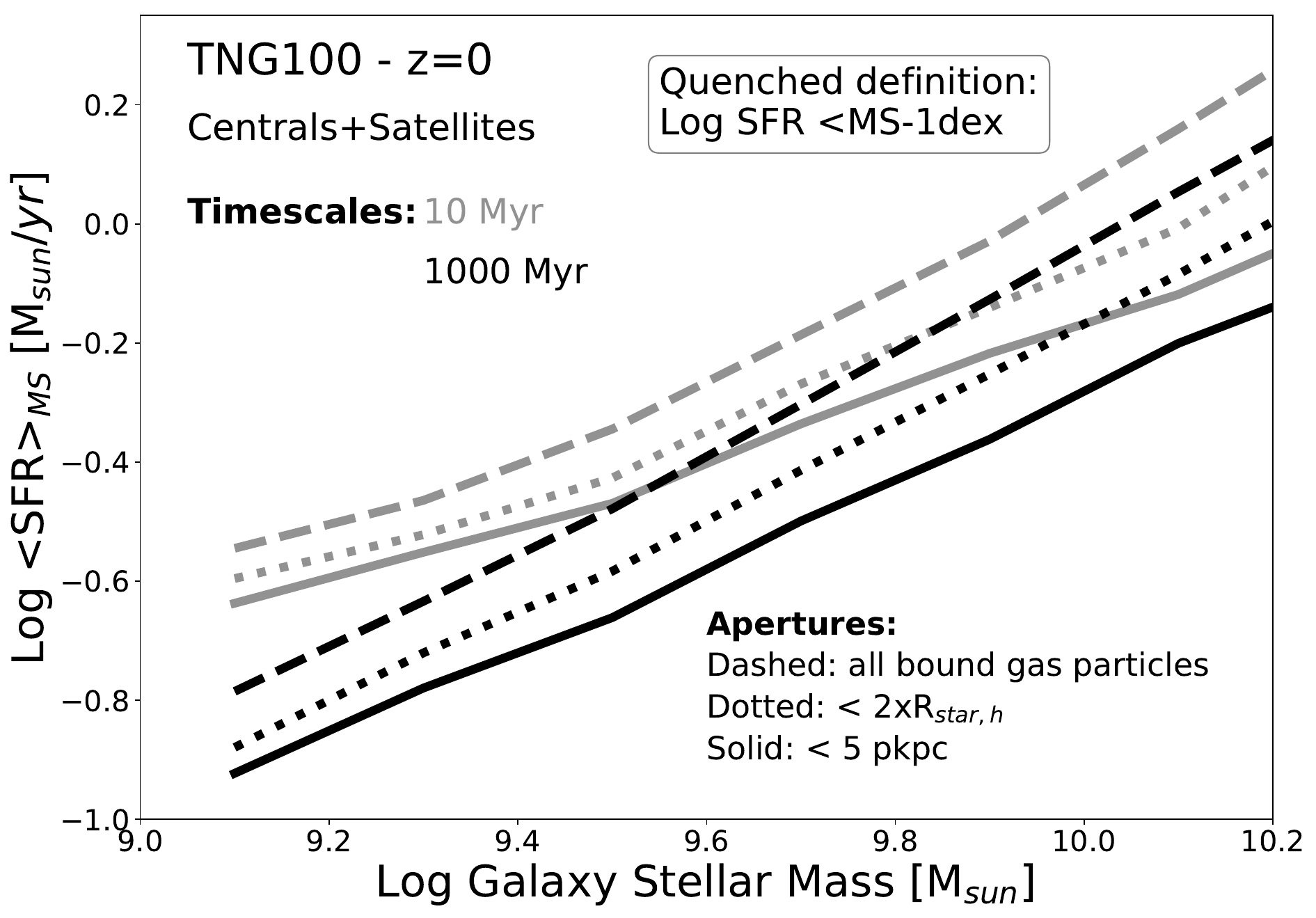}
\includegraphics[width=0.49\textwidth]{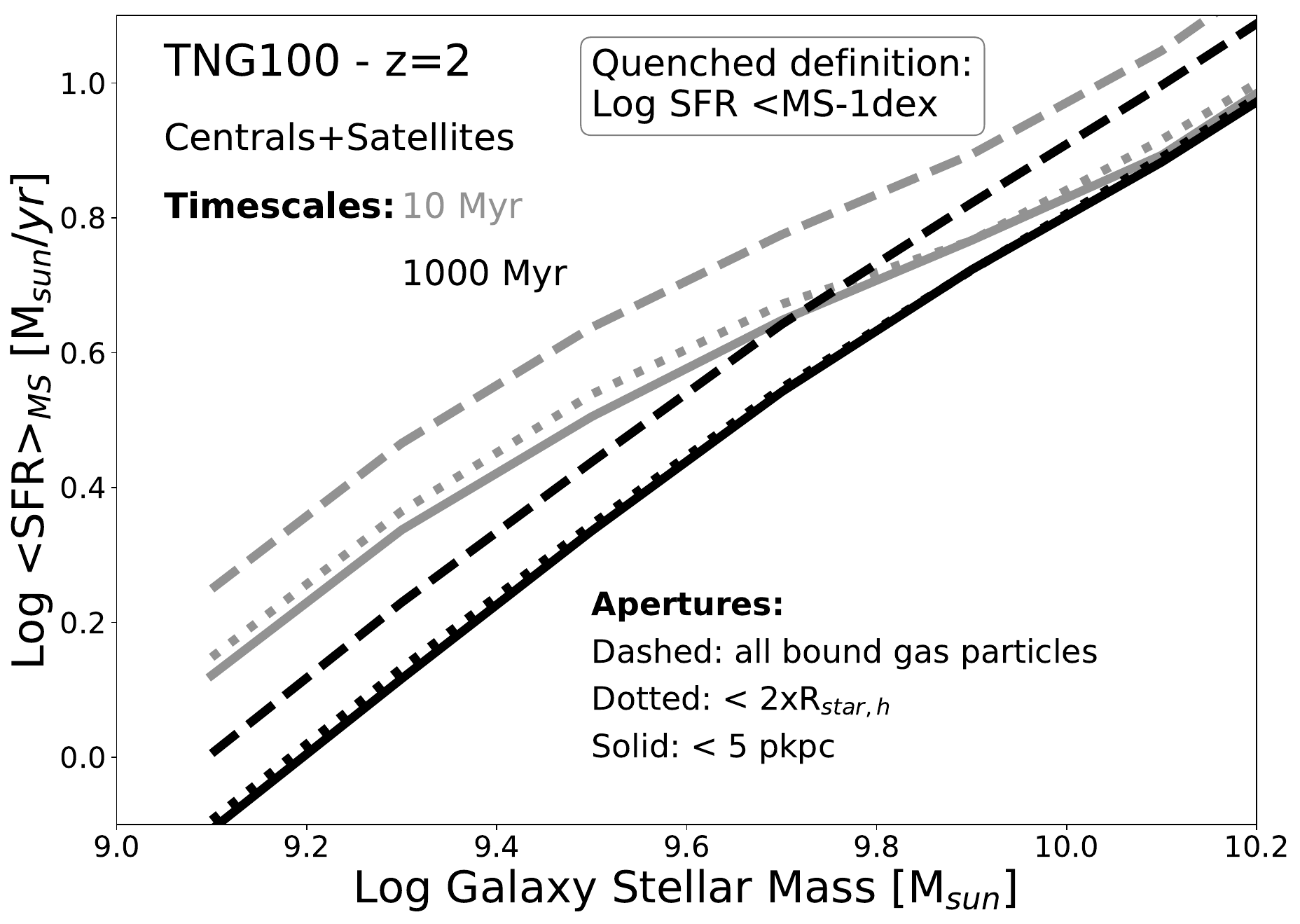}
\includegraphics[width=0.49\textwidth]{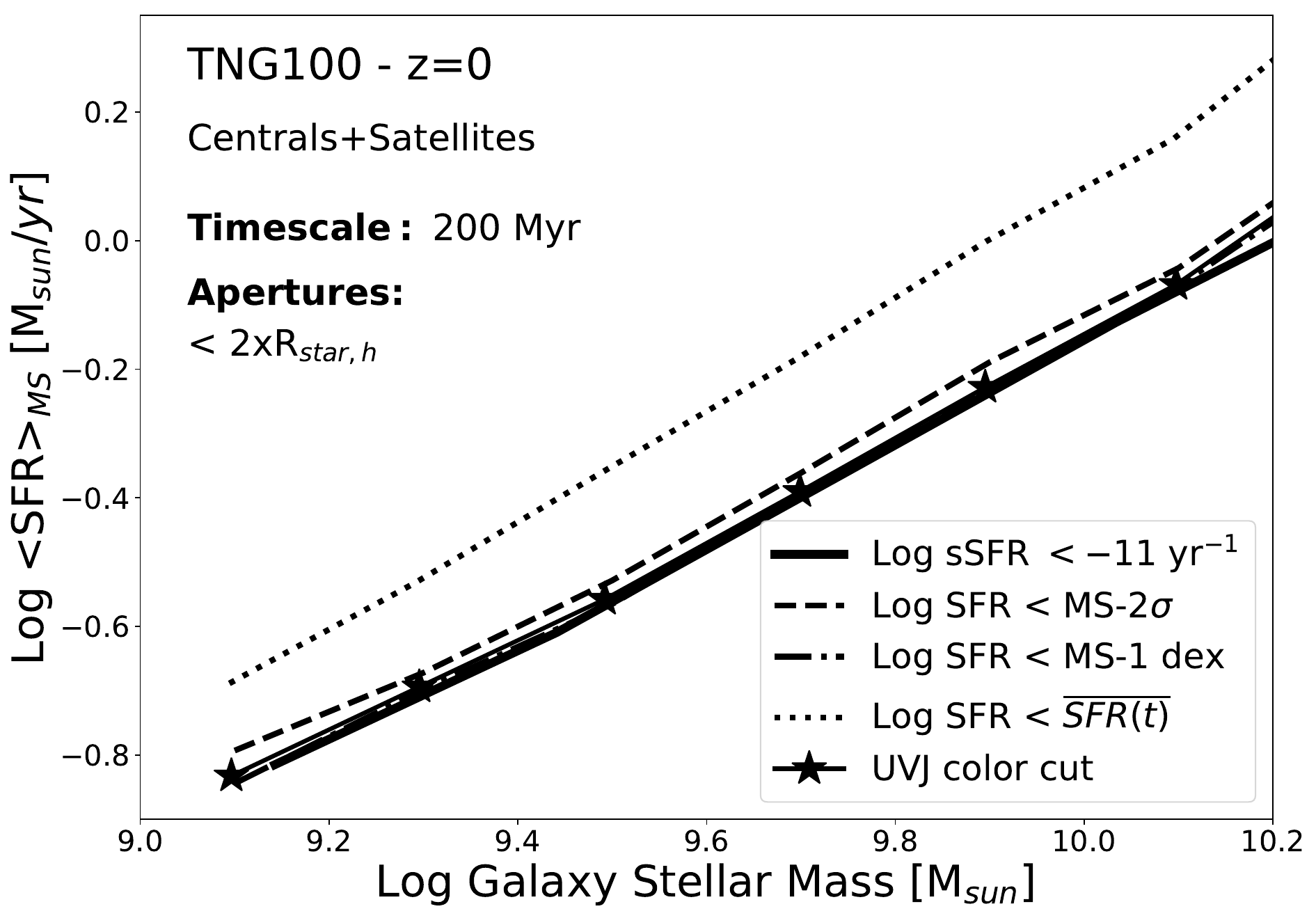}
\includegraphics[width=0.49\textwidth]{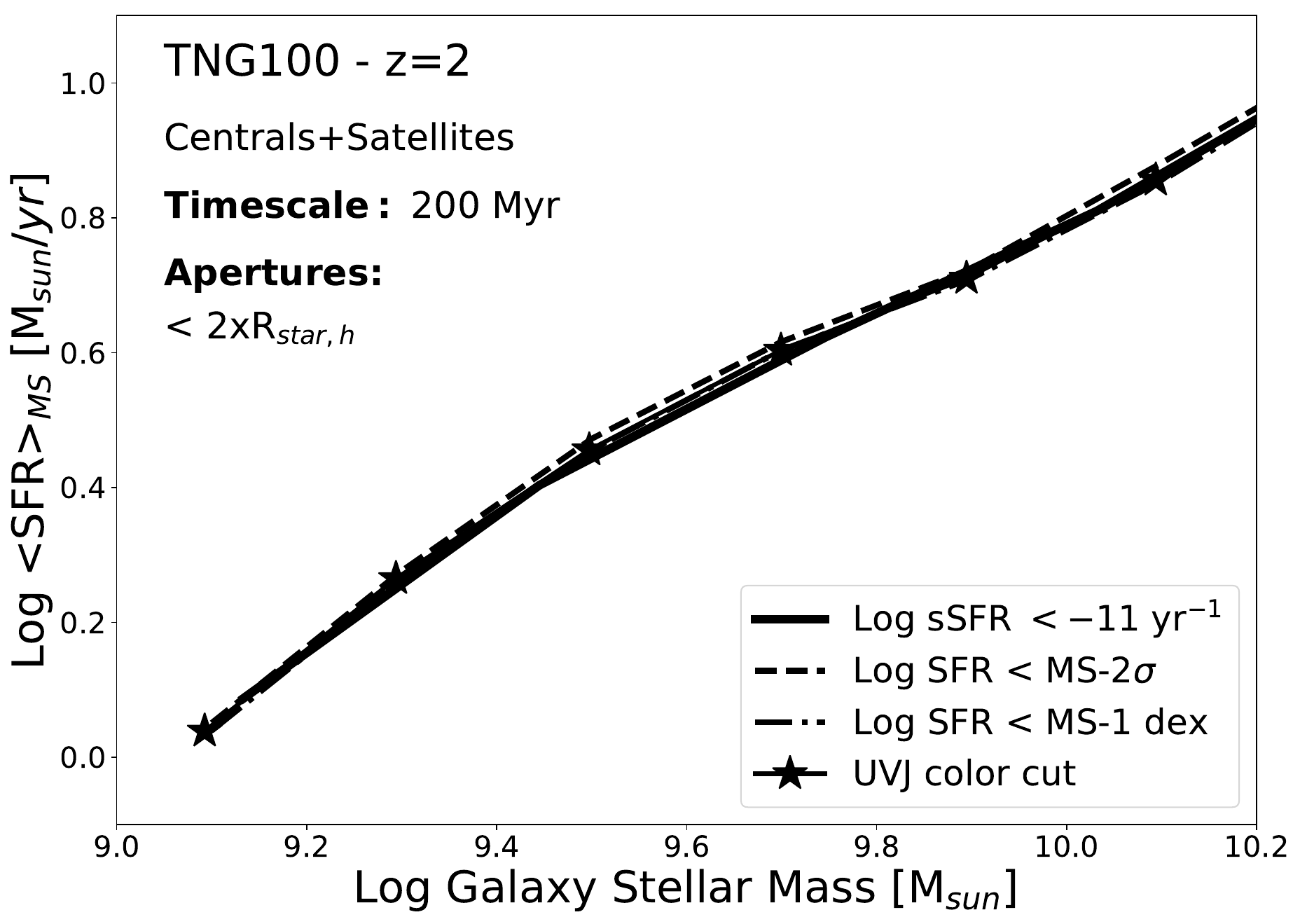}

\caption{\label{fig:systematics} Median of the MS in TNG100 at $z=0$ (left column) and $z=2$ (right column). Here we consider different systematics, as explained in the following.
Top panels: comparison among SFRs measured over different averaging timescales (from 10 Myr, gray curves, to 1000 Myr, black curves) and the instantaneous one (red curves). Here quenched galaxies have been removed if their SFR is below 1 dex from the MS. The larger deviation at the high-mass end in the top right panel (dashed part of 10 Myr curve) is likely an effect of numerical resolution.
Central panels: comparison across two SFR-averaging timescales (10 Myr, gray curves, and 1000 Myr, black curves) and three different apertures: all gravitationally bound elements (dashed), within twice the stellar half mass radius ($\rm R_{star,h}$, dotted), and within 5 pkpc (solid curves). 
Bottom panels: comparison among different classifications of quenched vs. star-forming galaxies, as labeled in the legend.}
\end{figure*}

\begin{figure*}
\centering
\includegraphics[width=0.49\textwidth]{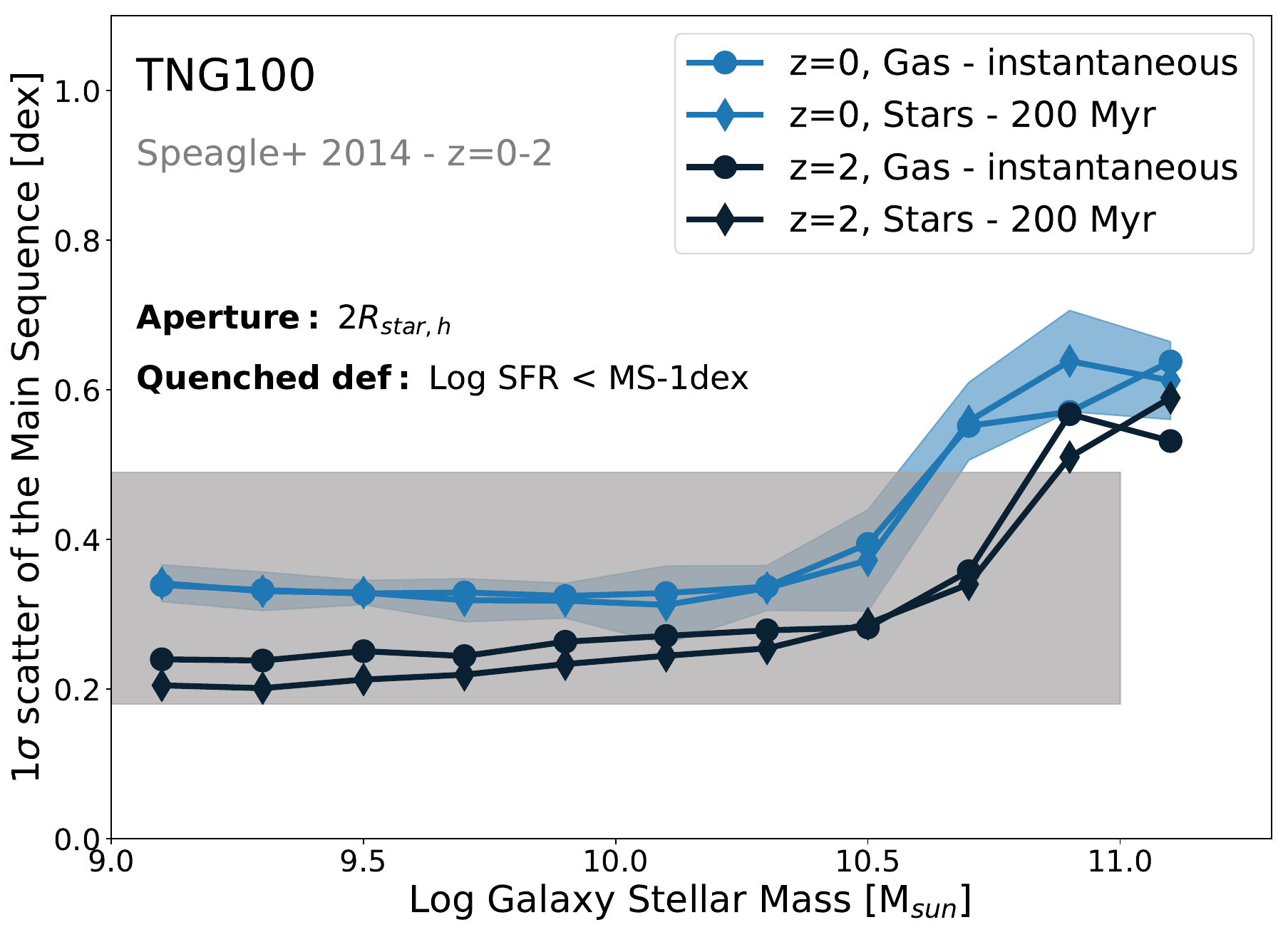}
\includegraphics[width=0.49\textwidth]{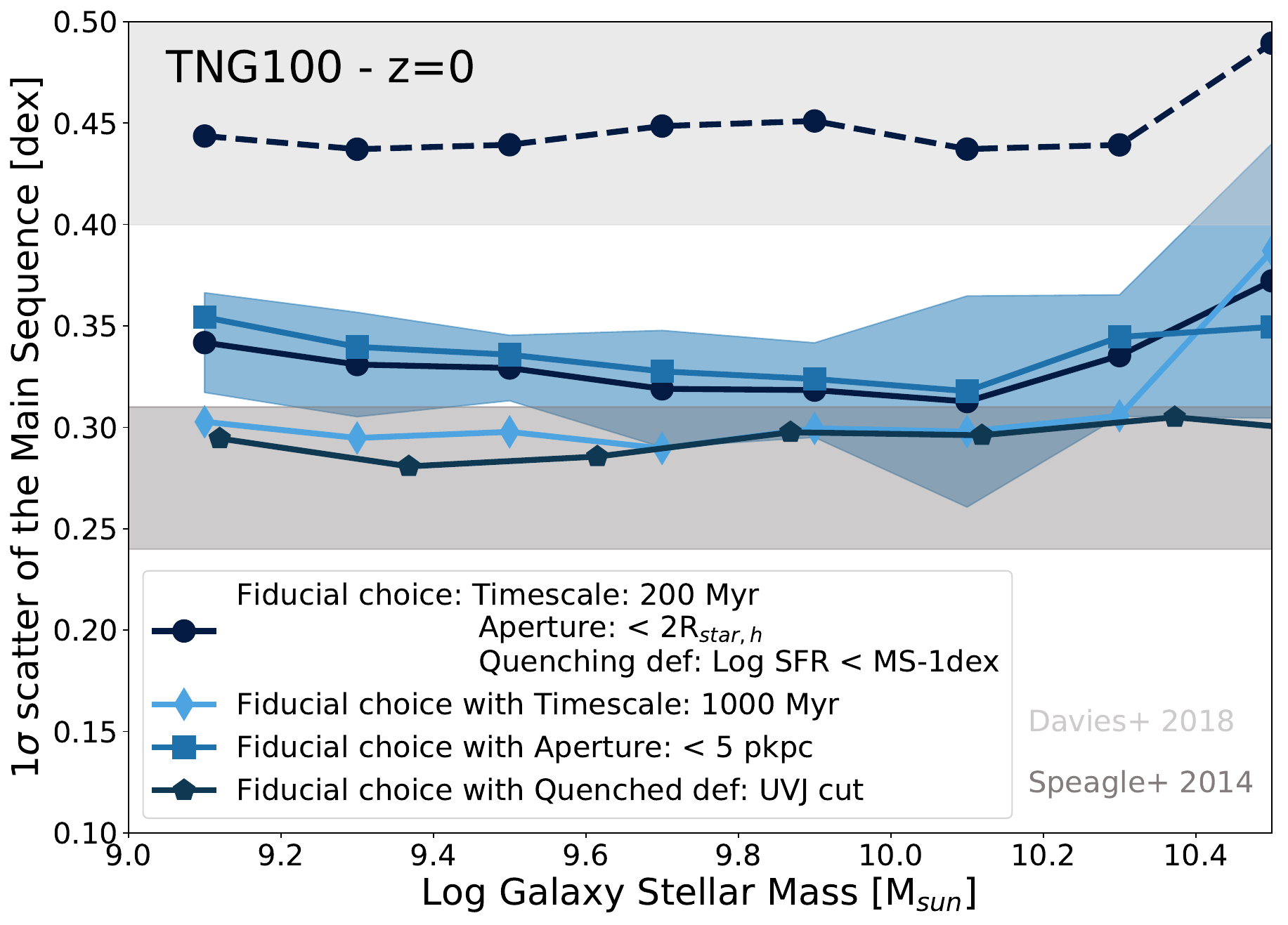}
\caption{\label{fig:scatter} Left panel: 1$\sigma$ scatter of the MS of TNG100 galaxies as a function of galaxy stellar mass, at $z=0$ (blue lines) and $z=2$ (dark blue lines). The SFRs are measured from the instantaneous SFRs of the gas (filled circles) and from the stars formed over the last 200 Myr (filled diamonds). A selection of observational estimates of the intrinsic scatter \citep[see Tab. 4 of][]{2014Speagle} is indicated in the gray box at $0\le z \le 2$. 
Right panel: scatter of the MS in the stellar mass range $10^9<\MS \, (\Ms) <10^{10.5}$. Once a fiducial choice (filled circles) is defined, we change one systematic at a time: timescale (filled pentagons), aperture (filled squares) and quenched definition (filled diamonds). In addition, the scatter computed with the fiducial choice is shown in concert with the effect of 0.2 and 0.3 dex uncertainties on the stellar mass and SFR, respectively (dashed, dark blue curve). The shaded regions in both panels indicate the additional uncertainty in TNG100 due to cosmic variance, computed at $z=0$ for our fiducial choices. The dark gray boxes indicate the observed scatter by \citealt{2014Speagle} (dark gray) and by \citealt{2018Davies} (light gray).}
\end{figure*}

As mentioned in Section \ref{intro}, different observational techniques adopted to infer stellar masses and SFRs of galaxies, along with uncertainties on the IMF and diverse galaxy sample selection functions, can lead to a difficult quantitative comparison between theoretical models and observations and, not rarely, also among observational results themselves \cite[see e.g.][and references therein]{2014Speagle}.
In this Section, we quantify how the different measurement choices introduced in Section \ref{systematics} affect the locus of the MS of TNG galaxies and hence provide plausible guidelines for comparison across different datasets.

In Fig.~\ref{fig:systematics}, we show the MS as a function of galaxy stellar mass in TNG100 at $z=0$ (left column) and $z=2$ (right column). Each row of the figure reports the study of a measurement choice: different timescales (top panels), apertures (central panels), definitions of quenched vs. star-forming galaxies (bottom panels).
 
At $z=0$ we find that, while the slope of the MS changes only slightly when different averaging timescales are adopted to determine SFR (different colors), its normalization shows a remarkable variation among them, being up to $\sim 0.1-0.15$ dex lower for 1000 Myr than for 10 Myr at $10^{9.5}\Ms$ \footnote{The larger deviation at lower masses is, we think, an effect of numerical resolution: see Appendix~\ref{appendix} for details.}. 
At $z=2$, the offset of the MS for the shortest SFR timescales is more than 0.2 dex higher with respect to the longest one ($10^9< \MS < 10^{9.6} \, \Ms$), the discrepancy being smaller ($\leq 0.1$ dex) at higher masses (top right panel).

In order to quantify the effects of different apertures (central panels), we measure the SFR of the whole galaxy -- by summing the SFRs of all gravitationally bound star particles -- (dashed curves), within $2 \times R_{\rm star,h}$ (dotted curves) and within 5pkpc (solid curves).
Moreover, the median SFR is measured from stars formed in the last 10 (gray curves) and 1000 Myr (black curves), to bracket the variations identified in the upper panels.
At $z=0$ and $z=2$ we find that, at fixed timescales, the normalization of the MS can be up to $\sim 0.1-0.15$ dex lower for smaller apertures at the low-mass end. At $z=0$ and for galaxies with $\MS > 10^{9.5} \, \Ms$, this difference rises up to $\sim 0.3$ dex whereas it remains $\sim 0.1$ dex for $z=2$, regardless of galaxy stellar mass. In all cases, we find that the slope of the MS is not affected by different apertures. 
Interestingly, the differences between the SFRs within $2\times R_{\rm star,h}$ and within 5 pkcp is more pronounced at $z=0$ ($\sim 0.05-0.1$ dex for both 10 Myr and 1000 Myr) than at $z=2$ (almost overlapping curves for both timescales).
These findings suggest that the star-formation regions of TNG galaxies extend beyond the nominal stellar mass or stellar light sizes, and more so for more massive galaxies at lower redshifts \citep[see also][]{2019Pillepich,2019Nelson}.

Finally, we compare the MS measured by selecting star-forming galaxies with different quenched definitions (bottom panels). Here, we average the SFRs over 200 Myr, within an aperture of $2\times R_{\rm star,h}$. 
The MS exhibits an offset $0.1$ dex higher for the $\overline{\rm SFR(t)}$-based selection in comparison to all other definitions (dotted curves in the bottom left panel).
Albeit this difference is not negligible, we stress here that this selection based on the star-formation histories of galaxies is purely theoretical, and we use it only at $z=0$ to compare with the other quenched definitions and not in comparison to observations.
On the other hand, the variations in the locus of the MS measure less than $0.05$ dex when we consider the population-based criteria, thus suggesting that the MS in TNG100 is well captured even when using different classification criteria. 
This is even more evident at $z=2$, when the MSs overlap one another regardless of the underlying adopted criterion (bottom right panel).

The results presented in this Section and visualized in Fig.~\ref{fig:systematics} can be used as practical guidelines to quantify biases and systematics also across observational datasets with differently-derived measures.

\subsection{The scatter of the TNG Main Sequence}
\label{sec:scatter}

Finally, we conclude our quantification of the TNG star-formation activities by measuring the scatter of the main sequence as a function of galaxy stellar mass for TNG100 galaxies.

We follow the operational definitions given in Eq.~(\ref{sigma}) and Section~\ref{MS and scatter}. Here, we adopt the set of fiducial choices (see Section~\ref{sec:fiducial}) where all galaxies with $\rm Log~SFR > MS - 1 ~\rm dex$ are classified as star-forming and hence determine the width of the star-forming MS. In Fig.~\ref{fig:scatter}, the intrinsic scatter of the TNG MS is given as a function of galaxy stellar mass and we quantify the effects of the measurement choices by changing them one at a time. 

In the stellar mass range $\MS = 10^9-10^{10.5} \, \Ms$, the scatter of the MS is constant with stellar mass and slightly decreases with increasing redshift, reading $\sim 0.35$ dex at $z=0$ and $\sim 0.2$ dex at $z=2$ (light and dark blue lines in the left panel). For TNG300, these values read $\sim 0.4$ dex and $\sim 0.25$ dex, respectively. 
Moreover, the difference between the instantaneous and 200-Myr SFRs is negligible at low redshift (overlapping light blue curves) but becomes more significant at $z=2$, albeit still smaller than $0.05$ dex.

On the other hand, at the high-mass end, we find that the scatter significantly rises up to $0.6$ dex for $\MS \sim 10^{11} \, \Ms$, likely due to an ill-defined main sequence in this mass range, as also outlined for the original Illustris simulation by \cite{2015Sparre}.

The effect of different measurement choices are shown in the right panel of Fig.~\ref{fig:scatter}.
With our fiducial choices as a reference (filled circles with solid curves), we find that the scatter decreases by up $\sim 0.05$ dex for longer timescales (filled diamonds) and it slightly increases if the SFR is measured within smaller apertures (filled squares). Selecting star-forming galaxies using a TNG UVJ cut instead of SFR-based cut, we find that the scatter is $\sim 0.05$ dex lower (filled pentagons) than our fiducial choice.
Due to the limited simulation volume, we account for the cosmic variance by measuring the scatter (at $z=0$ for our fiducial choices) in eight sub-boxes of $\sim$50 Mpc on a side and computing the \textit{jackknife} error (shaded areas in both panels).

All these are estimates for the {\it intrinsic} scatter. In Fig.~\ref{fig:scatter}, right panel, a dashed black curve denotes the 1$\sigma$ scatter when measurement uncertainties are taken into account: 0.2 dex uncertainties on the galaxy stellar mass and 0.3 dex on the SFRs. The effect of the measurement uncertainties necessarily turns into an increase of the scatter by about $\sim 0.1$ dex with respect the one computed with the raw data. Even if not shown, we find that the consequences of different choices quantified above hold also when mass and SFRs uncertainties are taken into account, and in general all these trends are preserved and slightly enhanced at high redshifts.

Observational estimates place the intrinsic scatter in the range $0.2-0.5$ dex \citep[see e.g.][and references therein]{2012Whitaker,2014Speagle,2015Schreiber,2018Pearson,2018Davies}, depending on the SFR measures and on the galaxy sample.
For the sake of clarity, we stress here that for the comparison to \cite{2018Davies} (right panel), since the authors use different methods to isolate star-forming and quiescent populations, we choose their scatter values measured using a $u-r$ color selection and the selection based on the sSFRs.
We find that, in the redshift range $0\le z \le 2$ the scatter of the MS in TNG100 is in good agreement with observations, indicated as grey boxes in both panels of Fig.~\ref{fig:scatter} but a more careful comparison once again requires a proper match or mocking of the adopted measurement choices.

\begin{figure*}
\centering
\includegraphics[width=0.33\textwidth]{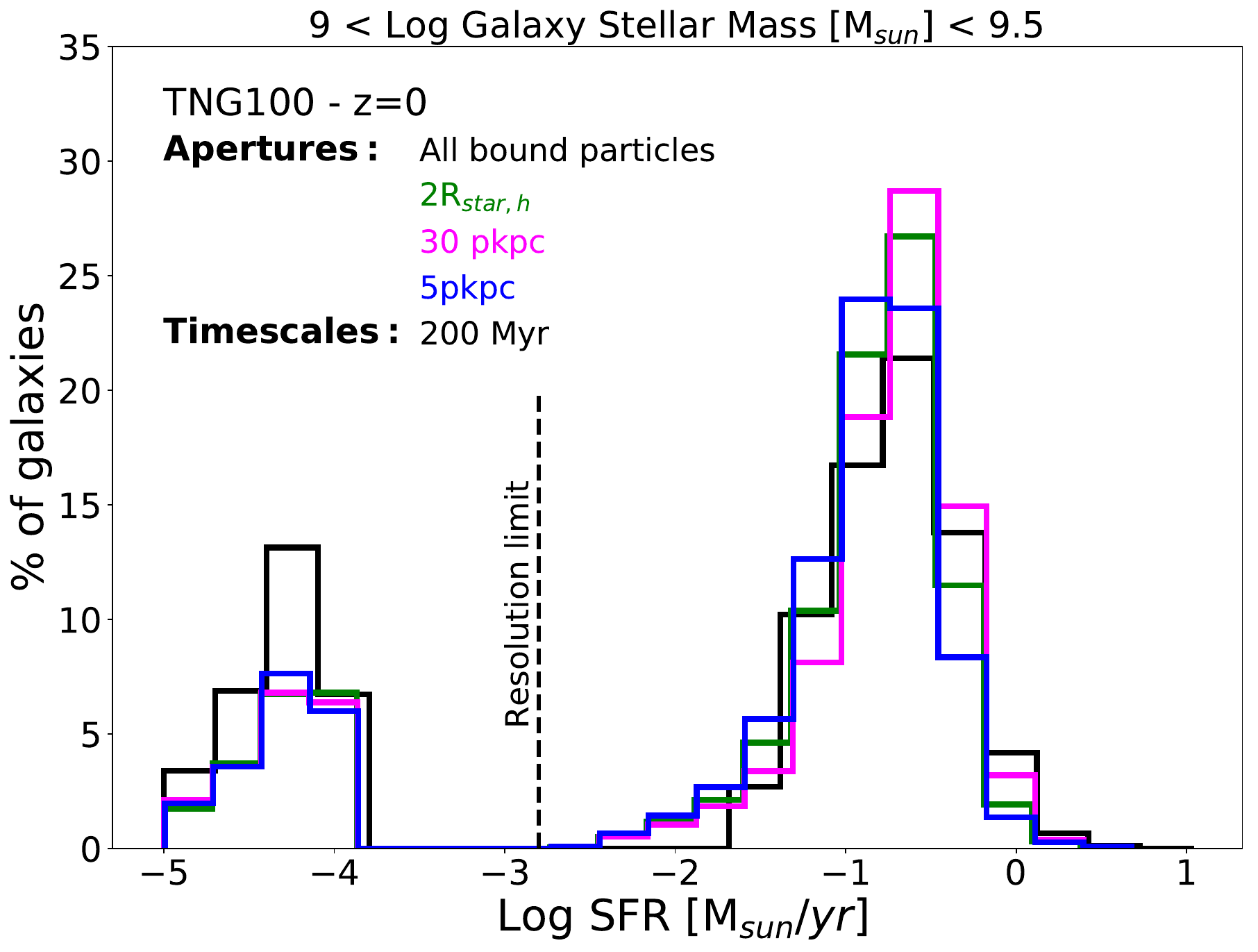}
\includegraphics[width=0.33\textwidth]{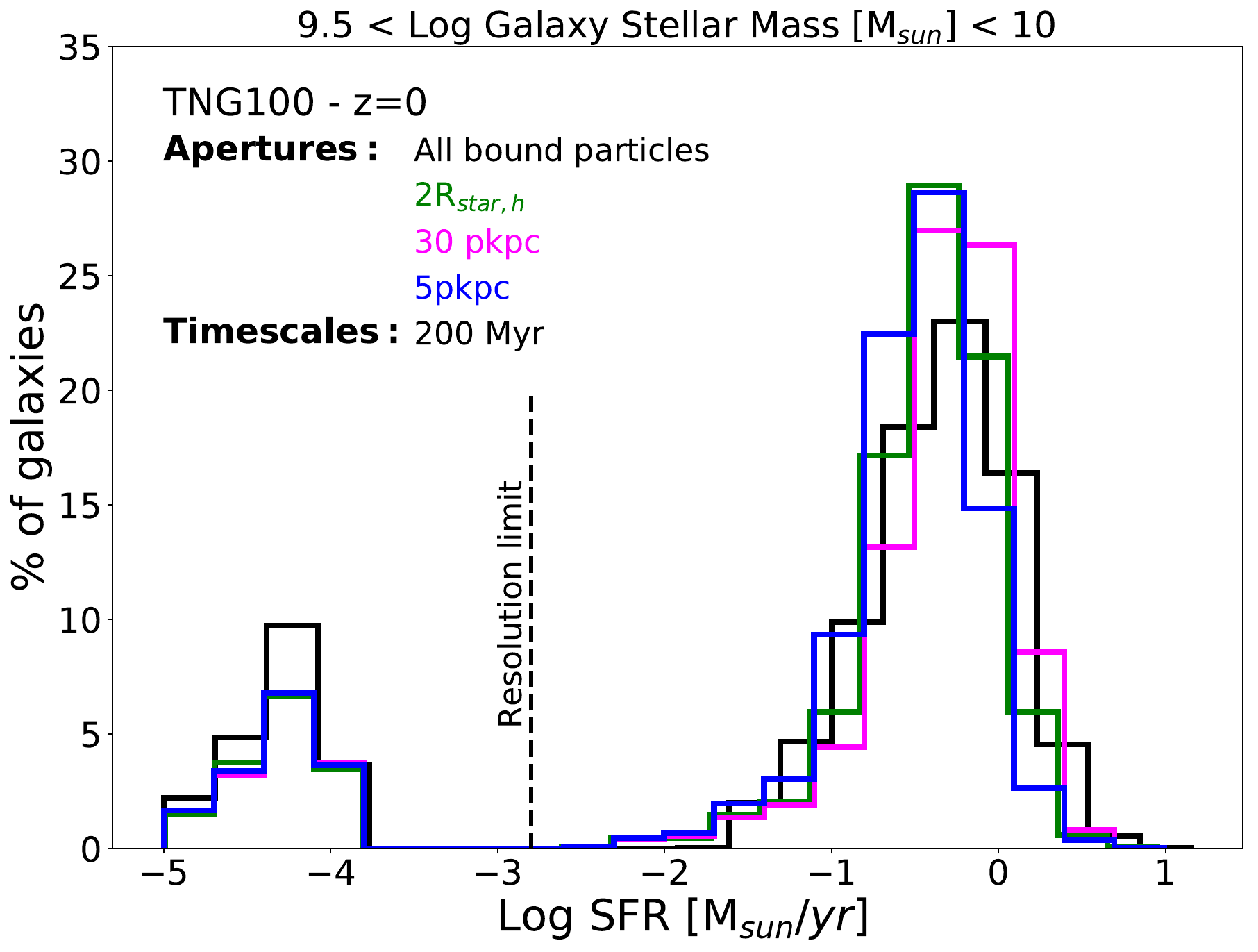}
\includegraphics[width=0.33\textwidth]{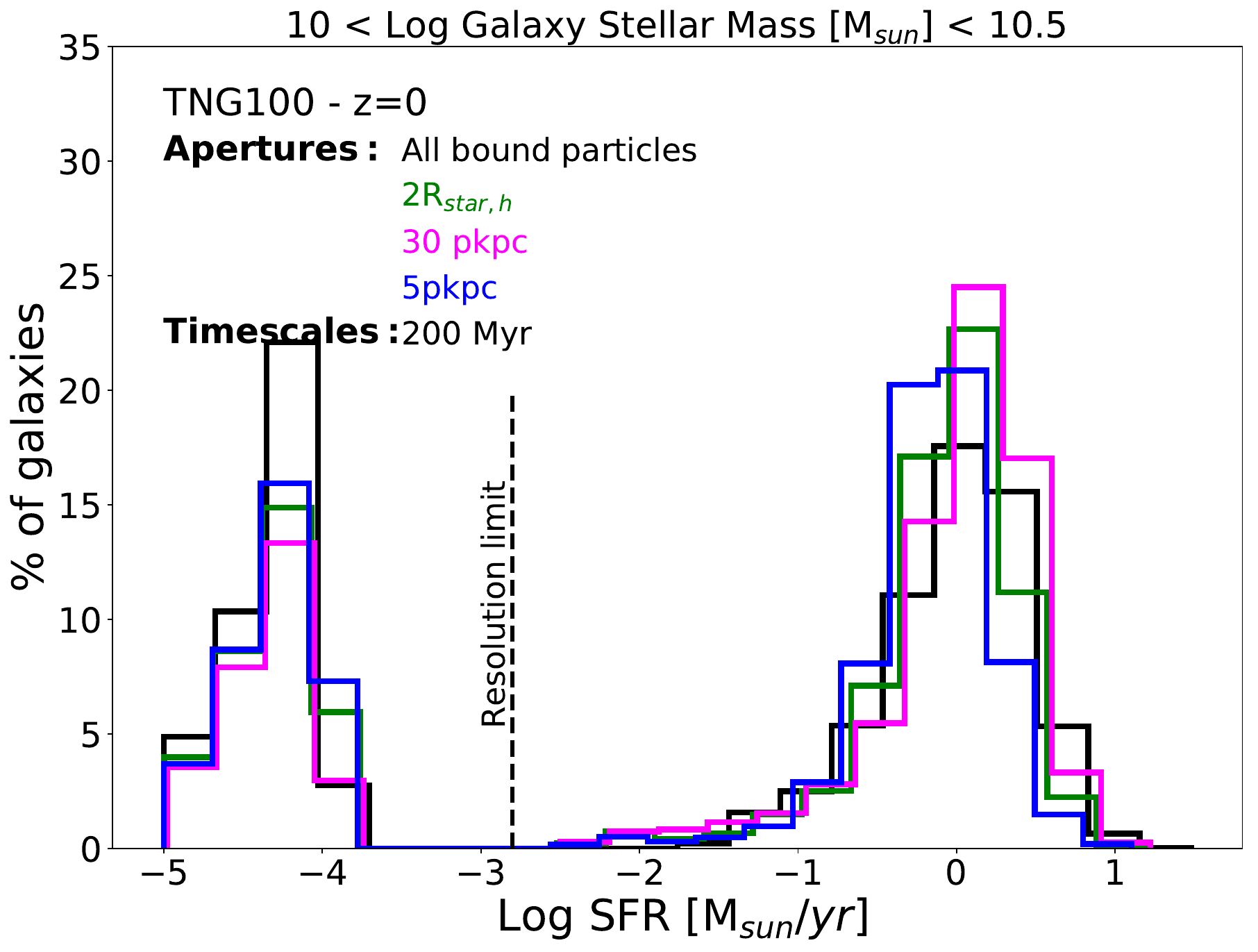}
\includegraphics[width=0.33\textwidth]{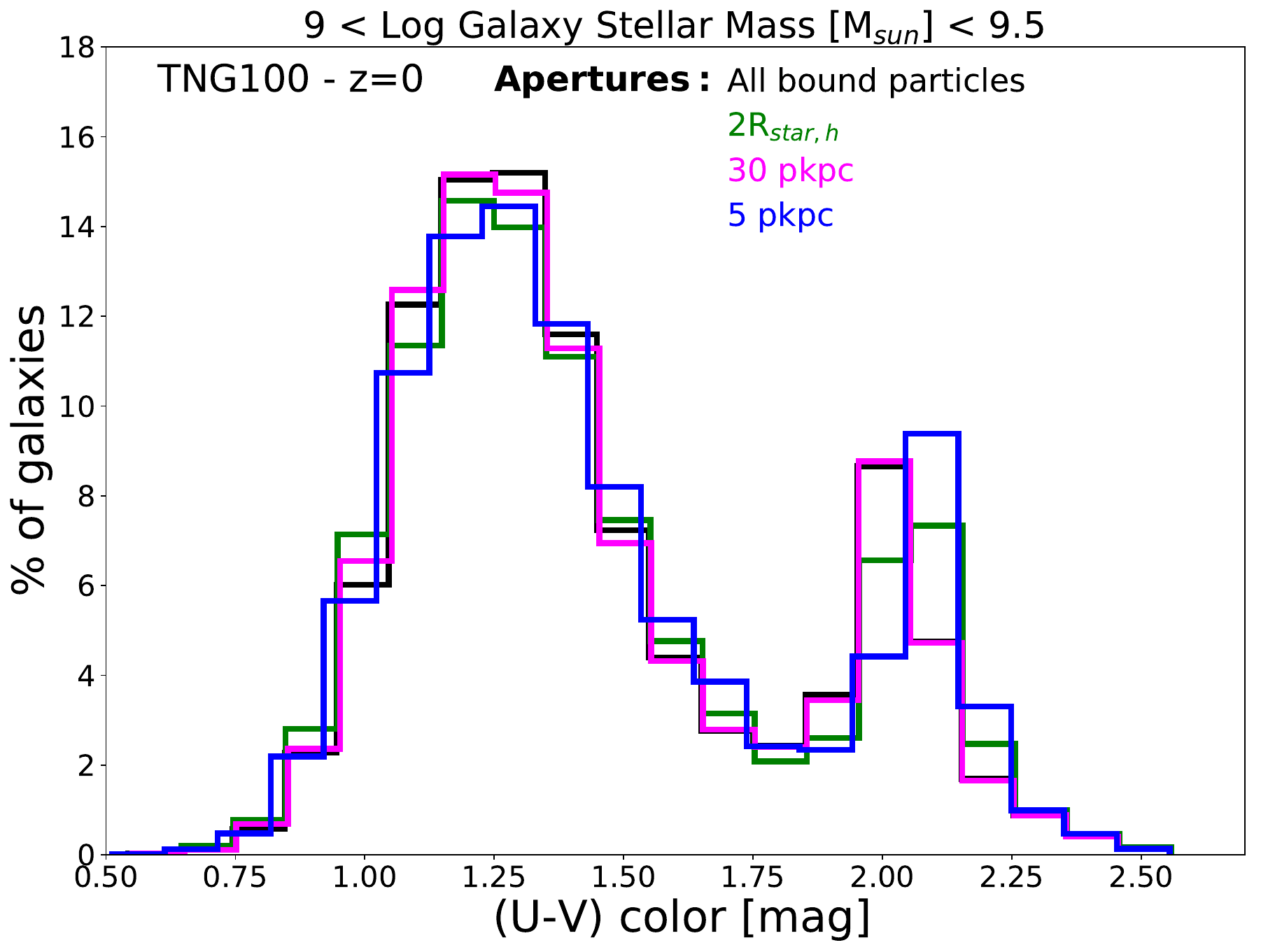}
\includegraphics[width=0.33\textwidth]{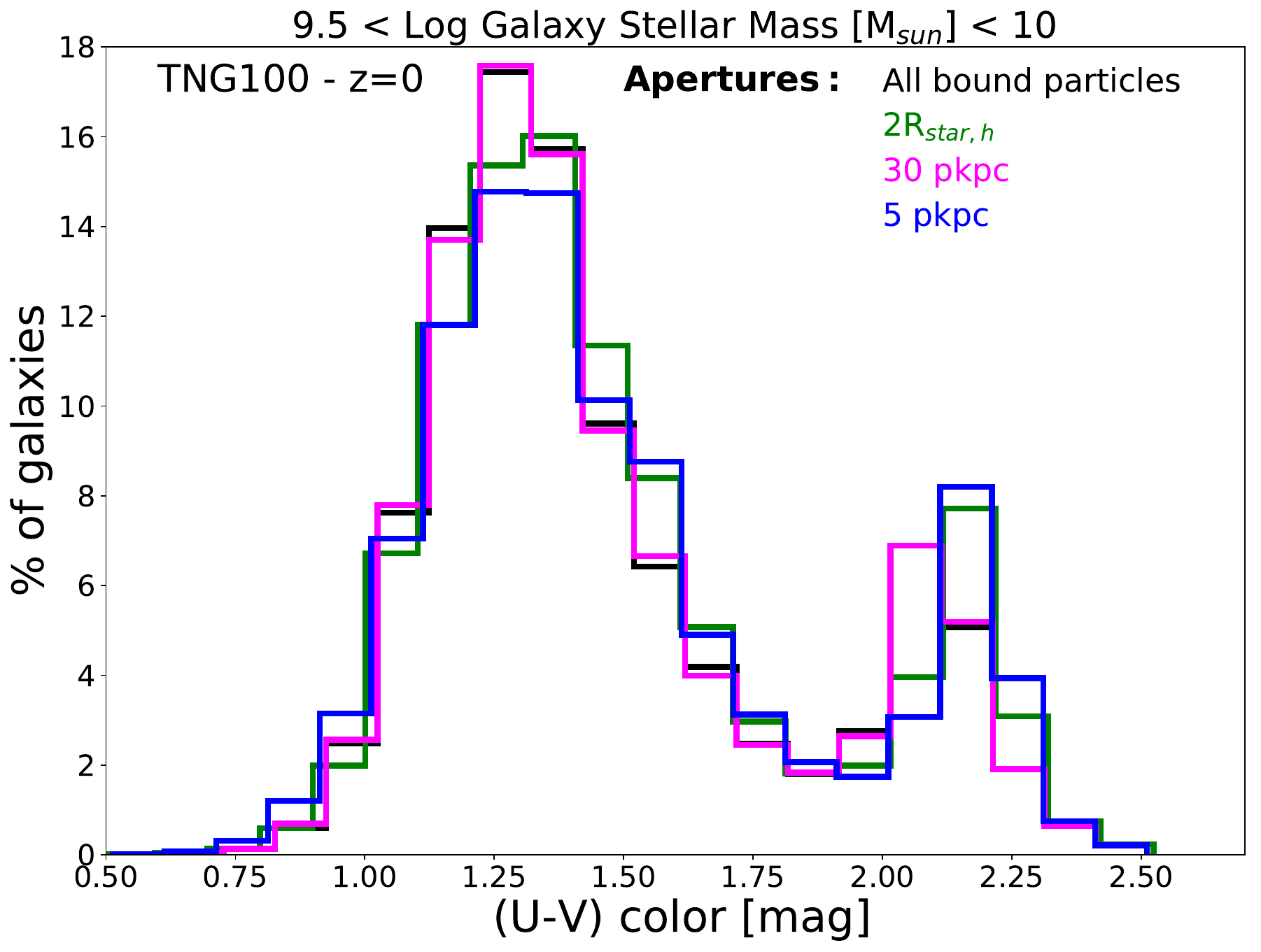}
\includegraphics[width=0.33\textwidth]{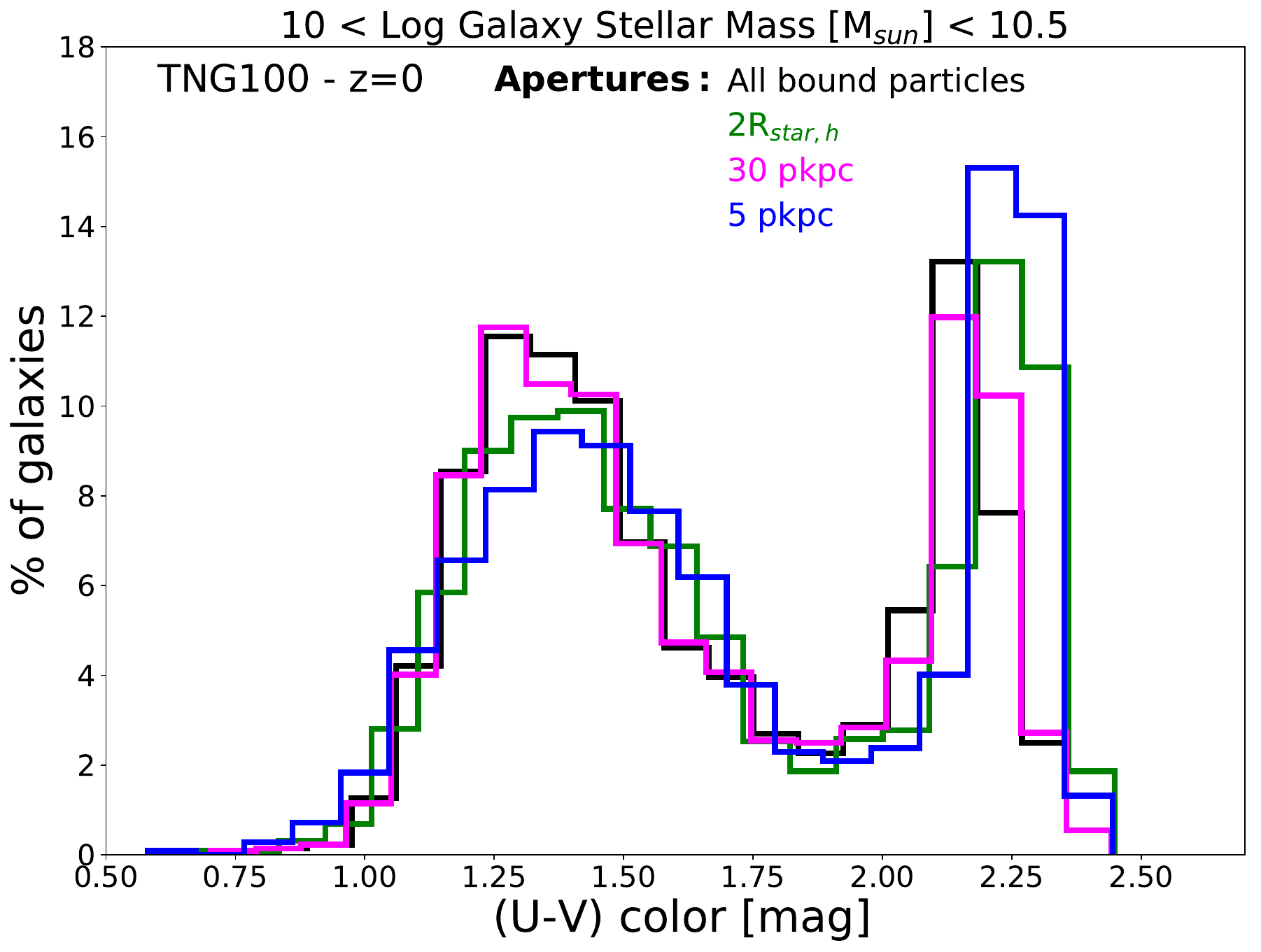}
\caption{\label{fig:bimodality} Distribution of the star-formation rate (top panels) and U-V colors (bottom panels) of TNG100 galaxies at $z=0$ for three stellar mass bins, as labeled in each panel. For the SFR distributions, all galaxies with $\rm Log \, SFR$ between -4 and -5 $\Ms \, \rm yr^{-1}$ are placed there arbitrarily, since they have unresolved SFR values and their actual distribution of SFRs is unknown (albeit firmly placed below the indicated resolution limit). In all panels, histograms are normalized to the total number in each stellar mass bins. A clear bimodality in color is associated to a distribution in SFR that, if galaxies with unresolved values of SFR are ignored, rather than being bimodal in any sense, is unimodal and asymmetric.}
\end{figure*}
\section{Discussion and implications}
\label{discussion}

\subsection {On the bimodality: color vs. SFR}
\label{sec:bimodality}

As mentioned in Section \ref{intro}, several galaxy properties are observed to be bimodal in narrow bins of galaxy stellar mass, including color, SFR, morphology. As a result, galaxies are traditionally classified into two main categories: those with red colors, which are mostly quiescent and exhibit early-type i.e. elliptical-like morphologies; on the other side, blue-cloud galaxies that are on the star-forming main sequence and with late-type i.e. disk-like morphologies \citep{2001Strateva,2003Kauffmann,2004Baldry,2004Balogh}. 

For example, \cite{2004Kauffmann} and \cite{2012Wetzel} have examined the number density distribution of SDSS galaxies in specific star-formation rate and found a break, a local minimum at $\rm sSFR = 10^{-11} \, \rm yr^{-1}$ at $0.04 < z < 0.06$ whose value is independent of galaxy stellar mass and host halo mass (at least for stellar masses around $10^{10}\Ms$ and above). Interestingly, as pointed out by the authors themselves, albeit the rise of the distribution below the break is real, the strong peak near $\rm sSFR =10^{-12} \, yr^{-1}$ is artificial, since all galaxies with undetectable SFR have been assigned arbitrary values in that range. Other authors over the years have questioned whether the galaxy distribution in SFR is effectively bimodal \citep[e.g.][]{2007Elbaz, 2011Mcgee}. Yet, the idea that low-redshift galaxies are separable into two groupings of star-forming vs. quiescent galaxies {\it because} their distribution of sSFR or SFR is bimodal in bins of stellar mass remains a foundational motif in galaxy evolution.

Recently, \cite{2017Feldmann} has questioned the existence of a bimodality of the galaxy population in SFR and argued that, if one neglects `dead' galaxies in numerical models (i.e. galaxies with unresolved SFR values), the logarithmic distribution of SFR is unimodal and that the observed bimodality is likely due to uncertainties in the SFR measurements.
In the same spirit, \cite{2017Eales} have stressed again that the tight galaxy red sequence in the optical color-magnitude diagram is the result of optical colors depending only very weakly on specific star-formation rate below a certain value. Therefore the red sequence is the result of the pile up at the same optical color of galaxies with very low sSFR rather than the representation of a distinct class of galaxies. Even more fundamentally, \cite{2018Eales} have argued that submm and far-IR Herschel data favor an sSFR vs. galaxy mass plane that is populated by a single ``Galaxy Sequence`` that smoothly extends towards very low values of sSFR rather than by two distinct, well-separated classes of galaxies.  They argue that the existence of a less densely-populated region denoted the `green valley' and residing between the star-forming and quenched populations is likely due to observational biases.

Here we show that numerical models like IllustrisTNG (and in fact also Illustris) naturally return a bimodal distribution in color even though a clear bimodality in the logarithm of SFR or sSFR in bins of galaxy stellar mass is not present. 

In Fig.~\ref{fig:bimodality}, we show the distribution of the logarithmic SFR averaged over 200 Myr (top) and (U-V) color (bottom) of TNG100 galaxies at $z=0$, stacked in three stellar mass bins, as labeled at the top of each panel. Histograms are normalized to the total number of object in each stellar mass bins.
In all cases, we show the distribution accounting for four different physical apertures, denoted by different colors. 
As already discussed in Section \ref{systematics}, small apertures shift the locus of the MS to lower values (see Fig.~\ref{fig:systematics}, central panels). Indeed, the top panels of Fig.~\ref{fig:bimodality} show that, while the tails of the distribution are not affected by different apertures, the peak -- which marks the normalization of the MS -- for 5 pkpc aperture is moved to lower values of the Log SFRs (blue histrograms).
As done throughout, all galaxies with Log SFR between $-4$ and $-5\, \Ms \, yr^{-1}$ are placed there artificially, since they have unresolved SFR values, i.e. values of SFR that fall below the resolution limit of the simulation: the latter is indicated with vertical dashed lines in each panel. It is important to notice that such resolution limit is one or two orders of magnitude lower than observational ones.

Once the galaxies with unresolved SFR values are ignored, the distribution of log SFR is clearly unimodal, albeit asymmetric and skewed towards low SFR values. All these findings are consistent with the aforementioned arguments proposed by e.g. \cite{2017Feldmann} and \cite{2018Eales}. In fact, the SFR distributions of Fig.~\ref{fig:bimodality}, top panels, could be characterized using a Gaussian mixture modeling, as recently proposed for e.g. the Illustris, EAGLE and Mufasa results by \cite{2018Hahn}, but it is unclear whether the medians of such multiple Gaussians would identify distinct classes of galaxies or simply represent a numerical over-fitting.
In order to quantify the asymmetry, we fit the log SFR distribution (above the resolution limit) with a zero-inflated gamma distribution proposed by \cite{2017Feldmann}. We find that the asymmetric log SFR distribution of Fig. 8 is well characterized by this description, more reliably than those based on the standard log-normality assumption.

The lower panels of Fig.~\ref{fig:bimodality} extend the analysis on TNG presented by \cite{2018Nelson}, who characterized the bimodal distribution of $(g-r)$ colors of TNG galaxies, in the stellar mass range $10^9 <\MS \, (\Ms) < 10^{12.5}$. There, a quantitative comparison to SDSS data has shown TNG galaxies to be in excellent agreement with $z\sim0$ observations, providing an empirical indirect validation to the underlying choices for AGN and stellar feedback adopted in TNG in comparison to the original Illustris. 
Here, in addition to such previous analysis, we explicitly examine the (U-V) color distribution of TNG galaxies in three bins of stellar mass (bottom panels of Fig.~\ref{fig:bimodality}). 
In the previous Sections, we have demonstrated that the UVJ diagram of TNG galaxies is populated by two groups of galaxies that in terms of number density and SFR values clearly separate on the UVJ plane. 
Consistent with this, a pronounced bimodality in (U-V) at fixed galaxy stellar mass is apparent, with a depression around 1.75 mag for low-mass galaxies between $10^9-10^{9.5} \, \Ms$ and at slightly redder colors for the more massive ones.
This again confirms that a strong color bimodality may coexist with a unimodal distribution of SFRs.

Even if not shown, we have repeated the bottom panel of Fig.~\ref{fig:bimodality}, by excluding galaxies with unresolved SFR values from the sample: in this case, the color bimodality is not in place. This could be interpreted similarly to what have been seen for the SFR distributions, that look like bimodal only when very low SFR galaxies are accounted before (i.e. by accumulating them at some arbitrary low SFR value). In fact, this result enhances and reinforces what stated above: color distributions can be bimodal even if SFR distributions are not. Galaxies with SFR=0 would distribute below the resolution limit but not necessarily creating a bimodality. Instead, colors are bimodal whenever well defined and realistic color values are considered. In the simulations, the galaxy colors are well captured also for those galaxies whose SFR values are below the resolution limit. Even though a fraction of galaxies have artificially null SFR values, the same galaxies have well defined and observationally consistent ($g-r$) and (U-V) colors, making the color bimodality of Fig.~\ref{fig:bimodality} robust.

Finally, although we do not show it, we have examined the color distribution of TNG100 galaxies also at higher redshifts, e.g. $z=1$ and $z=2$. 
Differently from the bottom panels of Fig.~\ref{fig:bimodality}, where the color bimodality is manifest for low-mass as well as high-mass galaxies, the color bimodality is still present at higher redshifts but only at higher galaxy stellar masses. Specifically, at $z=1$, two peaks are clearly visible for galaxies more massive than $10^{10} \, \Ms$ whereas at $z=2$ such bimodality starts to be apparent only for galaxies with stellar mass $>10^{10.5} \, \Ms$. 

\subsection {On the bending of the MS at the high-mass end}
\label{sec:bending}

\begin{figure*}
\centering
\includegraphics[width=14cm]{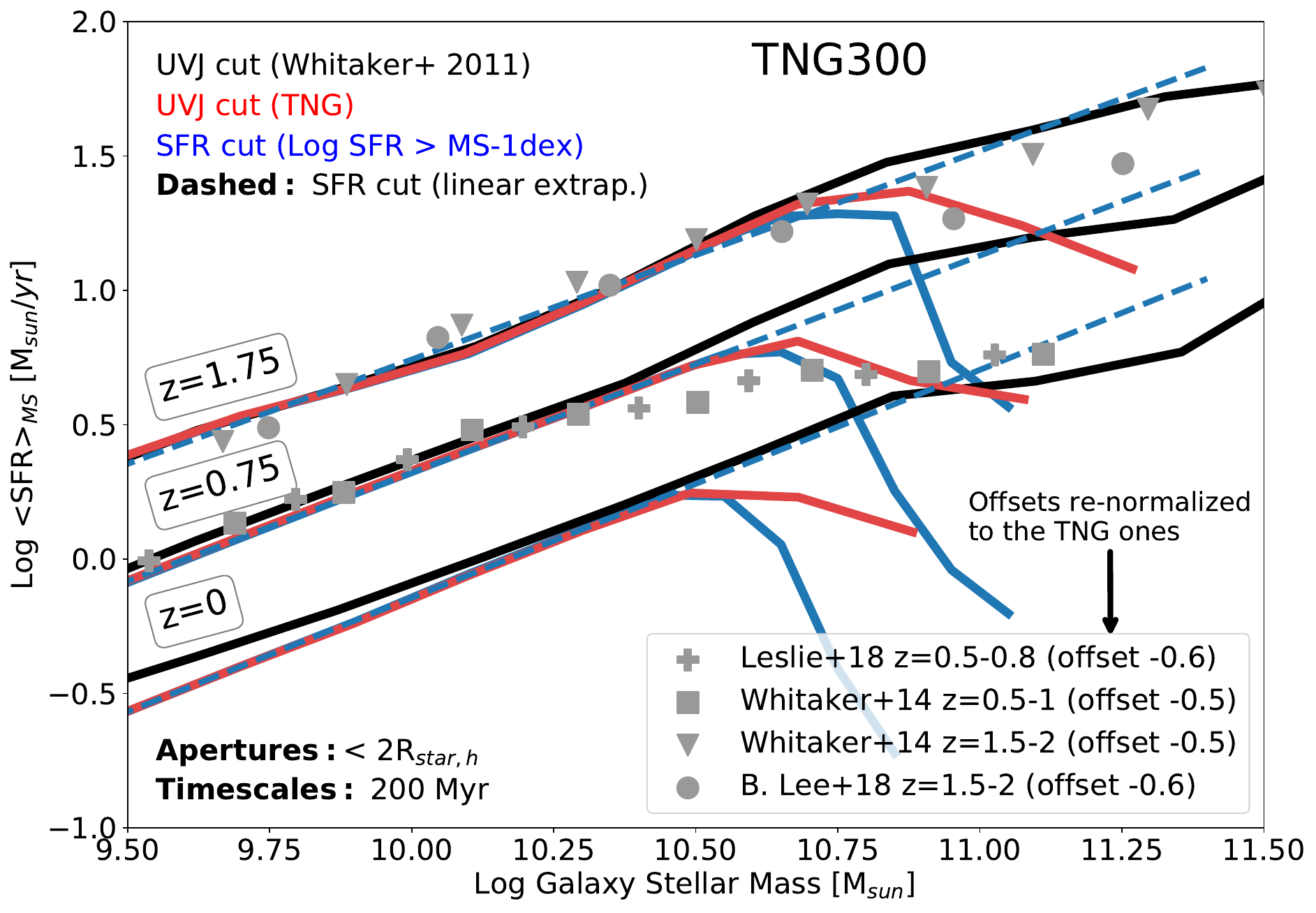}
\caption{\label{fig:bending} Median SFR on the MS of TNG300 galaxies at $z=0, 0.75, 1.75$. The aperture is $2 \times R_{\rm star,h}$ and the galaxy SFR is averaged over 200 Myr. Star-forming galaxies are identified in different ways: with the UVJ selection proposed by \citet{2011Whitaker} (solid black lines), with the TNG UVJ selection (solid red lines), and with a SFR cut using $\rm Log \, SFR > MS - 1 dex$ (dashed blue lines). Additionally, the linear extrapolation from low masses of the MS obtained via the SFR cut is shown (dotted line), for reference. Gray symbols are observational measurements, arbitrarily normalized to the TNG ones (as indicated in the legend), to highlight the turnover of the MS. Given the same TNG300 galaxy sample, the star-forming MS may or may not bend according to the criterion adopted to separate star-forming vs. quiescent galaxies.}
\end{figure*}

As demonstrated in this work and in several previous observational and theoretical studies, the MS of star-forming galaxies with stellar mass $\MS \lesssim 10^{10} \, \Ms$ follows a single power-law, whose slope and normalization values depend somewhat on the adopted IMF, galaxy sample selection, and SFR tracer \cite[see][and references therein]{2014Speagle}. 

On the other hand, at higher galaxy masses, several authors have observed a more complex scenario. From the local Universe to $z \sim 4$, some studies indicate the presence of a turnover, namely of a downwardly bending of the SFR-$\MS$ relation for galaxies with $\MS \gtrsim 10^{10}-10^{10.5} \, \Ms$, hence proposing a stellar mass-dependent slope as best fit instead of a single power-law \citep{2011Karim,2012Whitaker,2014Whitaker,2015NLee,2015Gavazzi,2016Tomczak,2018BLee}. Whether the star-forming MS bends or not, and by how much, is still a matter of debate in the observational literature. Moreover, its physical reason and the existence of a turnover mass are still unclear. Morphological studies of galaxies have been proposed to explain the bending \citep{2014Abramson,2014Schawinski}, albeit the most likely scenario is related to secular internal processes, like AGN feedback, that drive the quenching of massive galaxies.

As shown in Section \ref{sec:uvj-sfr} and Fig.~\ref{fig:SFMS}, the main sequence of TNG galaxies exhibits a bending {\it if} the classification of galaxies into star-forming and quiescent is done by means of a UVJ cut. There we found that the bending is somewhat more pronounced at lower redshifts and that the turnover occurs at large stellar masses ($\MS \gtrsim 10^{10.5}\Ms$), clearly revealing that massive galaxies classified as star-forming through a UVJ cut have lower SFRs in comparison to lower mass counterparts. 



We now show that, given the same galaxy population, the star-forming MS may bend or not according to the criterion adopted to separate star-forming vs. quiescent galaxies.
Fig. \ref{fig:bending} shows the median of the MS of TNG300 galaxies in the stellar mass range $\MS = 10^{9.5}-10^{11} \, \Ms$ at $z=0,0.75,1.75$. 
Star-forming galaxies are selected in three ways: according to their colors by using the UVJ cut proposed by \cite{2011Whitaker} (black curves), and the cut based on the TNG diagram itself (Eqs. \ref{TNG_cut} and \ref{TNG_cut2}: red curves), and according to their SFR values, $\rm Log \, SFR > MS - 1 dex$ (blue curves), without extrapolating the linear MS from lower masses.  Additionally, we show the linear extrapolation of the MS (dashed curves), for reference. 

For galaxies more massive than $\sim 10^{10.5} \, \Ms$ a bending of the MS is clearly noticeable if star-forming galaxies are identified via the SFRs-based criterion and via the TNG UVJ cut.
While a steep decline of the MS for massive galaxies is manifest at all redshifts, such a turnover is less prominent in the case of the TNG UVJ cut when compared to the SFR-based selection. 
In fact, when we adopt instead the UVJ cut defined in \cite{2011Whitaker}, we find that the turnover of the MS is much less pronounced than the other two methods, even though a deviation from the linear trend is recovered also in this case at the highest mass end.

For comparison, a selection of observational data points at $z=0.75$ and $z=1.75$ is included as grey symbols. They all adopt a UVJ cut to select for star-forming galaxies but for \textcolor{blue}{Leslie et al. (in prep)}, who use NUV-r-J selection. Here the normalization of the observationally-derived MSs is artificially adjusted to match the one from TNG, in order to facilitate the  comparison of the bending. 
In the linear, low-mass regime, irrespective of the adopted selection criteria, the slope of the MS in TNG is in good agreement with the observational data at both $z=0.75$ and $z=1.75$. Moreover, the TNG UVJ cut (red curve) reproduces quite well some of the observational results in terms of bending, at least at $z=0.75$. However, the level of bending at $z\sim1.75$ is also unclear across observational datasets, while we find that the TNG model returns a weak trend with redshift of the turnover mass, that moves from $\MS \simeq 10^{10.5} \, \Ms$ at $z=0$ to about $10^{10.7} \, \Ms$ at $z=1.75$.


In conclusion, the characterization of the shape of the MS at the highest-mass end depends on the criterion adopted to separate between star-forming and quiescent galaxies. This is the case at least for TNG galaxies but could 
also apply to the observed galaxy population. UVJ selections that minimize the number of quenched galaxies (and hence classify a larger number of galaxies as star forming: e.g. the TNG UVJ cut in Fig.~\ref{fig:UVJ} in comparison to the Whitaker ones) are destined to return a MS that is more bending than those that are more conservative in classifying galaxies as star forming. In the case of TNG galaxies, this ambiguity is strictly connected to the way galaxies occupy the SFR-$\MS$ plane at the high-mass end
\footnote {For the sake of clarity, we stress here that even if not shown, we have computed the main sequence also by removing galaxies with unresolved SFR values. Our findings suggest that neither the locus of the "linear" MS nor the bending are quantitatively affected by this choice.}. 
As can be seen in Figs. \ref{fig:SFR-UVJ}, \ref{fig:SFR-MSTAR} and \ref{fig:bimodality}, TNG galaxies populate the low-SFR regions below the high-density star-forming MS without producing an obvious depression or separation or green valley: see previous Section and the absence of a clear galaxy bimodality in SFR in TNG. This in turn makes the classification of galaxies into two classes somewhat arbitrary, albeit operationally still viable, and the characterization of the star-forming main sequence at the highest-mass end somewhat ill defined. We postpone to future work the task to connect the shape of the SFR-$\MS$ plane to the timescales of the underlying quenching mechanisms in the TNG model.


\begin{table*}
\centering
\begin{tabular}{l|c|l|c|c|c|c}
\hline
References & Redshift & SFR tracers & IMF & Galaxies selection &  Main Sequence & Log $\MS \, (\Ms)$ range\\
           &         &            &  & criteria & shape &  \\
\hline
\hline
\cite{2007salim}        & 0.005 - 0.22 & UV-SED	& \cite{2003Chabrier} & NUV-r    & Linear &  9-11.1\\
\cite{2007Elbaz}	    & 0.015 - 0.1 & H$\alpha$ & \cite{1955Salpeter} & U-g-M$_B$   & Linear    &  9.1-11.3 \\
\cite{2012Zahid}	    & 0.04 - 0.1  & H$\alpha$ & \cite{2003Chabrier} & ?      & Linear    &  8.5-10.4 \\
\cite{2010Oliver}	    & 0.0-0.2     & IR    & \cite{1955Salpeter} & optical/NIR SED      & Linear    &  9.1-11.6 \\
\cite{2015Chang}	    & 0.0-0.2     & 12-22 $\mu$m   & \cite{2003Chabrier} & u-r-z & Linear & 9-11.1\\
\cite{2011Karim}	    & 0.6-0.8     & Radio    & \cite{2003Chabrier} & NUV-r-J      & Linear    &  9.1-11.1 \\
\cite{2014Whitaker}     & 0.5 - 1& UV+IR	& \cite{2003Chabrier} & UVJ       & Bending   &  8.4-11.1 \\
\textcolor{blue}{Leslie et al. (in prep)} & 0.5 - 1& Radio-IR 	& \cite{2003Chabrier} & NUV-r-J  & Bending & 8.8-11.2\\
\cite{2015NLee}         & 0.8 - 1 &	IR 		& \cite{2003Chabrier} & NUV-r-J  & Bending   &  8.5-11  \\
\textcolor{blue}{Leslie et al. (in prep)} & 0.8 - 1.1& Radio-IR 	& \cite{2003Chabrier} & NUV-r-J  & Bending & 8.8-11.2\\
\cite{2018BLee}         & 1.5 - 2 & FUV+IR	& \cite{2003Chabrier} & UVJ     & Bending &  9-11.5\\
\cite{2014Whitaker}     & 1.5 - 2  & UV+IR	& \cite{2003Chabrier} & UVJ     & Bending & 9.2-11.5 \\
\cite{2009Santini}      & 1.5 - 2.5 & UV+IR	& \cite{1955Salpeter} & ?      & Linear & 8.7-11.5 \\
\cite{2011Rodighiero}   & 1.5 - 2.5& FUV	    & \cite{1955Salpeter} & BzK        & Linear & 8.7-11.3 \\
\cite{2009Pannella}     & 1  - 3   & Radio	    & \cite{1955Salpeter} & ?       & Linear & 10.1-11.2 \\
\cite{2014Whitaker}     & 2 - 2.5  & UV+IR	& \cite{2003Chabrier} & UVJ       & Bending   &  8.4-11.1 \\
\cite{2016Tomczak}      & 2 - 2.5 & UV+IR	& \cite{2003Chabrier} & UVJ       & Bending   &  8.5-11.5 \\
\hline
\cite{2014Speagle}      & 0,0.75,1.75,2 & Multiple tracers& \cite{2001kroupa} & Multiple criteria & Linear &  9.7-11.1 \\
\hline
\end{tabular}
\caption{\label{tab:obervations} Summary of the observational datasets adopted for comparisons in Figs.~\ref{fig:bending} and \ref{fig:SFR_tracers}. Columns read: 1) reference; 2) redshift range; 3) SFR indicator; 4) Assumed stellar initial mass function; 5) criterion used to separate quenched from star-forming galaxies; 6) shape of the MSs; 7) mass range observed in each sample.}
\end{table*}


\subsection {Comparison to observations}
\label{comparison}

As demonstrated in Section~\ref{systematics}, different measurement choices can influence to various degrees the locus of the star-forming MS. Now, with this awareness, we attempt to compare the MS of TNG100 galaxies to selected observational works. 

As noticeable from Table~\ref{tab:obervations}, different observational analyses adopt different methods or indicators to derive the SFRs of observed galaxies (H$\alpha$, UV, FUV, IR...), different IMFs and, unavoidably different implicitly-adopted physical apertures within the luminous bodies of galaxies. As discussed at length throughout the paper, such a diversity of methodologies can lead to non-negligible apparent discrepancies in the quantification of the normalization of the MS \cite[see e.g. Fig.~\ref{fig:systematics} and ][and references therein]{2014Speagle,2018Theios}.
At $z=0$, different observed MS normalizations taken at face value differ by up to $\sim$ 0.5-0.7 dex; this is the case also at $z=2$. Until a robust understanding of the observational systematics is available, we would be tempted to treat the comparison between observations and theoretical models by only focusing on the slope and mass trends of the MS and by ``arbitrarily'' matching its normalization, as we have done in Figure~\ref{fig:bending} and has been done with the evolution of the TNG galaxy mass-metallicity relation by e.g. \citealt{2018Torrey}. 
Indeed, a insightful comparison between simulated and observational results demands a deep understanding of how different observational SFR indicators compare to one another and effectively map into un-biased estimates of the SFRs of galaxies as directly available from simulations. In fact, here we can only proceed with a face-value comparison, by which we mean that we keep unchanged the observational data points and only adjust them to be consistent with the Chabrier IMF adopted in the simulations ($\rm Log \, \MS^{\rm Chabrier} = \rm Log \, \MS^{\rm Salpeter} -0.24$; $\rm Log \, SFR^{\rm Chabrier} = \rm Log \, SFR^{\rm Salpeter} -0.15$, see e.g. \citealt{2008Dave,2014Santini}). 
We focus on a selection of observational dataset based on commonly available redshift ranges and median redshifts.

\begin{figure}
\centering
\includegraphics[width=\columnwidth]{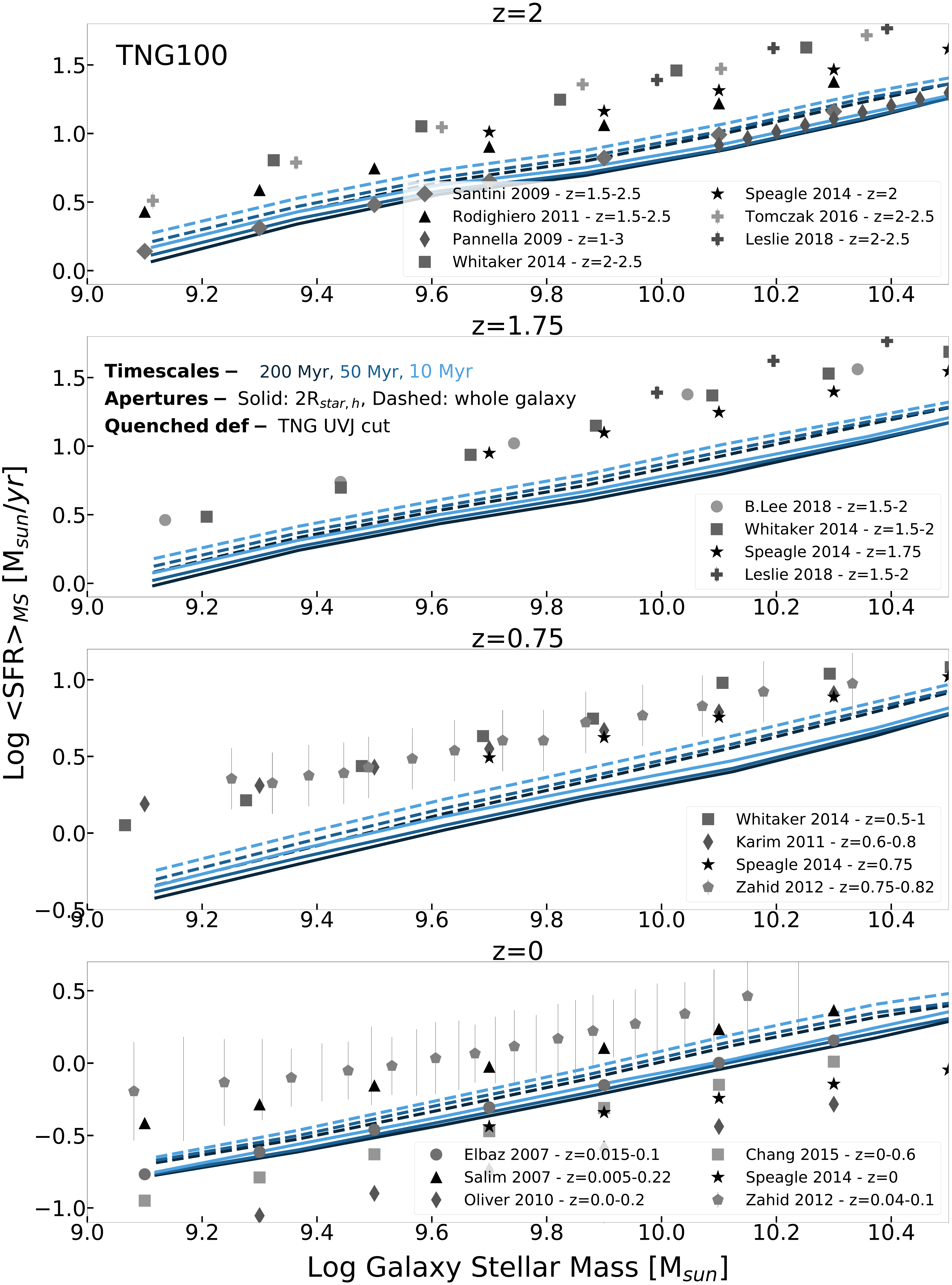}
\caption{\label{fig:SFR_tracers} Star-forming MS in TNG100 at $z=0,0.75,1.75,2$ (from bottom to top panels). The SFR is measured within $2\times R_{\rm star,h}$ (solid lines) and by summing the individual SFR of all stars gravitationally bound to the galaxy (dashed lines). The shades of blue curves denote three different timescales: 200 Myr (dark blue), 50 Myr (blue) and 10 Myr (light blue). The star-forming galaxies are selected using the TNG UVJ cut, as explained in Section \ref{systematics}. A selection of observational data is indicated in grey symbols.}
\end{figure}

In Fig.~\ref{fig:SFR_tracers}, we show the TNG100 MS in the stellar-mass range $\MS=10^9-10^{10.5} \, \Ms$. The SFRs are averaged over 200 Myr (dark blue curves), 50 Myr (blue) and 10 Myr (light blue), and measured within $2\times R_{\rm star,h}$ (solid) or by considering all the star particles gravitationally bound to the galaxy (dashed). All these variations are included to bracket the largest uncertainties due to the possible measurement choices. Star-forming galaxies are selected using the color cut based on the UVJ diagram of TNG100 galaxies discussed in Section \ref{method}. Each panel denotes a different redshift: $z=2$ (top), $z=1.75$ (top central), $z=0.75$ (bottom central) and $z=0$ (bottom). The selection of observationally-derived MS is indicated in grey symbols.

At first glance, a qualitative and quantitative agreement in the slope of the MSs between TNG and observed data is manifest.
Indeed, by fitting Eq. (\ref{ms}), we measured the MS slope of TNG galaxies in the range $\alpha_{\rm TNG} \sim 0.7-0.8$ (see Table \ref{tab:table_ms}). This settles within the observational range, $\alpha_{\rm OBS} \sim 0.6-1$, independent of redshift.

On the other hand, statements concerning the normalization of the star-forming MS and its comparison to observational data depend significantly on redshift.
At $z=0$, the MS of TNG galaxies lies in the ballpark of the observations, falling in between \cite{2010Oliver} and \cite{2012Zahid}. At intermediate redshifts -- $z=0.75$ and $z=1.75$ --, on the other hand, the TNG normalization is lower than any available observational result, by up to 0.5 dex and by at least 0.2 dex, even considering different measurement choices.
This discrepancy holds even at higher redshift ($z=2$, top panel), although to a lesser degree: the TNG normalization is $\sim0.25$ dex lower than the majority of the selected datasets. By allowing for larger apertures and smaller averaging timescales (light blue dashed curves), the TNG MS becomes progressively more consistent with the results by \cite{2009Santini} and \cite{2009Pannella}, in which galaxy SFRs are derived from UV+IR and from radio, respectively.

We stress here that the discrepancies between the TNG outcome and observations at intermediate redshifts persists even when the calibrated data from \cite{2014Speagle} are considered -- the latter being a recompilation of many observational results all converted to a common absolute calibration to overcome the limitations of the different adopted star-formation indicators.
For reference, we include these data as black stars in Fig.~\ref{fig:SFR_tracers}, by using their best MS fit \citep[Eq. 28 of][]{2014Speagle} and, as for the other observational data, by adjusting the IMF to be consistent with the Chabrier one ($\rm Log \, \MS^{\rm Chabrier} = \rm Log \, \MS^{\rm Kroupa} -0.025$; $\rm Log \, SFR^{\rm Chabrier} = \rm Log \, SFR^{\rm Kroupa} -0.15$).

In summary, despite a reasonable consistency between the TNG and observed MS at $z=0$ and despite controlling for a number of measurement choices or systematics, we find non-negligible tension at intermediate redshifts ($0.75 \leq z < 2$) in terms of normalization of the MS, with the simulated one being lower than observed. These at-face-value findings are in agreement with what outlined also for the original Illustris model (see e.g. \citealt{2015Sparre}, \citealt{2014Genel} and \citealt{2018Davidzon}), even though the Illustris' MS is higher than the TNG ones, and by several other works that compared the observed MS with theoretical models of galaxy formation \citep{2007daddi,2008Dave,2015somerville}. This could point towards some fundamental limitations of theoretical models. However, it is also important to keep in mind that, in observations, possible contamination from non-star-forming sources may have a notable impact on the derivation of the SFRs of galaxies, biasing high their estimates. For instance, the SFR of a galaxy inferred using H$\alpha$ emission can be contaminated by non star-forming sources. Indeed, this emission is due to ionized hydrogen gas and its ionization could be the result of different processes and not only due to the star formation regions. Moreover, in SDSS, the largest source of uncertainty in evaluating the galaxies SFR within the fibre is due to the possible contamination of their spectra by other sources of ionizing radiation \citep{2004Brinchmann}. For the emission in the IR band, the major contamination is thought to be the contribution of synchrotron radiation to the long-wavelength thermal dust emission: if not well-constrained, the latter can lead to an over estimation of the galaxies' SFR \citep{2001Archibald,2018Falkendal}.
In addition, galaxy SFRs can also be biased high by dust-heating from older stellar populations \cite[see e.g.][]{2010Bendo,2012Groves}.
Finally, in the UV band, especially for systems with very low SFRs, another source of contamination is from old post-AGB stars and the derived SFR should be intended only as an upper limit \citep{SFRtracers}.

\section{Summary and conclusions}
\label{summary}

A tight relation between the star-formation rate and the stellar mass of star-forming galaxies has been widely observed in the past decade, known as the star-forming \textit{main sequence}. However, debates about how it evolves in time, what drives its shape, about the existence of a downwardly bending at the high-mass end, and whether two well-distinct classes of galaxies effectively exist based on their SFRs (star-forming vs. quiescent) are still open. 

In this paper, we have used the two highest-resolution simulations currently available from the IllustrisTNG project (TNG100 and TNG300) to verify whether the underlying galaxy-physics hydrodynamical model naturally returns such a fundamentally observed property of the galaxy population as the relationship between star-formation activity and stellar mass.
In particular, we have selected all galaxies with $\MS > 10^9 \, \Ms$, sampling tens of thousands of objects between satellites and centrals, and we have quantified the SFR-$\MS$ plane, its star-forming MS, the fraction of quenched galaxies, and the UVJ diagram of the TNG simulated galaxies from $z=0$ to $z\sim2$.

Particular focus has been placed in quantifying some of the most notable measurement choices i.e. systematics that affect the derivation of star-formation rates from various tracers, and hence possibly bias the comparison between theoretical and observational results.
We have explored a number of ways to separate between star-forming and quiescent galaxies: color-color cuts in the UVJ diagram as well as selections based on the position of individual galaxies on the SFR-$\MS$ plane (see Table~\ref{tab:thresholds}). We have measured the SFR of galaxies based on the instantaneous SFR of their gas cells as well as from stars formed over the last 10, 50, 200 and 1000 Myr, in order to connect with the timescales probed by observational star-formation indicators. Moreover, we have accounted for the star formation occurring within different portions of the stellar bodies of galaxies: within 5 physical kpc, 30 physical kpc, within twice the stellar half-mass radius and throughout the whole gravitationally-bound matter: see Section \ref{systematics} for details.
We have therefore attempted some ``face-value'' comparisons with observational measurements.

Our main findings can be summarized as follows:

\begin{itemize}
\item The TNG model reproduces the general qualitative features of the observed UVJ diagram at both $z=0$ and $2$ (Fig. \ref{fig:UVJ}), with a tight clump of low-SFR galaxies residing in the top region at redder (U-V) colors and a broader cloud of star-forming galaxies that populate the bluer (U-V) colors. A low-density region, where fewer galaxies lie, separates the two populations, thus allowing us to adopt a color-color cut based on the number density distribution of TNG galaxies on the UVJ plane (Eqs.~\ref{TNG_cut}-\ref{TNG_cut2}). In fact, also observationally-proposed UVJ cuts (e.g. by \citealt{2011Whitaker} and \citealt{2009Williams}) reasonably intersect this depression, providing a validation of the TNG model in comparison to observations (see Section \ref{sec:uvj-sfr}).
\\
\item TNG galaxies populate the SFR-$\MS$ plane also in ways that are qualitatively consistent with observations: a dense region exists, where the SFR of galaxies tightly correlates with their stellar mass in the range $\MS = 10^9-10^{10.5} \, \Ms$. Such a well-defined star-forming main sequence is present at both low ($z=0$) and high ($z=2$) redshifts (see Fig.~\ref{fig:SFR-MSTAR}) and independently from the method adopted to measure the SFRs (see Appendix \ref{appendix} and Fig.~\ref{fig:SFR-MSTAR_timescales}). 
\\
\item A number of galaxies seem to {\it fall off} from the star-forming MS, by asymmetrically and continuously broadening it towards lower values of SFR. However, because of the finite numerical resolution of the simulations, galaxy SFRs are resolved only down to a minimum value, depending on the averaging timescales. For example, for 1000 Myr, the minimum resolved SFR value reads about $10^{-2.6}~\Ms$yr$^{-1}$ for TNG300 and $10^{-3.2}~\Ms$yr$^{-1}$ for TNG100, see Appendix \ref{appendix}. Below this limit, all simulated galaxies are labeled with SFR $\equiv0$. Such ``completely quenched'' galaxies are distinctly detached from the star-forming MS but feature as a distinct class of objects on the simulated SFR-$\MS$ only because of the simulation resolution limit.
\\
\item The relative fractions of quenched galaxies depend on stellar mass, redshift, and operational definition (see Fig.~\ref{fig:Qfrac_methods}). The fraction of quenched galaxies in general increases from high to low redshifts, and more so at larger masses. At $z=0$, the TNG model predicts about 80 per cent of galaxies in the $10^{10.5} \le \MS / \Ms \le 10^{11}$ range to be quiescent. This is largely due to the adopted AGN feedback. However, this estimate can vary anywhere between 70 and 85 per cent according to the operational definition of what constitutes quenched galaxies. Interestingly, within the TNG simulations, a properly identified cut in the UVJ diagram returns quenched fractions in the redshift range $0\le z \le2$ that are in agreement at the percentage level with those obtained by labeling quenched those galaxies whose SFR is 1 dex below the locus of the linearly extrapolated MS (see Section \ref{sec:qfracts} and Fig. \ref{fig:SFR-UVJ} and \ref{fig:Qfrac_methods}, top panel). 
\\
\item At $z=1.75$, the TNG model returns a fraction of massive quenched galaxies
of $\approx$ 65 per cent ($M_{\rm stars}=10^{11}-10^{11.5} ~ \Ms$). In general, we obtain a larger quenched fractions at progressively larger masses and lower redshifts. 
However, effects of resolution are not negligible in the resulting quenched fractions: at $z=0$, TNG300 returns a fractions of quenched galaxies of about 10-20 per cent  larger than TNG100, for the whole stellar mass range; at higher redshift  ($z=0.75$) the resolution effect is apparent for galaxies more massive than $10^{10.5} \, \Ms$, being $\sim$ 20 per cent larger for TNG300 with respect to TNG100. In the same mass regime, at $z=1.75$, the discrepancy between the models is less than 10 per cent.
Despite such differences between TNG100 and TNG300, the TNG model returns systematically larger quenched fractions than the original Illustris and other models: this brings the TNG galaxies overall to a better agreement with observational measurements (see Section \ref{sec:qfracts} and Fig. \ref{fig:Qfrac_methods}, bottom panels) 
\\

\item The TNG model qualitatively reproduces the salient features of the observed star-forming MS: in the mass range $\MS = 10^9-10^{10.5} \, \Ms$, the MS can be described by a single power-law (Eq. \ref{ms}); the normalization increases with redshift by $\sim 0.6$ dex from $z=0$ to $z=1$ and by $\sim 0.4$ dex from $z=1$ to $z=2$, independent of galaxy stellar mass, while the MS slope appears to be quite constant across cosmic times. At the high-mass end, if a UVJ cut is adopted to identify star-forming galaxies, the TNG MS exhibits a bending, with a shallower slope for galaxies with $\MS > 10^{10.5} \, \Ms$ (see Section \ref{general properties} and Fig. \ref{fig:SFMS}, top panel).\\

\item Differently from the quenched fractions, a comparison between the output of the TNG100 and TNG300 simulations reveals instead an excellent agreement between the two runs in terms of locus and shape of the star-forming MS, despite the different mass resolutions and sizes of the simulated volumes. On the other hand, the MS of Illustris galaxies exhibits a steeper slope with respect to TNG and a normalization that is $\sim 0.1$ dex higher than that of TNG galaxies with $\MS > 10^{10} \, \Ms$.  This scenario is related to the different recipes for feedbacks implemented in the two simulation projects, being less efficient at regulating or even halting the star formation in massive galaxies in Illustris than in TNG (see Section \ref{general properties} and Fig. \ref{fig:SFMS}, bottom panels).
\\
\item The precise locus of the MS depends on the adopted measurement choices. Shorter timescales for the averaging of the SFR estimates return higher normalizations: the MS normalization obtained by averaging the SFRs over 10 and 50 Myr (that can be associated to UV, H$\alpha$ and H$\beta$ observational tracers) is found to be $\sim 0.1$ dex ($\sim 0.2$ dex) at $z=0$ (at $z=2$) higher than that obtained by measuring SFRs over longer timescales like 200 and 1000 Myr (comparable to those of radio and IR indicators): see Fig. \ref{fig:systematics}, top panels. Smaller physical apertures imply systematically lower SFRs and a lower normalization of the MS, with discrepancies by up to 0.3 dex between e.g. the 5 pkpc aperture and larger ones for massive galaxies at low redshifts (Fig. \ref{fig:systematics}, centrals panels). Different methods to classify star-forming vs. quiescent galaxies, on the other hand, lead to negligible variations in the MS offset, thus suggesting that the MS of TNG galaxies is well captured independently from the adopted selection criterion (see Section \ref{sec:MS_systematics} and Fig. \ref{fig:systematics}, bottom panels).
\\
\item The intrinsic scatter of the star-forming MS is constant with galaxy stellar mass ($\lesssim 10^{10.5}\Ms$) and decreases with increasing redshift, reading $\sim 0.35$ dex at $z=0$ and $\sim 0.2$ dex at $z=2$ in our fiducial choices (Fig. \ref{fig:scatter}, left panel). Longer averaging timescales lead to smaller levels of scatter (e.g. 0.30 vs. 0.35 dex for 1000 vs. 200 Myr timescales), while the effects of apertures are negligible and the definition of quenched galaxies affects the quantitative assessment of the MS scatter only at the high-mass end (Fig. \ref{fig:scatter}, right panel). By accounting for measurement uncertainties in stellar mass ($\sim 0.2$ dex) and SFR ($\sim 0.3$ dex), the MS scatter is inflated by an additional 0.1 dex, making the width of the simulated MS overall consistent with observational findings (see Section \ref{sec:scatter}).\\

\item The TNG model predicts a clear and strong bimodality in the color distributions of galaxies that is visible in both the (U-V) distributions in bins of galaxy stellar mass (see Section \ref{sec:bimodality} and Fig. \ref{fig:bimodality}, bottom) and in the morphology of the UVJ diagram (Fig. \ref{fig:UVJ}), at both $z=0$ and $z=2$. However, such a color bimodality is in place even though a bimodality in the log SFR distributions is much less obvious, if not non-existent. Once galaxies with unresolved SFR values are neglected, the logarithmic SFR distributions at SFR $\gtrsim 10^{-2}~ \Ms$~yr$^{-1}$ at $z=0$ can be well described as unimodal, with a peak at the star-forming MS and a long asymmetric tail extending towards lower SFR values (see Section \ref{sec:bimodality} and Fig. \ref{fig:bimodality}, top).\\

\item Within our simulated samples of galaxies, whether the MS bends or not at the high-mass end, and by how much, depends on the criterion adopted to separate star-forming vs. quiescent galaxies. This is related to the shape of the SFR-$\MS$ plane in the TNG simulations, where a clear well-defined MS ceases to be in place at stellar masses in the range of $10^{10.7-11.5}\Ms$. When star-forming galaxies are selected based on their SFR values (e.g. SFR $>$ MS $-1$ dex, where the MS is {\it not} linearly extrapolated), the TNG MS exhibits a well-defined turnover mass: $10^{10.5}-10^{10.8} \, \Ms$, depending on the redshift. TNG UVJ selections also return a downward bending MS, albeit less pronounced than the SFR-based selection. In both cases, we recover a weak trend of larger turnover masses at larger redshifts and overall a broad agreement with observed bending (see Section \ref{sec:bending} and Fig. \ref{fig:bending}).\\

\item However, when we compare the TNG results on the MS with observational estimates taken at face value, we find that while the TNG model reproduces, qualitatively and quantitatively, the observed slope of the MS, the redshift evolution of its normalization in the stellar mass range $10^9-10^{10.5} \, \Ms$ is in tension with observations. While the simulated MS falls in the ballpark of $z=0$ observations, its normalization is systematically lower than observations imply at intermediate redshifts ($0.75 \le z \lesssim 2$), by up to $0.2-0.5$ dex and despite allowing for a number of measurement choices and systematics (see Section \ref{comparison} and Fig. \ref{fig:SFR_tracers}).
\end{itemize}

In conclusion, in this work we have demonstrated that the TNG model naturally returns the qualitative basic features of the UVJ and SFR-$\MS$ planes and the existence of a star-forming MS and of two classes of galaxies (star-forming vs. quiescent) that separate clearly at least in their color distributions. Importantly, we have demonstrated that the TNG model is capable of quenching massive galaxies at intermediate redshifts ($z\sim1-2$) in fractions that are in better agreement with observational findings than previous numerical model. At the same time, and most probably for different reasons, an apparent underestimation of the locus of the star-forming main sequence is still in place in the TNG model at $0.75 \le z \lesssim 2$ in comparison to observations. We have quantified the effects of a number of measurement choices in relation to SFR timescales, galaxy apertures, and quenched definitions, and hence provided guidelines for the systematic biases that may affect comparisons across datasets, whether simulated or observed. Such systematics can affect the locus of the star-forming MS by up to $0.2-0.3$ dex. For example, the fact that smaller physical apertures shift the locus of the MS towards lower values is an indication that star formation can occur in non-negligible amounts also in regions that extend far beyond the main bright body of galaxies. 

The findings outlined in this paper in relation to the comparison to observations might indicate the following learning point: the TNG model captures reasonably well the effects of the dominant quenching mechanism at the high-mass end (i.e. the AGN feedback) but might require important adjustments in terms of the feedback mechanisms that regulate (without necessarily halting) star formation in $\gtrsim 10^9\Ms$ galaxies at the cosmic noon (i.e. stellar feedback). However, two important conceptual caveats remain before conclusive messages can be derived. Firstly, in this paper we have not ventured into a thorough discussion of the biases that may affect observational estimates of the SFRs of galaxies. From the theoretical view point, the SFRs we can measure from the simulated galaxies are unbiased, while non-negligible uncertainties affect the mapping between observed light of certain star-formation indicators and SFRs, because of contamination or dust reddening. For example,  \cite{2018Theios} have added to a vast body of literature and argued that observed SFR estimates can be afflicted by an additional 0.5 dex of systematic uncertainties, making previously published MS relations possibly biased high (at the intermediate redshifts of interest here). 
In \cite{2018Sorba} is showed that observational indicators of galaxy stellar mass may be biased, causing a systematic effect on the SSFR. Just recently, \cite{2018Leja} have re-inferred the SFRs and galaxy stellar masses of 3D-HST galaxies at $0.5<z<2.5$ taking advantage of a new high-dimensional model and have shown that, among other findings, the newly-derived galaxy SFRs can be 0.1-1 dex lower than the UV+IR-based ones. In fact, it is also important to notice that, while the SFRs of galaxies on the TNG MS appear low around e.g. $z\sim1$, a zeroth-order comparison of the TNG galaxy stellar mass function to observational findings does not raise any obvious flag: see Fig. 14 of \cite{2018Pillepich}. The one fundamental advantage in simulations is their self-consistency across times: the galaxy mass functions at certain times are the result of the time integral of the SFRs at previous times and smaller masses. If anything, the TNG galaxy mass functions at $0.5<z<2$ in certain mass regimes (just left of the knee) are somewhat higher than observational constraints. On the other hand, tensions between the observed evolution of the galaxy mass or luminosity functions and the observed sSFRs -- particularly at $z\gtrsim2$ -- have been pointed out by several authors \citep[see e.g][and references therein]{2015Leja,2016yu,2018Behroozi}, albeit the actual reason of these tensions is still unclear. We postpone to future work the task of placing into a coherent and consistent picture all these statements and the quantification of the effects of environment on the star formation activity of TNG galaxies. 

%

\section*{Aknowledgement}
The authors would like to thank Kate Whitaker, Camilla Pacifici, Sandy Faber, Arjen van der Wel, Bruno Henriques, Adam Muzzin, and Robert Feldmann for useful discussions and input and all the participants of the Lorentz center workshop on `A Decade of the Star-Forming Main Sequence' for inspiring conversations. 
The authors are also grateful to Martin Sparre, Gandhali Joshi, Gerg{\"o} Popping and Manuel Arca Sedda for comments on a draft version of this paper and to Sarah Leslie for sharing her data.
SG, through the Flatiron Institute, is supported by the Simons Foundation. 
MV acknowledges support through an MIT RSC award, a Kavli Research Investment Fund, NASA ATP grant NNX17AG29G, and NSF grants AST-1814053 and AST-1814259.
The flagship simulations of the IllustrisTNG
project used in this work have been run on the HazelHen Cray
XC40-system at the High Performance Computing Center Stuttgart
as part of project GCS-ILLU of the Gauss centres for Super-computing(GCS). Ancillary and test runs of the project were also run on the Stampede supercomputer at TACC/XSEDE (allocation
AST140063), at the Hydra and Draco supercomputers at the Max
Planck Computing and Data Facility, and on the MIT/Harvard computing facilities supported by FAS and MIT MKI.


\bibliographystyle{mn2e}
\bibliography{biblio}

\begin{thebibliography}{}
\makeatletter
\relax
\def\mn@urlcharsother{\let\do\@makeother \do\$\do\&\do\#\do\^\do\_\do\%\do\~}
\def\mn@doi{\begingroup\mn@urlcharsother \@ifnextchar [ {\mn@doi@}
  {\mn@doi@[]}}
\def\mn@doi@[#1]#2{\def\@tempa{#1}\ifx\@tempa\@empty \href
  {http://dx.doi.org/#2} {doi:#2}\else \href {http://dx.doi.org/#2} {#1}\fi
  \endgroup}
\def\mn@eprint#1#2{\mn@eprint@#1:#2::\@nil}
\def\mn@eprint@arXiv#1{\href {http://arxiv.org/abs/#1} {{\tt arXiv:#1}}}
\def\mn@eprint@dblp#1{\href {http://dblp.uni-trier.de/rec/bibtex/#1.xml}
  {dblp:#1}}
\def\mn@eprint@#1:#2:#3:#4\@nil{\def\@tempa {#1}\def\@tempb {#2}\def\@tempc
  {#3}\ifx \@tempc \@empty \let \@tempc \@tempb \let \@tempb \@tempa \fi \ifx
  \@tempb \@empty \def\@tempb {arXiv}\fi \@ifundefined
  {mn@eprint@\@tempb}{\@tempb:\@tempc}{\expandafter \expandafter \csname
  mn@eprint@\@tempb\endcsname \expandafter{\@tempc}}}

\bibitem[\protect\citeauthoryear{{Abramson}, {Kelson}, {Dressler}, {Poggianti},
  {Gladders}, {Oemler}  \& {Vulcani}}{{Abramson} et~al.}{2014}]{2014Abramson}
{Abramson} L.~E.,  {Kelson} D.~D.,  {Dressler} A.,  {Poggianti} B.,  {Gladders}
  M.~D.,  {Oemler} Jr. A.,   {Vulcani} B.,  2014, \mn@doi [\apjl]
  {10.1088/2041-8205/785/2/L36}, \href
  {http://adsabs.harvard.edu/abs/2014ApJ...785L..36A} {785, L36}

\bibitem[\protect\citeauthoryear{{Archibald}, {Dunlop}, {Hughes}, {Rawlings},
  {Eales}  \& {Ivison}}{{Archibald} et~al.}{2001}]{2001Archibald}
{Archibald} E.~N.,  {Dunlop} J.~S.,  {Hughes} D.~H.,  {Rawlings} S.,  {Eales}
  S.~A.,   {Ivison} R.~J.,  2001, \mn@doi [\mnras]
  {10.1046/j.1365-8711.2001.04188.x}, \href
  {https://ui.adsabs.harvard.edu/#abs/2001MNRAS.323..417A} {323, 417}

\bibitem[\protect\citeauthoryear{{Bah{\'e}} \& {McCarthy}}{{Bah{\'e}} \&
  {McCarthy}}{2015}]{2015Bahe}
{Bah{\'e}} Y.~M.,  {McCarthy} I.~G.,  2015, \mn@doi [\mnras]
  {10.1093/mnras/stu2293}, \href
  {http://adsabs.harvard.edu/abs/2015MNRAS.447..969B} {447, 969}

\bibitem[\protect\citeauthoryear{{Baldry}, {Balogh}, {Bower}, {Glazebrook}  \&
  {Nichol}}{{Baldry} et~al.}{2004}]{2004Baldry}
{Baldry} I.~K.,  {Balogh} M.~L.,  {Bower} R.,  {Glazebrook} K.,   {Nichol}
  R.~C.,  2004, in {Allen} R.~E.,  {Nanopoulos} D.~V.,   {Pope} C.~N.,  eds,
  American Institute of Physics Conference Series Vol. 743, The New Cosmology:
  Conference on Strings and Cosmology. pp 106--119 (\mn@eprint {}
  {astro-ph/0410603}), \mn@doi{10.1063/1.1848322}

\bibitem[\protect\citeauthoryear{{Balogh}, {Baldry}, {Nichol}, {Miller},
  {Bower}  \& {Glazebrook}}{{Balogh} et~al.}{2004}]{2004Balogh}
{Balogh} M.~L.,  {Baldry} I.~K.,  {Nichol} R.,  {Miller} C.,  {Bower} R.,
  {Glazebrook} K.,  2004, \mn@doi [\apjl] {10.1086/426079}, \href
  {http://adsabs.harvard.edu/abs/2004ApJ...615L.101B} {615, L101}

\bibitem[\protect\citeauthoryear{{Behroozi}, {Wechsler}, {Hearin}  \&
  {Conroy}}{{Behroozi} et~al.}{2018}]{2018Behroozi}
{Behroozi} P.,  {Wechsler} R.,  {Hearin} A.,   {Conroy} C.,  2018, arXiv
  e-prints, \href {http://adsabs.harvard.edu/abs/2018arXiv180607893B} {}

\bibitem[\protect\citeauthoryear{{Bendo} et~al.,}{{Bendo}
  et~al.}{2010}]{2010Bendo}
{Bendo} G.~J.,  et~al., 2010, \mn@doi [\aap] {10.1051/0004-6361/201014568},
  \href {http://adsabs.harvard.edu/abs/2010A%26A...518L..65B} {518, L65}

\bibitem[\protect\citeauthoryear{{Bouch{\'e}} et~al.,}{{Bouch{\'e}}
  et~al.}{2010}]{2010Bouche}
{Bouch{\'e}} N.,  et~al., 2010, \mn@doi [\apj] {10.1088/0004-637X/718/2/1001},
  \href {https://ui.adsabs.harvard.edu/\#abs/2010ApJ...718.1001B} {718, 1001}

\bibitem[\protect\citeauthoryear{{Brinchmann}, {Charlot}, {White}, {Tremonti},
  {Kauffmann}, {Heckman}  \& {Brinkmann}}{{Brinchmann}
  et~al.}{2004}]{2004Brinchmann}
{Brinchmann} J.,  {Charlot} S.,  {White} S.~D.~M.,  {Tremonti} C.,  {Kauffmann}
  G.,  {Heckman} T.,   {Brinkmann} J.,  2004, \mn@doi [\mnras]
  {10.1111/j.1365-2966.2004.07881.x}, \href
  {http://adsabs.harvard.edu/abs/2004MNRAS.351.1151B} {351, 1151}

\bibitem[\protect\citeauthoryear{{Chabrier}}{{Chabrier}}{2003}]{2003Chabrier}
{Chabrier} G.,  2003, \mn@doi [\pasp] {10.1086/376392}, \href
  {http://adsabs.harvard.edu/abs/2003PASP..115..763C} {115, 763}

\bibitem[\protect\citeauthoryear{{Chang}, {van der Wel}, {da Cunha}  \&
  {Rix}}{{Chang} et~al.}{2015}]{2015Chang}
{Chang} Y.-Y.,  {van der Wel} A.,  {da Cunha} E.,   {Rix} H.-W.,  2015, \mn@doi
  [The Astrophysical Journal Supplement Series] {10.1088/0067-0049/219/1/8},
  \href {https://ui.adsabs.harvard.edu/#abs/2015ApJS..219....8C} {219, 8}

\bibitem[\protect\citeauthoryear{{Conroy} \& {Gunn}}{{Conroy} \&
  {Gunn}}{2010}]{2010Conroy}
{Conroy} C.,  {Gunn} J.~E.,  2010, \mn@doi [\apj]
  {10.1088/0004-637X/712/2/833}, \href
  {http://adsabs.harvard.edu/abs/2010ApJ...712..833C} {712, 833}

\bibitem[\protect\citeauthoryear{{Conroy}, {Gunn}  \& {White}}{{Conroy}
  et~al.}{2009}]{2009Conroy}
{Conroy} C.,  {Gunn} J.~E.,   {White} M.,  2009, \mn@doi [\apj]
  {10.1088/0004-637X/699/1/486}, \href
  {http://adsabs.harvard.edu/abs/2009ApJ...699..486C} {699, 486}

\bibitem[\protect\citeauthoryear{{Daddi} et~al.,}{{Daddi}
  et~al.}{2007}]{2007daddi}
{Daddi} E.,  et~al., 2007, \mn@doi [\apj] {10.1086/521818}, \href
  {https://ui.adsabs.harvard.edu/#abs/2007ApJ...670..156D} {670, 156}

\bibitem[\protect\citeauthoryear{{Darvish}, {Mobasher}, {Sobral}, {Rettura},
  {Scoville}, {Faisst}  \& {Capak}}{{Darvish} et~al.}{2016}]{2016darvish}
{Darvish} B.,  {Mobasher} B.,  {Sobral} D.,  {Rettura} A.,  {Scoville} N.,
  {Faisst} A.,   {Capak} P.,  2016, \mn@doi [\apj]
  {10.3847/0004-637X/825/2/113}, \href
  {http://adsabs.harvard.edu/abs/2016ApJ...825..113D} {825, 113}

\bibitem[\protect\citeauthoryear{{Darvish}, {Mobasher}, {Martin}, {Sobral},
  {Scoville}, {Stroe}, {Hemmati}  \& {Kartaltepe}}{{Darvish}
  et~al.}{2017}]{2017Darvish}
{Darvish} B.,  {Mobasher} B.,  {Martin} D.~C.,  {Sobral} D.,  {Scoville} N.,
  {Stroe} A.,  {Hemmati} S.,   {Kartaltepe} J.,  2017, \mn@doi [\apj]
  {10.3847/1538-4357/837/1/16}, \href
  {http://adsabs.harvard.edu/abs/2017ApJ...837...16D} {837, 16}

\bibitem[\protect\citeauthoryear{{Dav{\'e}}}{{Dav{\'e}}}{2008}]{2008Dave}
{Dav{\'e}} R.,  2008, \mn@doi [\mnras] {10.1111/j.1365-2966.2008.12866.x},
  \href {https://ui.adsabs.harvard.edu/#abs/2008MNRAS.385..147D} {385, 147}

\bibitem[\protect\citeauthoryear{{Dav{\'e}}, {Finlator}  \&
  {Oppenheimer}}{{Dav{\'e}} et~al.}{2012}]{2012Dave}
{Dav{\'e}} R.,  {Finlator} K.,   {Oppenheimer} B.~D.,  2012, \mn@doi [\mnras]
  {10.1111/j.1365-2966.2011.20148.x}, \href
  {https://ui.adsabs.harvard.edu/\#abs/2012MNRAS.421...98D} {421, 98}

\bibitem[\protect\citeauthoryear{{Dav{\'e}}, {Rafieferantsoa}  \&
  {Thompson}}{{Dav{\'e}} et~al.}{2017}]{2017Dave}
{Dav{\'e}} R.,  {Rafieferantsoa} M.~H.,   {Thompson} R.~J.,  2017, \mn@doi
  [\mnras] {10.1093/mnras/stx1693}, \href
  {http://adsabs.harvard.edu/abs/2017MNRAS.471.1671D} {471, 1671}

\bibitem[\protect\citeauthoryear{{Davidzon}, {Ilbert}, {Faisst}, {Sparre}  \&
  {Capak}}{{Davidzon} et~al.}{2018}]{2018Davidzon}
{Davidzon} I.,  {Ilbert} O.,  {Faisst} A.~L.,  {Sparre} M.,   {Capak} P.~L.,
  2018, \mn@doi [\apj] {10.3847/1538-4357/aaa19e}, \href
  {https://ui.adsabs.harvard.edu/\#abs/2018ApJ...852..107D} {852, 107}

\bibitem[\protect\citeauthoryear{{Davies} et~al.,}{{Davies}
  et~al.}{2017}]{2017Davies}
{Davies} L.~J.~M.,  et~al., 2017, \mn@doi [\mnras] {10.1093/mnras/stw3080},
  \href {https://ui.adsabs.harvard.edu/\#abs/2017MNRAS.466.2312D} {466, 2312}

\bibitem[\protect\citeauthoryear{{Davies} et~al.,}{{Davies}
  et~al.}{2018}]{2018Davies}
{Davies} L.~J.~M.,  et~al., 2018, \mn@doi [\mnras] {10.1093/mnras/sty2957},
  \href {http://adsabs.harvard.edu/abs/2018MNRAS.tmp.2825D} {}

\bibitem[\protect\citeauthoryear{{Davis}, {Efstathiou}, {Frenk}  \&
  {White}}{{Davis} et~al.}{1985}]{1985Davis}
{Davis} M.,  {Efstathiou} G.,  {Frenk} C.~S.,   {White} S.~D.~M.,  1985,
  \mn@doi [\apj] {10.1086/163168}, \href
  {http://adsabs.harvard.edu/abs/1985ApJ...292..371D} {292, 371}

\bibitem[\protect\citeauthoryear{{De Lucia}, {Weinmann}, {Poggianti},
  {Arag{\'o}n-Salamanca}  \& {Zaritsky}}{{De Lucia} et~al.}{2012}]{2012Delucia}
{De Lucia} G.,  {Weinmann} S.,  {Poggianti} B.~M.,  {Arag{\'o}n-Salamanca} A.,
   {Zaritsky} D.,  2012, \mn@doi [\mnras] {10.1111/j.1365-2966.2012.20983.x},
  \href {http://adsabs.harvard.edu/abs/2012MNRAS.423.1277D} {423, 1277}

\bibitem[\protect\citeauthoryear{{Eales}, {de Vis}, {Smith}, {Appah}, {Ciesla},
  {Duffield}  \& {Schofield}}{{Eales} et~al.}{2017}]{2017Eales}
{Eales} S.,  {de Vis} P.,  {Smith} M.~W.~L.,  {Appah} K.,  {Ciesla} L.,
  {Duffield} C.,   {Schofield} S.,  2017, \mn@doi [\mnras]
  {10.1093/mnras/stw2875}, \href
  {http://adsabs.harvard.edu/abs/2017MNRAS.465.3125E} {465, 3125}

\bibitem[\protect\citeauthoryear{{Eales} et~al.,}{{Eales}
  et~al.}{2018}]{2018Eales}
{Eales} S.,  et~al., 2018, \mn@doi [\mnras] {10.1093/mnras/stx2548}, \href
  {http://adsabs.harvard.edu/abs/2018MNRAS.473.3507E} {473, 3507}

\bibitem[\protect\citeauthoryear{{Elbaz} et~al.,}{{Elbaz}
  et~al.}{2007}]{2007Elbaz}
{Elbaz} D.,  et~al., 2007, \mn@doi [\aap] {10.1051/0004-6361:20077525}, \href
  {http://adsabs.harvard.edu/abs/2007A%26A...468...33E} {468, 33}

\bibitem[\protect\citeauthoryear{{Falkendal} et~al.,}{{Falkendal}
  et~al.}{2018}]{2018Falkendal}
{Falkendal} T.,  et~al., 2018, preprint, \href
  {http://adsabs.harvard.edu/abs/2018arXiv180909427F} {} (\mn@eprint {arXiv}
  {1809.09427})

\bibitem[\protect\citeauthoryear{{Fang} et~al.,}{{Fang}
  et~al.}{2018}]{2018Fang}
{Fang} J.~J.,  et~al., 2018, \mn@doi [\apj] {10.3847/1538-4357/aabcba}, \href
  {http://adsabs.harvard.edu/abs/2018ApJ...858..100F} {858, 100}

\bibitem[\protect\citeauthoryear{{Feldmann}}{{Feldmann}}{2017}]{2017Feldmann}
{Feldmann} R.,  2017, \mn@doi [\mnras] {10.1093/mnrasl/slx073}, \href
  {http://adsabs.harvard.edu/abs/2017MNRAS.470L..59F} {470, L59}

\bibitem[\protect\citeauthoryear{{Foreman-Mackey}, {Sick}  \&
  {Johnson}}{{Foreman-Mackey} et~al.}{2014}]{2014ForemanMackey}
{Foreman-Mackey} D.,  {Sick} J.,   {Johnson} B.,  2014, Technical report,
  python-fsps: Python bindings to FSPS (v0.1.1) (Version v0.1.1). Zenodo.

\bibitem[\protect\citeauthoryear{{Furlong} et~al.,}{{Furlong}
  et~al.}{2015}]{2015Furlong}
{Furlong} M.,  et~al., 2015, \mn@doi [\mnras] {10.1093/mnras/stv852}, \href
  {https://ui.adsabs.harvard.edu/#abs/2015MNRAS.450.4486F} {450, 4486}

\bibitem[\protect\citeauthoryear{{Gavazzi} et~al.,}{{Gavazzi}
  et~al.}{2015}]{2015Gavazzi}
{Gavazzi} G.,  et~al., 2015, \mn@doi [\aap] {10.1051/0004-6361/201425351},
  \href {http://adsabs.harvard.edu/abs/2015A%26A...580A.116G} {580, A116}

\bibitem[\protect\citeauthoryear{{Genel} et~al.,}{{Genel}
  et~al.}{2014}]{2014Genel}
{Genel} S.,  et~al., 2014, \mn@doi [\mnras] {10.1093/mnras/stu1654}, \href
  {http://adsabs.harvard.edu/abs/2014MNRAS.445..175G} {445, 175}

\bibitem[\protect\citeauthoryear{{Genel} et~al.,}{{Genel}
  et~al.}{2018}]{2018Genel}
{Genel} S.,  et~al., 2018, \mn@doi [\mnras] {10.1093/mnras/stx3078}, \href
  {http://adsabs.harvard.edu/abs/2018MNRAS.474.3976G} {474, 3976}

\bibitem[\protect\citeauthoryear{{Groves} et~al.,}{{Groves}
  et~al.}{2012}]{2012Groves}
{Groves} B.,  et~al., 2012, \mn@doi [\mnras]
  {10.1111/j.1365-2966.2012.21696.x}, \href
  {http://adsabs.harvard.edu/abs/2012MNRAS.426..892G} {426, 892}

\bibitem[\protect\citeauthoryear{{Guidi}, {Scannapieco}, {Walcher}  \&
  {Gallazzi}}{{Guidi} et~al.}{2016}]{2016Guidi}
{Guidi} G.,  {Scannapieco} C.,  {Walcher} J.,   {Gallazzi} A.,  2016, \mn@doi
  [\mnras] {10.1093/mnras/stw1790}, \href
  {http://adsabs.harvard.edu/abs/2016MNRAS.462.2046G} {462, 2046}

\bibitem[\protect\citeauthoryear{{Habouzit} et~al.,}{{Habouzit}
  et~al.}{2018}]{2018Habouzit}
{Habouzit} M.,  et~al., 2018, preprint, \href
  {http://adsabs.harvard.edu/abs/2018arXiv180905588H} {} (\mn@eprint {arXiv}
  {1809.05588})

\bibitem[\protect\citeauthoryear{{Hahn} et~al.,}{{Hahn}
  et~al.}{2018}]{2018Hahn}
{Hahn} C.,  et~al., 2018, preprint, \href
  {http://adsabs.harvard.edu/abs/2018arXiv180901665H} {} (\mn@eprint {arXiv}
  {1809.01665})

\bibitem[\protect\citeauthoryear{{Henriques}, {White}, {Thomas}, {Angulo},
  {Guo}, {Lemson}  \& {Wang}}{{Henriques} et~al.}{2017}]{2017Henriques}
{Henriques} B.~M.~B.,  {White} S.~D.~M.,  {Thomas} P.~A.,  {Angulo} R.~E.,
  {Guo} Q.,  {Lemson} G.,   {Wang} W.,  2017, \mn@doi [\mnras]
  {10.1093/mnras/stx1010}, \href
  {http://adsabs.harvard.edu/abs/2017MNRAS.469.2626H} {469, 2626}

\bibitem[\protect\citeauthoryear{{Hirschmann}, {De Lucia}, {Wilman},
  {Weinmann}, {Iovino}, {Cucciati}, {Zibetti}  \& {Villalobos}}{{Hirschmann}
  et~al.}{2014}]{2014Hirschmann}
{Hirschmann} M.,  {De Lucia} G.,  {Wilman} D.,  {Weinmann} S.,  {Iovino} A.,
  {Cucciati} O.,  {Zibetti} S.,   {Villalobos} {\'A}.,  2014, \mn@doi [\mnras]
  {10.1093/mnras/stu1609}, \href
  {http://adsabs.harvard.edu/abs/2014MNRAS.444.2938H} {444, 2938}

\bibitem[\protect\citeauthoryear{{Ilbert}, {McCracken}, {Le F{\`e}vre}, {Capak}
   \& {et al.}}{{Ilbert} et~al.}{2013}]{2013Ilbert}
{Ilbert} O.,  {McCracken} H.~J.,  {Le F{\`e}vre} O.,  {Capak} P.,   {et al.}
  2013, \mn@doi [\aap] {10.1051/0004-6361/201321100}, \href
  {http://adsabs.harvard.edu/abs/2013A%26A...556A..55I} {556, A55}

\bibitem[\protect\citeauthoryear{{Ilbert} et~al.,}{{Ilbert}
  et~al.}{2015}]{2015Ilbert}
{Ilbert} O.,  et~al., 2015, \mn@doi [\aap] {10.1051/0004-6361/201425176}, \href
  {http://adsabs.harvard.edu/abs/2015A%26A...579A...2I} {579, A2}

\bibitem[\protect\citeauthoryear{{Jian} et~al.,}{{Jian}
  et~al.}{2018}]{2018Jian}
{Jian} H.-Y.,  et~al., 2018, \mn@doi [\pasj] {10.1093/pasj/psx096}, \href
  {http://adsabs.harvard.edu/abs/2018PASJ...70S..23J} {70, S23}

\bibitem[\protect\citeauthoryear{{Kang} \& {van den Bosch}}{{Kang} \& {van den
  Bosch}}{2008}]{2008Kang}
{Kang} X.,  {van den Bosch} F.~C.,  2008, \mn@doi [\apjl] {10.1086/587620},
  \href {http://adsabs.harvard.edu/abs/2008ApJ...676L.101K} {676, L101}

\bibitem[\protect\citeauthoryear{{Karim} et~al.,}{{Karim}
  et~al.}{2011}]{2011Karim}
{Karim} A.,  et~al., 2011, \mn@doi [\apj] {10.1088/0004-637X/730/2/61}, \href
  {http://adsabs.harvard.edu/abs/2011ApJ...730...61K} {730, 61}

\bibitem[\protect\citeauthoryear{{Katsianis} et~al.,}{{Katsianis}
  et~al.}{2017}]{2017Katsianis}
{Katsianis} A.,  et~al., 2017, \mn@doi [\mnras] {10.1093/mnras/stx2020}, \href
  {http://adsabs.harvard.edu/abs/2017MNRAS.472..919K} {472, 919}

\bibitem[\protect\citeauthoryear{{Kauffmann} et~al.,}{{Kauffmann}
  et~al.}{2003}]{2003Kauffmann}
{Kauffmann} G.,  et~al., 2003, \mn@doi [\mnras]
  {10.1046/j.1365-8711.2003.06292.x}, \href
  {http://adsabs.harvard.edu/abs/2003MNRAS.341...54K} {341, 54}

\bibitem[\protect\citeauthoryear{{Kauffmann}, {White}, {Heckman}, {M{\'e}nard},
  {Brinchmann}, {Charlot}, {Tremonti}  \& {Brinkmann}}{{Kauffmann}
  et~al.}{2004}]{2004Kauffmann}
{Kauffmann} G.,  {White} S.~D.~M.,  {Heckman} T.~M.,  {M{\'e}nard} B.,
  {Brinchmann} J.,  {Charlot} S.,  {Tremonti} C.,   {Brinkmann} J.,  2004,
  \mn@doi [\mnras] {10.1111/j.1365-2966.2004.08117.x}, \href
  {http://adsabs.harvard.edu/abs/2004MNRAS.353..713K} {353, 713}

\bibitem[\protect\citeauthoryear{{Kennicutt}}{{Kennicutt}}{1989}]{1989Kennicutt}
{Kennicutt} Jr. R.~C.,  1989, \mn@doi [\apj] {10.1086/167834}, \href
  {http://adsabs.harvard.edu/abs/1989ApJ...344..685K} {344, 685}

\bibitem[\protect\citeauthoryear{{Kennicutt} \& {Evans}}{{Kennicutt} \&
  {Evans}}{2012}]{2012Kennicutt}
{Kennicutt} R.~C.,  {Evans} N.~J.,  2012, \mn@doi [\araa]
  {10.1146/annurev-astro-081811-125610}, \href
  {http://adsabs.harvard.edu/abs/2012ARA%26A..50..531K} {50, 531}

\bibitem[\protect\citeauthoryear{Kennicutt Jr. R.~C.}{Kennicutt Jr.
  R.~C.}{1998}]{SFRtracers}
Kennicutt Jr. R.~C. Schweizer~F. B. J.~E.,  1998, Galaxies: Interactions and
  Induced Star Formation.
Springer

\bibitem[\protect\citeauthoryear{{Kewley}, {Jansen}  \& {Geller}}{{Kewley}
  et~al.}{2005}]{2005Kewley}
{Kewley} L.~J.,  {Jansen} R.~A.,   {Geller} M.~J.,  2005, \mn@doi [\pasp]
  {10.1086/428303}, \href {http://adsabs.harvard.edu/abs/2005PASP..117..227K}
  {117, 227}

\bibitem[\protect\citeauthoryear{{Kroupa}}{{Kroupa}}{2001}]{2001kroupa}
{Kroupa} P.,  2001, \mn@doi [\mnras] {10.1046/j.1365-8711.2001.04022.x}, \href
  {http://adsabs.harvard.edu/abs/2001MNRAS.322..231K} {322, 231}

\bibitem[\protect\citeauthoryear{{Lee} et~al.,}{{Lee} et~al.}{2015}]{2015NLee}
{Lee} N.,  et~al., 2015, \mn@doi [\apj] {10.1088/0004-637X/801/2/80}, \href
  {http://adsabs.harvard.edu/abs/2015ApJ...801...80L} {801, 80}

\bibitem[\protect\citeauthoryear{{Lee} et~al.,}{{Lee} et~al.}{2018}]{2018BLee}
{Lee} B.,  et~al., 2018, \mn@doi [\apj] {10.3847/1538-4357/aaa40f}, \href
  {http://adsabs.harvard.edu/abs/2018ApJ...853..131L} {853, 131}

\bibitem[\protect\citeauthoryear{{Leja}, {van Dokkum}, {Franx}  \&
  {Whitaker}}{{Leja} et~al.}{2015}]{2015Leja}
{Leja} J.,  {van Dokkum} P.~G.,  {Franx} M.,   {Whitaker} K.~E.,  2015, \mn@doi
  [\apj] {10.1088/0004-637X/798/2/115}, \href
  {http://adsabs.harvard.edu/abs/2015ApJ...798..115L} {798, 115}

\bibitem[\protect\citeauthoryear{{Leja} et~al.,}{{Leja}
  et~al.}{2018}]{2018Leja}
{Leja} J.,  et~al., 2018, arXiv e-prints, \href
  {https://ui.adsabs.harvard.edu/\#abs/2018arXiv181205608L} {p.
  arXiv:1812.05608}

\bibitem[\protect\citeauthoryear{{Lilly}, {Carollo}, {Pipino}, {Renzini}  \&
  {Peng}}{{Lilly} et~al.}{2013}]{2013Lilly}
{Lilly} S.~J.,  {Carollo} C.~M.,  {Pipino} A.,  {Renzini} A.,   {Peng} Y.,
  2013, \mn@doi [\apj] {10.1088/0004-637X/772/2/119}, \href
  {https://ui.adsabs.harvard.edu/\#abs/2013ApJ...772..119L} {772, 119}

\bibitem[\protect\citeauthoryear{{Lin} et~al.,}{{Lin} et~al.}{2014}]{2014Lin}
{Lin} L.,  et~al., 2014, \mn@doi [\apj] {10.1088/0004-637X/782/1/33}, \href
  {http://adsabs.harvard.edu/abs/2014ApJ...782...33L} {782, 33}

\bibitem[\protect\citeauthoryear{{Lovell} et~al.,}{{Lovell}
  et~al.}{2018}]{2018Lovell}
{Lovell} M.~R.,  et~al., 2018, \mn@doi [\mnras] {10.1093/mnras/sty2339}, \href
  {http://adsabs.harvard.edu/abs/2018MNRAS.481.1950L} {481, 1950}

\bibitem[\protect\citeauthoryear{{Marinacci} et~al.,}{{Marinacci}
  et~al.}{2018}]{2018Marinacci}
{Marinacci} F.,  et~al., 2018, \mn@doi [\mnras] {10.1093/mnras/sty2206}, \href
  {http://adsabs.harvard.edu/abs/2018MNRAS.480.5113M} {480, 5113}

\bibitem[\protect\citeauthoryear{{McGee}, {Balogh}, {Wilman}, {Bower},
  {Mulchaey}, {Parker}  \& {Oemler}}{{McGee} et~al.}{2011}]{2011Mcgee}
{McGee} S.~L.,  {Balogh} M.~L.,  {Wilman} D.~J.,  {Bower} R.~G.,  {Mulchaey}
  J.~S.,  {Parker} L.~C.,   {Oemler} A.,  2011, \mn@doi [\mnras]
  {10.1111/j.1365-2966.2010.18189.x}, \href
  {http://adsabs.harvard.edu/abs/2011MNRAS.413..996M} {413, 996}

\bibitem[\protect\citeauthoryear{{Mistani} et~al.,}{{Mistani}
  et~al.}{2016}]{2016Mistani}
{Mistani} P.~A.,  et~al., 2016, \mn@doi [\mnras] {10.1093/mnras/stv2435}, \href
  {http://adsabs.harvard.edu/abs/2016MNRAS.455.2323M} {455, 2323}

\bibitem[\protect\citeauthoryear{{Mitra}, {Dav{\'e}}  \& {Finlator}}{{Mitra}
  et~al.}{2015}]{2015Mitra}
{Mitra} S.,  {Dav{\'e}} R.,   {Finlator} K.,  2015, \mn@doi [\mnras]
  {10.1093/mnras/stv1387}, \href
  {https://ui.adsabs.harvard.edu/\#abs/2015MNRAS.452.1184M} {452, 1184}

\bibitem[\protect\citeauthoryear{{Mitra}, {Dav{\'e}}, {Simha}  \&
  {Finlator}}{{Mitra} et~al.}{2017}]{2017Mitra}
{Mitra} S.,  {Dav{\'e}} R.,  {Simha} V.,   {Finlator} K.,  2017, \mn@doi
  [\mnras] {10.1093/mnras/stw2527}, \href
  {https://ui.adsabs.harvard.edu/\#abs/2017MNRAS.464.2766M} {464, 2766}

\bibitem[\protect\citeauthoryear{{Moustakas} et~al.,}{{Moustakas}
  et~al.}{2013}]{2013moustakas}
{Moustakas} J.,  et~al., 2013, \mn@doi [\apj] {10.1088/0004-637X/767/1/50},
  \href {http://adsabs.harvard.edu/abs/2013ApJ...767...50M} {767, 50}

\bibitem[\protect\citeauthoryear{{Muzzin}, {Marchesini}, {Stefanon}, {Franx}
  \& {et al.}}{{Muzzin} et~al.}{2013}]{2013Muzzin}
{Muzzin} A.,  {Marchesini} D.,  {Stefanon} M.,  {Franx} M.,   {et al.} 2013,
  \mn@doi [\apj] {10.1088/0004-637X/777/1/18}, \href
  {http://adsabs.harvard.edu/abs/2013ApJ...777...18M} {777, 18}

\bibitem[\protect\citeauthoryear{{Naiman} et~al.,}{{Naiman}
  et~al.}{2018}]{2018Naiman}
{Naiman} J.~P.,  et~al., 2018, \mn@doi [\mnras] {10.1093/mnras/sty618}, \href
  {http://adsabs.harvard.edu/abs/2018MNRAS.477.1206N} {477, 1206}

\bibitem[\protect\citeauthoryear{{Nelson} et~al.,}{{Nelson}
  et~al.}{2015}]{2015Nelson}
{Nelson} D.,  et~al., 2015, \mn@doi [Astronomy and Computing]
  {10.1016/j.ascom.2015.09.003}, \href
  {http://adsabs.harvard.edu/abs/2015A%26C....13...12N} {13, 12}

\bibitem[\protect\citeauthoryear{{Nelson} et~al.,}{{Nelson}
  et~al.}{2018a}]{2018Nelson}
{Nelson} D.,  et~al., 2018a, \mn@doi [\mnras] {10.1093/mnras/stx3040}, \href
  {http://adsabs.harvard.edu/abs/2018MNRAS.475..624N} {475, 624}

\bibitem[\protect\citeauthoryear{{Nelson} et~al.,}{{Nelson}
  et~al.}{2018b}]{2018NelsonB}
{Nelson} D.,  et~al., 2018b, \mn@doi [\mnras] {10.1093/mnras/sty656}, \href
  {http://adsabs.harvard.edu/abs/2018MNRAS.477..450N} {477, 450}

\bibitem[\protect\citeauthoryear{{Nelson} et~al.,}{{Nelson}
  et~al.}{2019}]{2019Nelson}
{Nelson} D.,  et~al., 2019, arXiv e-prints, \href
  {http://adsabs.harvard.edu/abs/2019arXiv190205554N} {}

\bibitem[\protect\citeauthoryear{{Noeske}, {Weiner}, {Faber}, {Papovich}  \&
  {et al.}}{{Noeske} et~al.}{2007}]{2007Noeske}
{Noeske} K.~G.,  {Weiner} B.~J.,  {Faber} S.~M.,  {Papovich} C.,   {et al.}
  2007, \mn@doi [\apjl] {10.1086/517926}, \href
  {http://adsabs.harvard.edu/abs/2007ApJ...660L..43N} {660, L43}

\bibitem[\protect\citeauthoryear{{Oliver} et~al.,}{{Oliver}
  et~al.}{2010}]{2010Oliver}
{Oliver} S.,  et~al., 2010, \mn@doi [\mnras]
  {10.1111/j.1365-2966.2010.16643.x}, \href
  {http://adsabs.harvard.edu/abs/2010MNRAS.405.2279O} {405, 2279}

\bibitem[\protect\citeauthoryear{{Paccagnella} et~al.,}{{Paccagnella}
  et~al.}{2016}]{2016paccagnella}
{Paccagnella} A.,  et~al., 2016, \mn@doi [\apjl] {10.3847/2041-8205/816/2/L25},
  \href {http://adsabs.harvard.edu/abs/2016ApJ...816L..25P} {816, L25}

\bibitem[\protect\citeauthoryear{{Pannella} et~al.,}{{Pannella}
  et~al.}{2009}]{2009Pannella}
{Pannella} M.,  et~al., 2009, \mn@doi [\apjl] {10.1088/0004-637X/698/2/L116},
  \href {http://adsabs.harvard.edu/abs/2009ApJ...698L.116P} {698, L116}

\bibitem[\protect\citeauthoryear{{Patel}, {Holden}, {Kelson}, {Franx}, {van der
  Wel}  \& {Illingworth}}{{Patel} et~al.}{2012}]{2012Patel}
{Patel} S.~G.,  {Holden} B.~P.,  {Kelson} D.~D.,  {Franx} M.,  {van der Wel}
  A.,   {Illingworth} G.~D.,  2012, \mn@doi [\apjl]
  {10.1088/2041-8205/748/2/L27}, \href
  {http://adsabs.harvard.edu/abs/2012ApJ...748L..27P} {748, L27}

\bibitem[\protect\citeauthoryear{{Pearson} et~al.,}{{Pearson}
  et~al.}{2018}]{2018Pearson}
{Pearson} W.~J.,  et~al., 2018, preprint, \href
  {http://adsabs.harvard.edu/abs/2018arXiv180403482P} {} (\mn@eprint {arXiv}
  {1804.03482})

\bibitem[\protect\citeauthoryear{{Peng}, {Lilly}, {Kova{\v c}}, {Bolzonella}
  \& {et al.}}{{Peng} et~al.}{2010}]{2010Peng}
{Peng} Y.-j.,  {Lilly} S.~J.,  {Kova{\v c}} K.,  {Bolzonella} M.,   {et al.}
  2010, \mn@doi [\apj] {10.1088/0004-637X/721/1/193}, \href
  {http://adsabs.harvard.edu/abs/2010ApJ...721..193P} {721, 193}

\bibitem[\protect\citeauthoryear{{Pillepich} et~al.,}{{Pillepich}
  et~al.}{2018a}]{2018Pillepich_A}
{Pillepich} A.,  et~al., 2018a, \mn@doi [\mnras] {10.1093/mnras/stx2656}, \href
  {http://adsabs.harvard.edu/abs/2018MNRAS.473.4077P} {473, 4077}

\bibitem[\protect\citeauthoryear{{Pillepich} et~al.,}{{Pillepich}
  et~al.}{2018b}]{2018Pillepich}
{Pillepich} A.,  et~al., 2018b, \mn@doi [\mnras] {10.1093/mnras/stx3112}, \href
  {http://adsabs.harvard.edu/abs/2018MNRAS.475..648P} {475, 648}

\bibitem[\protect\citeauthoryear{{Pillepich} et~al.,}{{Pillepich}
  et~al.}{2019}]{2019Pillepich}
{Pillepich} A.,  et~al., 2019, arXiv e-prints, \href
  {http://adsabs.harvard.edu/abs/2019arXiv190205553P} {}

\bibitem[\protect\citeauthoryear{{Planck Collaboration} et~al.,}{{Planck
  Collaboration} et~al.}{2016}]{2016planck}
{Planck Collaboration} et~al., 2016, \aap, 594, A13

\bibitem[\protect\citeauthoryear{{Quadri}, {Williams}, {Franx}  \&
  {Hildebrandt}}{{Quadri} et~al.}{2012}]{2012Quadri}
{Quadri} R.~F.,  {Williams} R.~J.,  {Franx} M.,   {Hildebrandt} H.,  2012,
  \mn@doi [\apj] {10.1088/0004-637X/744/2/88}, \href
  {http://adsabs.harvard.edu/abs/2012ApJ...744...88Q} {744, 88}

\bibitem[\protect\citeauthoryear{{Reddy}, {Steidel}, {Pettini}, {Adelberger},
  {Shapley}, {Erb}  \& {Dickinson}}{{Reddy} et~al.}{2008}]{2008Reddy}
{Reddy} N.~A.,  {Steidel} C.~C.,  {Pettini} M.,  {Adelberger} K.~L.,  {Shapley}
  A.~E.,  {Erb} D.~K.,   {Dickinson} M.,  2008, \mn@doi [\apjs]
  {10.1086/521105}, \href {http://adsabs.harvard.edu/abs/2008ApJS..175...48R}
  {175, 48}

\bibitem[\protect\citeauthoryear{{Renzini} \& {Peng}}{{Renzini} \&
  {Peng}}{2015}]{2015Renzini}
{Renzini} A.,  {Peng} Y.-j.,  2015, \mn@doi [\apjl]
  {10.1088/2041-8205/801/2/L29}, \href
  {http://adsabs.harvard.edu/abs/2015ApJ...801L..29R} {801, L29}

\bibitem[\protect\citeauthoryear{{Richards} et~al.,}{{Richards}
  et~al.}{2016}]{2016Richards}
{Richards} S.~N.,  et~al., 2016, \mn@doi [\mnras] {10.1093/mnras/stv2453},
  \href {https://ui.adsabs.harvard.edu/#abs/2016MNRAS.455.2826R} {455, 2826}

\bibitem[\protect\citeauthoryear{{Rodighiero}, {Daddi}, {Baronchelli},
  {Cimatti}  \& {et al.}}{{Rodighiero} et~al.}{2011}]{2011Rodighiero}
{Rodighiero} G.,  {Daddi} E.,  {Baronchelli} I.,  {Cimatti} A.,   {et al.}
  2011, \mn@doi [\apjl] {10.1088/2041-8205/739/2/L40}, \href
  {http://adsabs.harvard.edu/abs/2011ApJ...739L..40R} {739, L40}

\bibitem[\protect\citeauthoryear{{Rodighiero}, {Renzini}, {Daddi},
  {Baronchelli}, {Berta}, {Cresci}, {Franceschini}  \& {et al.}}{{Rodighiero}
  et~al.}{2014}]{2014Rodighiero}
{Rodighiero} G.,  {Renzini} A.,  {Daddi} E.,  {Baronchelli} I.,  {Berta} S.,
  {Cresci} G.,  {Franceschini} A.,   {et al.} 2014, \mn@doi [\mnras]
  {10.1093/mnras/stu1110}, \href
  {http://adsabs.harvard.edu/abs/2014MNRAS.443...19R} {443, 19}

\bibitem[\protect\citeauthoryear{{Salim} et~al.,}{{Salim}
  et~al.}{2007}]{2007salim}
{Salim} S.,  et~al., 2007, \mn@doi [\apjs] {10.1086/519218}, \href
  {http://adsabs.harvard.edu/abs/2007ApJS..173..267S} {173, 267}

\bibitem[\protect\citeauthoryear{{Salpeter}}{{Salpeter}}{1955}]{1955Salpeter}
{Salpeter} E.~E.,  1955, \mn@doi [\apj] {10.1086/145971}, \href
  {http://adsabs.harvard.edu/abs/1955ApJ...121..161S} {121, 161}

\bibitem[\protect\citeauthoryear{{Santini} et~al.,}{{Santini}
  et~al.}{2009}]{2009Santini}
{Santini} P.,  et~al., 2009, \mn@doi [\aap] {10.1051/0004-6361/200811434},
  \href {http://adsabs.harvard.edu/abs/2009A%26A...504..751S} {504, 751}

\bibitem[\protect\citeauthoryear{{Santini} et~al.,}{{Santini}
  et~al.}{2014}]{2014Santini}
{Santini} P.,  et~al., 2014, \mn@doi [\aap] {10.1051/0004-6361/201322835},
  \href {http://adsabs.harvard.edu/abs/2014A%26A...562A..30S} {562, A30}

\bibitem[\protect\citeauthoryear{{Santini} et~al.,}{{Santini}
  et~al.}{2017}]{2017Santini}
{Santini} P.,  et~al., 2017, \mn@doi [\apj] {10.3847/1538-4357/aa8874}, \href
  {http://adsabs.harvard.edu/abs/2017ApJ...847...76S} {847, 76}

\bibitem[\protect\citeauthoryear{{Schawinski} et~al.,}{{Schawinski}
  et~al.}{2014}]{2014Schawinski}
{Schawinski} K.,  et~al., 2014, \mn@doi [\mnras] {10.1093/mnras/stu327}, \href
  {http://adsabs.harvard.edu/abs/2014MNRAS.440..889S} {440, 889}

\bibitem[\protect\citeauthoryear{{Schreiber}, {Pannella}, {Elbaz},
  {B{\'e}thermin}  \& {et al.}}{{Schreiber} et~al.}{2015}]{2015Schreiber}
{Schreiber} C.,  {Pannella} M.,  {Elbaz} D.,  {B{\'e}thermin} M.,   {et al.}
  2015, \mn@doi [\aap] {10.1051/0004-6361/201425017}, \href
  {http://adsabs.harvard.edu/abs/2015A%26A...575A..74S} {575, A74}

\bibitem[\protect\citeauthoryear{{Schreiber}, {Elbaz}, {Pannella}, {Ciesla},
  {Wang}, {Koekemoer}, {Rafelski}  \& {Daddi}}{{Schreiber}
  et~al.}{2016}]{2016Schreiber}
{Schreiber} C.,  {Elbaz} D.,  {Pannella} M.,  {Ciesla} L.,  {Wang} T.,
  {Koekemoer} A.,  {Rafelski} M.,   {Daddi} E.,  2016, \mn@doi [\aap]
  {10.1051/0004-6361/201527200}, \href
  {http://adsabs.harvard.edu/abs/2016A%26A...589A..35S} {589, A35}

\bibitem[\protect\citeauthoryear{{Shivaei} et~al.,}{{Shivaei}
  et~al.}{2015}]{2015Shivaei}
{Shivaei} I.,  et~al., 2015, \mn@doi [\apj] {10.1088/0004-637X/815/2/98}, \href
  {https://ui.adsabs.harvard.edu/\#abs/2015ApJ...815...98S} {815, 98}

\bibitem[\protect\citeauthoryear{{Sijacki}, {Vogelsberger}, {Genel},
  {Springel}, {Torrey}, {Snyder}, {Nelson}  \& {Hernquist}}{{Sijacki}
  et~al.}{2015}]{2015Sijacki}
{Sijacki} D.,  {Vogelsberger} M.,  {Genel} S.,  {Springel} V.,  {Torrey} P.,
  {Snyder} G.~F.,  {Nelson} D.,   {Hernquist} L.,  2015, \mn@doi [\mnras]
  {10.1093/mnras/stv1340}, \href
  {http://adsabs.harvard.edu/abs/2015MNRAS.452..575S} {452, 575}

\bibitem[\protect\citeauthoryear{{Somerville} \& {Dav{\'e}}}{{Somerville} \&
  {Dav{\'e}}}{2015}]{2015somerville}
{Somerville} R.~S.,  {Dav{\'e}} R.,  2015, \mn@doi [Annual Review of Astronomy
  and Astrophysics] {10.1146/annurev-astro-082812-140951}, \href
  {https://ui.adsabs.harvard.edu/#abs/2015ARA&A..53...51S} {53, 51}

\bibitem[\protect\citeauthoryear{{Sorba} \& {Sawicki}}{{Sorba} \&
  {Sawicki}}{2018}]{2018Sorba}
{Sorba} R.,  {Sawicki} M.,  2018, \mn@doi [\mnras] {10.1093/mnras/sty186},
  \href {http://adsabs.harvard.edu/abs/2018MNRAS.476.1532S} {476, 1532}

\bibitem[\protect\citeauthoryear{{Sparre} et~al.,}{{Sparre}
  et~al.}{2015}]{2015Sparre}
{Sparre} M.,  et~al., 2015, \mn@doi [\mnras] {10.1093/mnras/stu2713}, \href
  {http://adsabs.harvard.edu/abs/2015MNRAS.447.3548S} {447, 3548}

\bibitem[\protect\citeauthoryear{{Speagle}, {Steinhardt}, {Capak}  \&
  {Silverman}}{{Speagle} et~al.}{2014}]{2014Speagle}
{Speagle} J.~S.,  {Steinhardt} C.~L.,  {Capak} P.~L.,   {Silverman} J.~D.,
  2014, \mn@doi [\apjs] {10.1088/0067-0049/214/2/15}, \href
  {http://adsabs.harvard.edu/abs/2014ApJS..214...15S} {214, 15}

\bibitem[\protect\citeauthoryear{{Springel}}{{Springel}}{2010}]{2010springel}
{Springel} V.,  2010, \mn@doi [\mnras] {10.1111/j.1365-2966.2009.15715.x},
  \href {http://adsabs.harvard.edu/abs/2010MNRAS.401..791S} {401, 791}

\bibitem[\protect\citeauthoryear{{Springel} \& {Hernquist}}{{Springel} \&
  {Hernquist}}{2003}]{2003Springel}
{Springel} V.,  {Hernquist} L.,  2003, \mn@doi [\mnras]
  {10.1046/j.1365-8711.2003.06206.x}, \href
  {http://adsabs.harvard.edu/abs/2003MNRAS.339..289S} {339, 289}

\bibitem[\protect\citeauthoryear{{Springel}, {White}, {Tormen}  \&
  {Kauffmann}}{{Springel} et~al.}{2001}]{2001springel}
{Springel} V.,  {White} S.~D.~M.,  {Tormen} G.,   {Kauffmann} G.,  2001,
  \mn@doi [\mnras] {10.1046/j.1365-8711.2001.04912.x}, \href
  {http://adsabs.harvard.edu/abs/2001MNRAS.328..726S} {328, 726}

\bibitem[\protect\citeauthoryear{{Springel} et~al.,}{{Springel}
  et~al.}{2018}]{2018Springel}
{Springel} V.,  et~al., 2018, \mn@doi [\mnras] {10.1093/mnras/stx3304}, \href
  {http://adsabs.harvard.edu/abs/2018MNRAS.475..676S} {475, 676}

\bibitem[\protect\citeauthoryear{{Strateva} et~al.,}{{Strateva}
  et~al.}{2001}]{2001Strateva}
{Strateva} I.,  et~al., 2001, \mn@doi [\aj] {10.1086/323301}, \href
  {http://adsabs.harvard.edu/abs/2001AJ....122.1861S} {122, 1861}

\bibitem[\protect\citeauthoryear{{Tasca}, {Le F{\`e}vre}, {Hathi}, {Schaerer},
   \& {et al.}}{{Tasca} et~al.}{2015}]{2015Tasca}
{Tasca} L.~A.~M.,  {Le F{\`e}vre} O.,  {Hathi} N.~P.,  {Schaerer} D.,    {et
  al.} 2015, \mn@doi [\aap] {10.1051/0004-6361/201425379}, \href
  {http://adsabs.harvard.edu/abs/2015A%26A...581A..54T} {581, A54}

\bibitem[\protect\citeauthoryear{{Theios}, {Steidel}, {Strom}, {Rudie},
  {Trainor}  \& {Reddy}}{{Theios} et~al.}{2018}]{2018Theios}
{Theios} R.~L.,  {Steidel} C.~C.,  {Strom} A.~L.,  {Rudie} G.~C.,  {Trainor}
  R.~F.,   {Reddy} N.~A.,  2018, preprint, \href
  {http://adsabs.harvard.edu/abs/2018arXiv180500016T} {} (\mn@eprint {arXiv}
  {1805.00016})

\bibitem[\protect\citeauthoryear{{Tomczak}, {Quadri}, {Tran}, {Labb{\'e}}  \&
  {et al.}}{{Tomczak} et~al.}{2014}]{2014tomczak}
{Tomczak} A.~R.,  {Quadri} R.~F.,  {Tran} K.-V.~H.,  {Labb{\'e}} I.,   {et al.}
  2014, \mn@doi [\apj] {10.1088/0004-637X/783/2/85}, \href
  {http://adsabs.harvard.edu/abs/2014ApJ...783...85T} {783, 85}

\bibitem[\protect\citeauthoryear{{Tomczak} et~al.,}{{Tomczak}
  et~al.}{2016}]{2016Tomczak}
{Tomczak} A.~R.,  et~al., 2016, \mn@doi [\apj] {10.3847/0004-637X/817/2/118},
  \href {http://adsabs.harvard.edu/abs/2016ApJ...817..118T} {817, 118}

\bibitem[\protect\citeauthoryear{{Torrey}, {Vogelsberger}, {Genel}, {Sijacki},
  {Springel}  \& {Hernquist}}{{Torrey} et~al.}{2014}]{2014Torrey}
{Torrey} P.,  {Vogelsberger} M.,  {Genel} S.,  {Sijacki} D.,  {Springel} V.,
  {Hernquist} L.,  2014, \mn@doi [\mnras] {10.1093/mnras/stt2295}, \href
  {http://adsabs.harvard.edu/abs/2014MNRAS.438.1985T} {438, 1985}

\bibitem[\protect\citeauthoryear{{Torrey} et~al.,}{{Torrey}
  et~al.}{2018}]{2018Torrey}
{Torrey} P.,  et~al., 2018, \mn@doi [\mnras] {10.1093/mnrasl/sly031}, \href
  {http://adsabs.harvard.edu/abs/2018MNRAS.477L..16T} {477, L16}

\bibitem[\protect\citeauthoryear{{Trayford}, {Theuns}, {Bower}, {Crain},
  {Lagos}, {Schaller}  \& {Schaye}}{{Trayford} et~al.}{2016}]{2016Trayford}
{Trayford} J.~W.,  {Theuns} T.,  {Bower} R.~G.,  {Crain} R.~A.,  {Lagos} C.
  d.~P.,  {Schaller} M.,   {Schaye} J.,  2016, \mn@doi [\mnras]
  {10.1093/mnras/stw1230}, \href
  {https://ui.adsabs.harvard.edu/#abs/2016MNRAS.460.3925T} {460, 3925}

\bibitem[\protect\citeauthoryear{{Vogelsberger}, {Genel}, {Sijacki}, {Torrey},
  {Springel}  \& {Hernquist}}{{Vogelsberger} et~al.}{2013}]{2013Vogelsberger}
{Vogelsberger} M.,  {Genel} S.,  {Sijacki} D.,  {Torrey} P.,  {Springel} V.,
  {Hernquist} L.,  2013, \mn@doi [\mnras] {10.1093/mnras/stt1789}, \href
  {http://adsabs.harvard.edu/abs/2013MNRAS.436.3031V} {436, 3031}

\bibitem[\protect\citeauthoryear{{Vogelsberger} et~al.,}{{Vogelsberger}
  et~al.}{2014a}]{2014MNRASVogel}
{Vogelsberger} M.,  et~al., 2014a, \mn@doi [\mnras] {10.1093/mnras/stu1536},
  \href {http://adsabs.harvard.edu/abs/2014MNRAS.444.1518V} {444, 1518}

\bibitem[\protect\citeauthoryear{{Vogelsberger} et~al.,}{{Vogelsberger}
  et~al.}{2014b}]{2014vogel}
{Vogelsberger} M.,  et~al., 2014b, \mn@doi [\nat] {10.1038/nature13316}, \href
  {http://adsabs.harvard.edu/abs/2014Natur.509..177V} {509, 177}

\bibitem[\protect\citeauthoryear{{Vogelsberger} et~al.,}{{Vogelsberger}
  et~al.}{2018}]{2018Vogelsberger}
{Vogelsberger} M.,  et~al., 2018, \mn@doi [\mnras] {10.1093/mnras/stx2955},
  \href {http://adsabs.harvard.edu/abs/2018MNRAS.474.2073V} {474, 2073}

\bibitem[\protect\citeauthoryear{{Vulcani}, {Poggianti}, {Finn}, {Rudnick},
  {Desai}  \& {Bamford}}{{Vulcani} et~al.}{2010}]{2010Vulcani}
{Vulcani} B.,  {Poggianti} B.~M.,  {Finn} R.~A.,  {Rudnick} G.,  {Desai} V.,
  {Bamford} S.,  2010, \mn@doi [\apjl] {10.1088/2041-8205/710/1/L1}, \href
  {http://adsabs.harvard.edu/abs/2010ApJ...710L...1V} {710, L1}

\bibitem[\protect\citeauthoryear{{Vulcani} et~al.,}{{Vulcani}
  et~al.}{2012}]{2012Vulcani}
{Vulcani} B.,  et~al., 2012, \mn@doi [\mnras]
  {10.1111/j.1365-2966.2011.20135.x}, \href
  {http://adsabs.harvard.edu/abs/2012MNRAS.420.1481V} {420, 1481}

\bibitem[\protect\citeauthoryear{{Weinberger} et~al.,}{{Weinberger}
  et~al.}{2018}]{2018Weinberger}
{Weinberger} R.,  et~al., 2018, \mn@doi [\mnras] {10.1093/mnras/sty1733}, \href
  {http://adsabs.harvard.edu/abs/2018MNRAS.479.4056W} {479, 4056}

\bibitem[\protect\citeauthoryear{{Wetzel}, {Tinker}  \& {Conroy}}{{Wetzel}
  et~al.}{2012}]{2012Wetzel}
{Wetzel} A.~R.,  {Tinker} J.~L.,   {Conroy} C.,  2012, \mn@doi [\mnras]
  {10.1111/j.1365-2966.2012.21188.x}, \href
  {http://adsabs.harvard.edu/abs/2012MNRAS.424..232W} {424, 232}

\bibitem[\protect\citeauthoryear{{Wetzel}, {Tinker}, {Conroy}  \& {van den
  Bosch}}{{Wetzel} et~al.}{2013}]{2013Wetzel}
{Wetzel} A.~R.,  {Tinker} J.~L.,  {Conroy} C.,   {van den Bosch} F.~C.,  2013,
  \mn@doi [\mnras] {10.1093/mnras/stt469}, \href
  {http://adsabs.harvard.edu/abs/2013MNRAS.432..336W} {432, 336}

\bibitem[\protect\citeauthoryear{{Whitaker} et~al.,}{{Whitaker}
  et~al.}{2010}]{2010Whitaker}
{Whitaker} K.~E.,  et~al., 2010, \mn@doi [\apj] {10.1088/0004-637X/719/2/1715},
  \href {http://adsabs.harvard.edu/abs/2010ApJ...719.1715W} {719, 1715}

\bibitem[\protect\citeauthoryear{{Whitaker} et~al.,}{{Whitaker}
  et~al.}{2011}]{2011Whitaker}
{Whitaker} K.~E.,  et~al., 2011, \mn@doi [\apj] {10.1088/0004-637X/735/2/86},
  \href {http://adsabs.harvard.edu/abs/2011ApJ...735...86W} {735, 86}

\bibitem[\protect\citeauthoryear{{Whitaker}, {van Dokkum}, {Brammer}  \&
  {Franx}}{{Whitaker} et~al.}{2012}]{2012Whitaker}
{Whitaker} K.~E.,  {van Dokkum} P.~G.,  {Brammer} G.,   {Franx} M.,  2012,
  \mn@doi [\apjl] {10.1088/2041-8205/754/2/L29}, \href
  {http://adsabs.harvard.edu/abs/2012ApJ...754L..29W} {754, L29}

\bibitem[\protect\citeauthoryear{{Whitaker} et~al.,}{{Whitaker}
  et~al.}{2014}]{2014Whitaker}
{Whitaker} K.~E.,  et~al., 2014, \mn@doi [\apj] {10.1088/0004-637X/795/2/104},
  \href {http://adsabs.harvard.edu/abs/2014ApJ...795..104W} {795, 104}

\bibitem[\protect\citeauthoryear{{Williams}, {Quadri}, {Franx}, {van Dokkum}
  \& {Labb{\'e}}}{{Williams} et~al.}{2009}]{2009Williams}
{Williams} R.~J.,  {Quadri} R.~F.,  {Franx} M.,  {van Dokkum} P.,   {Labb{\'e}}
  I.,  2009, \mn@doi [\apj] {10.1088/0004-637X/691/2/1879}, \href
  {http://adsabs.harvard.edu/abs/2009ApJ...691.1879W} {691, 1879}

\bibitem[\protect\citeauthoryear{{Williams}, {Quadri}, {Franx}, {van Dokkum},
  {Toft}, {Kriek}  \& {Labb{\'e}}}{{Williams} et~al.}{2010}]{2010Williams}
{Williams} R.~J.,  {Quadri} R.~F.,  {Franx} M.,  {van Dokkum} P.,  {Toft} S.,
  {Kriek} M.,   {Labb{\'e}} I.,  2010, \mn@doi [\apj]
  {10.1088/0004-637X/713/2/738}, \href
  {http://adsabs.harvard.edu/abs/2010ApJ...713..738W} {713, 738}

\bibitem[\protect\citeauthoryear{{Wuyts}, {F{\"o}rster Schreiber}, {van der
  Wel}, {Magnelli}  \& {et al.}}{{Wuyts} et~al.}{2011}]{2011Wuyts}
{Wuyts} S.,  {F{\"o}rster Schreiber} N.~M.,  {van der Wel} A.,  {Magnelli} B.,
   {et al.} 2011, \mn@doi [\apj] {10.1088/0004-637X/742/2/96}, \href
  {http://adsabs.harvard.edu/abs/2011ApJ...742...96W} {742, 96}

\bibitem[\protect\citeauthoryear{{Yu} \& {Wang}}{{Yu} \& {Wang}}{2016}]{2016yu}
{Yu} H.,  {Wang} F.~Y.,  2016, \mn@doi [\apj] {10.3847/0004-637X/820/2/114},
  \href {https://ui.adsabs.harvard.edu/\#abs/2016ApJ...820..114Y} {820, 114}

\bibitem[\protect\citeauthoryear{{Zahid}, {Dima}, {Kewley}, {Erb}  \&
  {Dav{\'e}}}{{Zahid} et~al.}{2012}]{2012Zahid}
{Zahid} H.~J.,  {Dima} G.~I.,  {Kewley} L.~J.,  {Erb} D.~K.,   {Dav{\'e}} R.,
  2012, \mn@doi [\apj] {10.1088/0004-637X/757/1/54}, \href
  {http://adsabs.harvard.edu/abs/2012ApJ...757...54Z} {757, 54}

\makeatother
\end{thebibliography}

\appendix
\section{SFRs over different timescales and effects of numerical resolution}
\label{appendix}

\begin{figure*}
\centering
\includegraphics[width=0.42\textwidth]{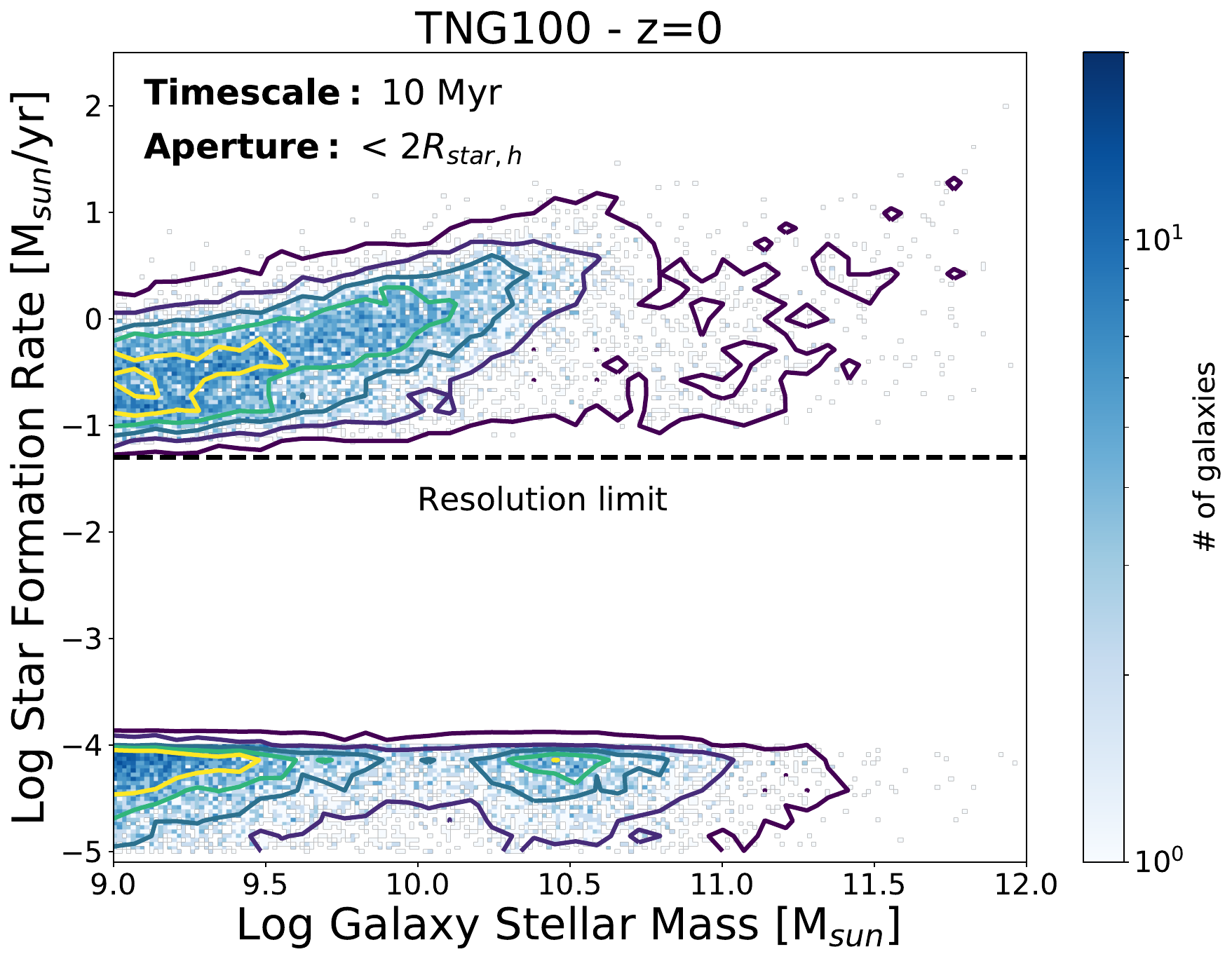}
\includegraphics[width=0.42\textwidth]{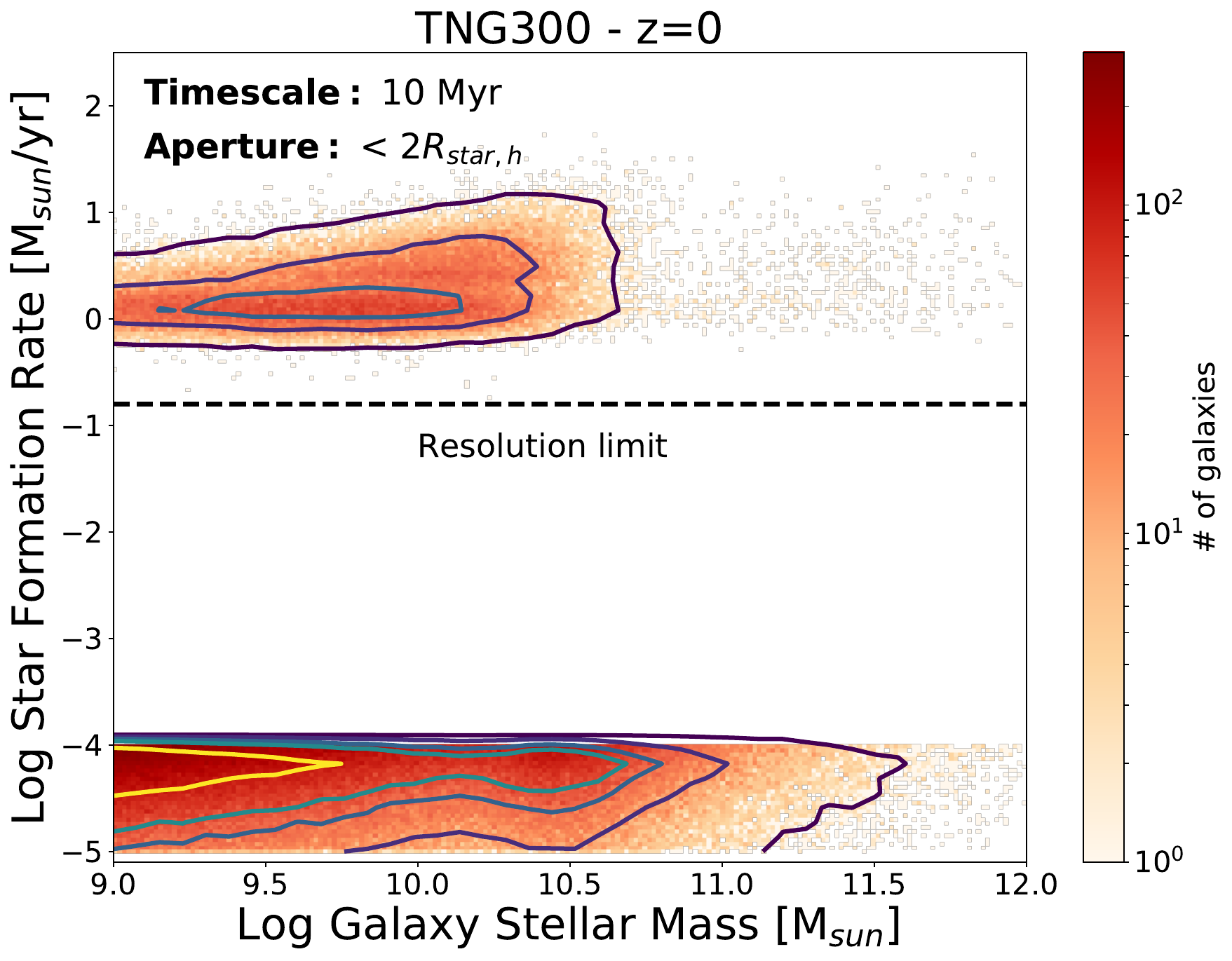}
\includegraphics[width=0.42\textwidth]{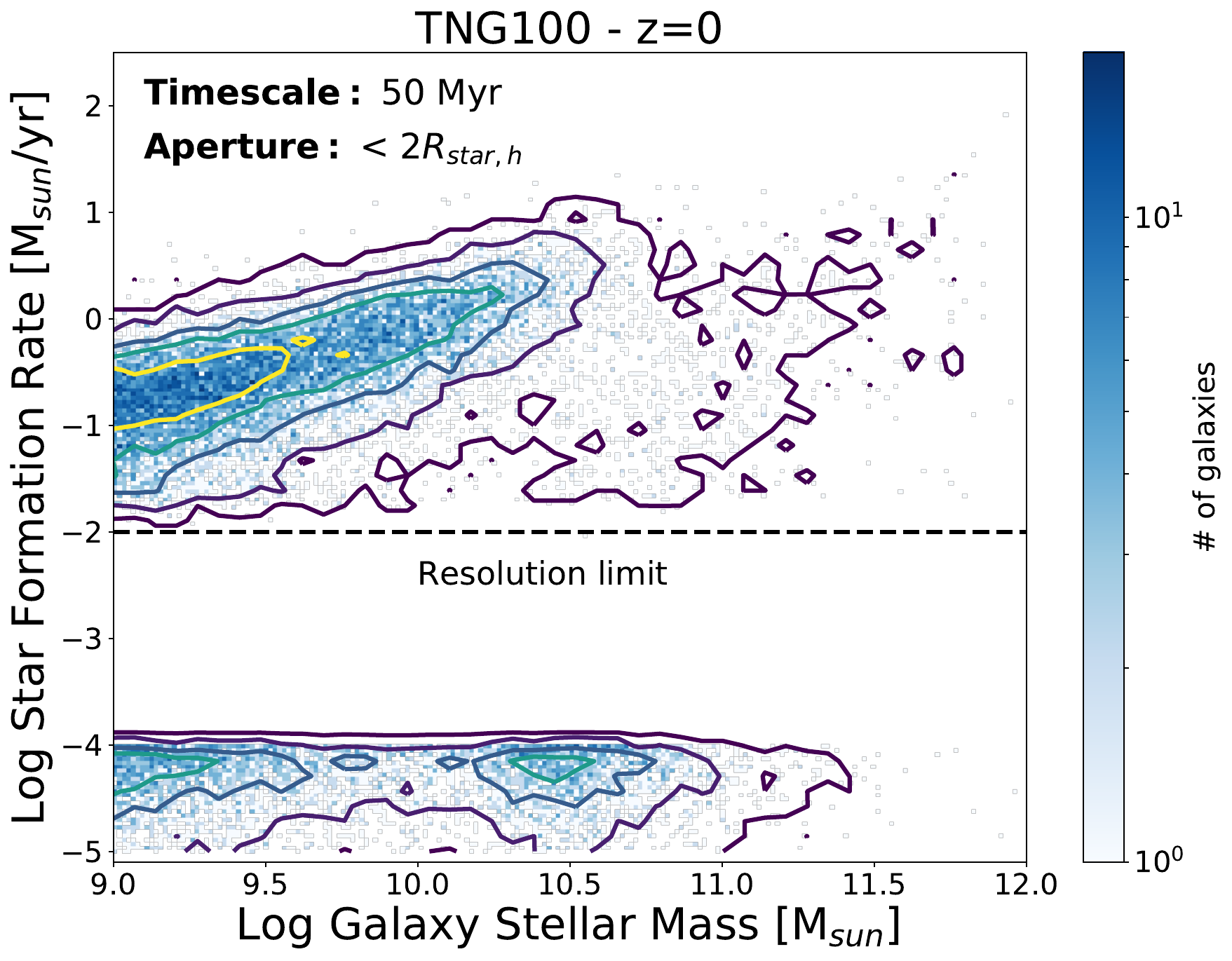}
\includegraphics[width=0.42\textwidth]{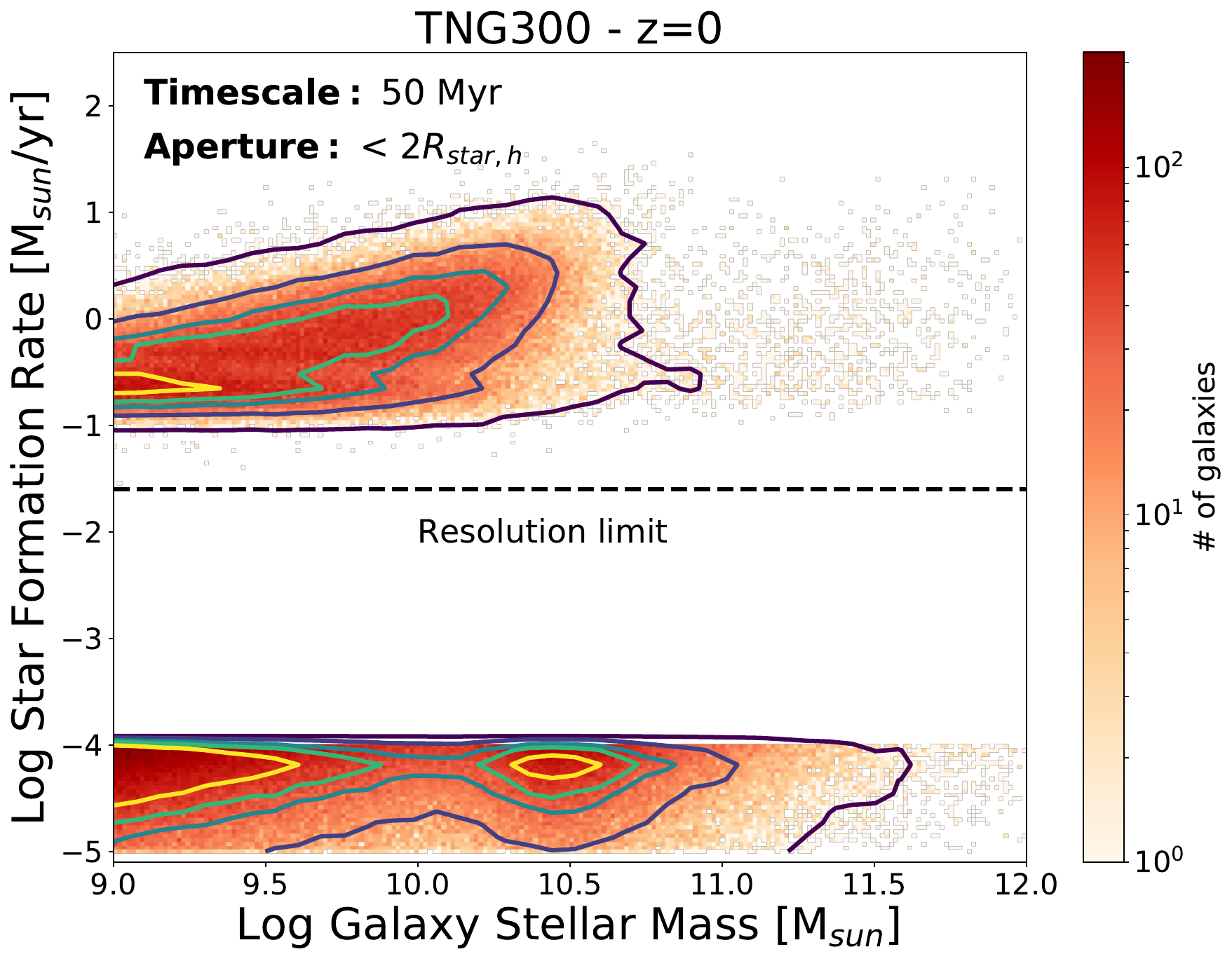}
\includegraphics[width=0.42\textwidth]{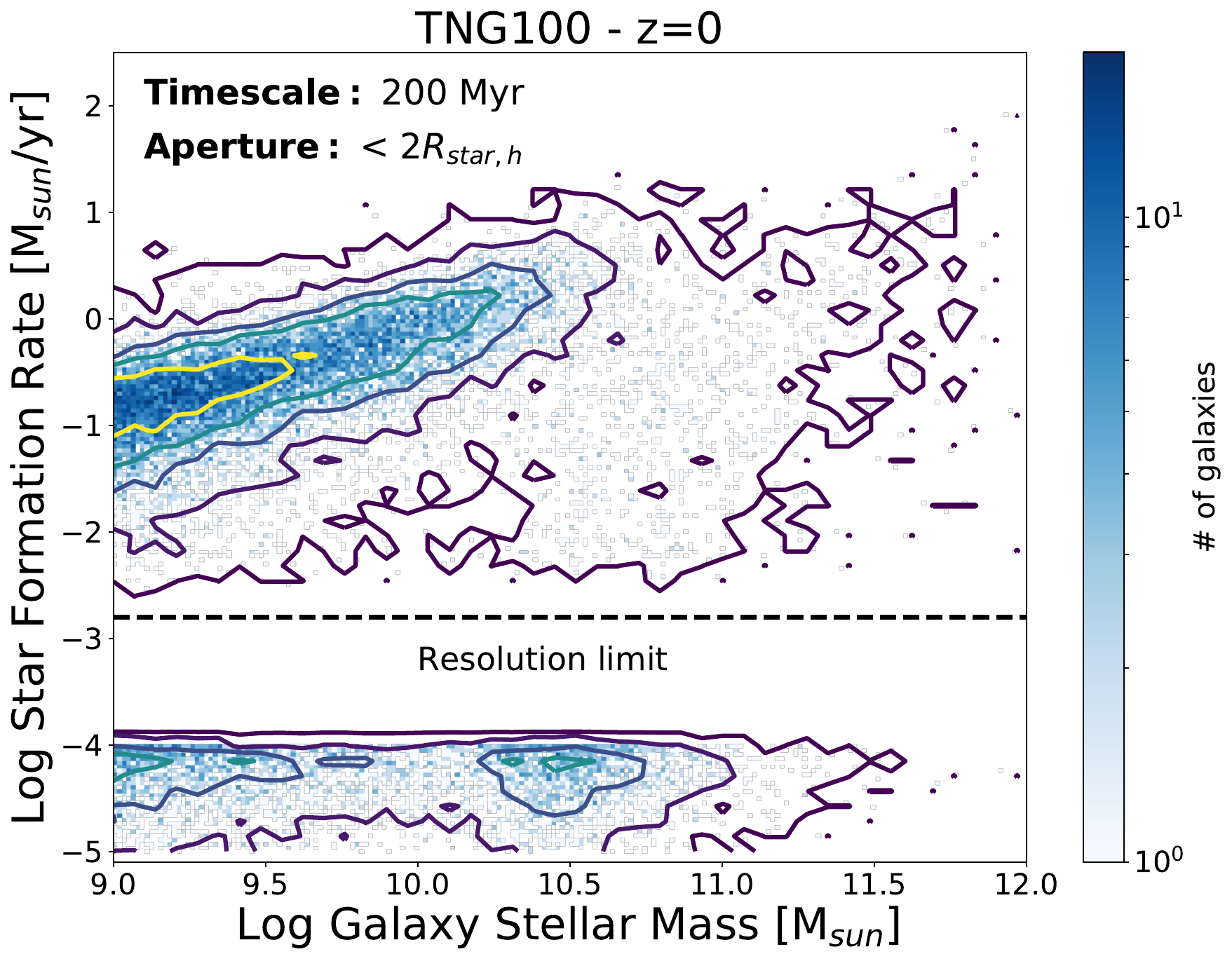}
\includegraphics[width=0.42\textwidth]{TNG300_z0_200Myr.pdf}
\includegraphics[width=0.42\textwidth]{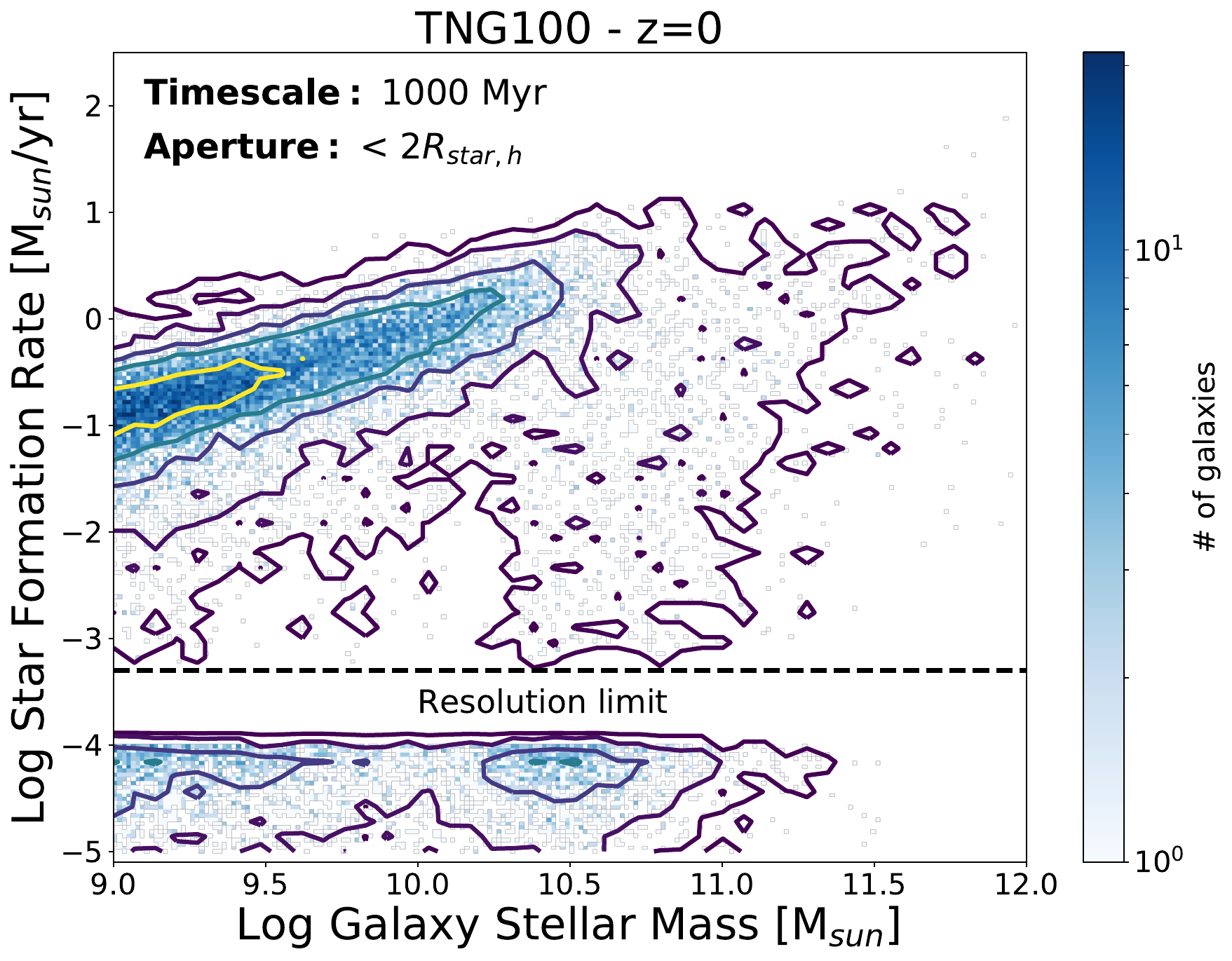}
\includegraphics[width=0.42\textwidth]{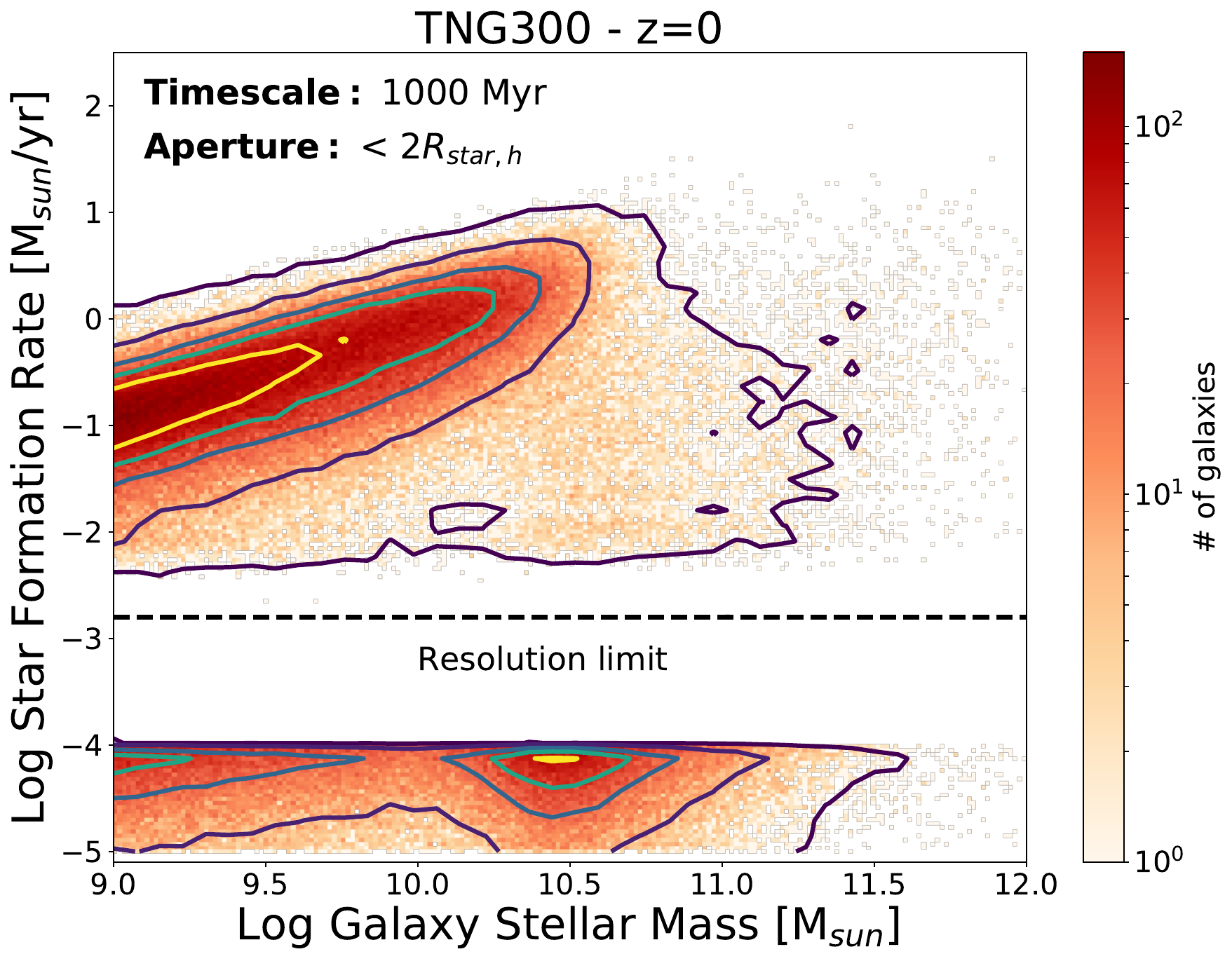}
\caption{\label{fig:SFR-MSTAR_timescales} SFR as a function of stellar mass $\MS$ of TNG300 (left column) and TNG100 (right column) galaxies. From top to bottom panels the SFR is averaged over 10, 50, 200 and 1000 Myrs and measured within 2$\times R_{\rm star,h}$. The color scale indicates the number of galaxies and the solid contours encompass 20, 50, 70, 90, and 99 $\%$ of the galaxies. To all galaxies with unresolved SFR values we assign a random values in the range $\rm  SFR = 10^{-5} - 10^{-4} \, \Ms \rm \, yr^{-1}$.}
\end{figure*}

In this paper we use two simulations of the IllustrisTNG project: TNG100 and TNG300, characterized by a factor of 8 (2) difference in mass (spatial) resolution (see Table \ref{tab:tng} and www.tng-project.org). We measure the SFRs of simulated galaxies by averaging the stellar mass production rate over different time scales. In this section, we discuss how our results are affected by different numerical resolutions and for different SFR estimates.

Firstly, Fig.~\ref{fig:SFR-MSTAR_timescales} shows explicitly the SFR-$\MS$ plane for TNG100 (left) and TNG300 (right) for increasingly longer SFR averaging timescales (from top to bottom).
The galaxy SFRs in all panels are measured within 2$\times R_{\rm star,h}$.
As a reminder, TNG100 has the same initial conditions, cosmological box volume ($\sim 100$ Mpc on-a-side) and similar resolution of the original Illustris run, with a baryonic target mass of $m_{\rm baryon} \sim 1.4 \times 10^6 \, \Ms$ and $\sim 1.26 \times 10^6 \, \Ms$, respectively.  On the other hand, the volume of TNG300 is $\sim 300$ Mpc on-a-side, with the baryonic target mass of $1.1 \times 10^7 \, \Ms$. The mass of each gas cells is kept within a factor of 2 of the target mass \citep{2018Pillepich_A} and so is the initial mass of the star particles at birth.

All galaxies considered in this work have a minimum stellar mass of $10^9 \, \Ms$, meaning that each galaxy in TNG100 has about at least one thousand star particles in total. In the extreme case of small galaxies ($\MS = 10^9 \, \Ms$) in the TNG300 simulated volume, the number of star particles counted in each galaxy is of the order of one hundred, forcing us to be careful in using TNG300 in this limiting case.

Moreover, the different mass resolutions affect our estimates for the galaxy SFRs, especially if it is averaged over different timescales. From the minimum baryonic particle masses of each simulation we can estimate the lowest measurable value of the SFR, $\rm SFR_{min}$.  For TNG100, the smallest non-vanishing SFR values read $\rm Log \, SFR_{min} ~ (\Ms \, \rm yr^{-1}) \simeq -1.15, -1.85, -2.46, -3.15 $ for SFRs averaged over 10, 50, 200 and 1000 Myr, respectively. Analog considerations for the TNG300 run imply a minimum resolved values for SFR of  $\rm Log \, SFR_{min} ~ (\Ms \, \rm yr^{-1}) \simeq -0.6, -1.3, -1.9, -2.6 $, for the same averaging timescales.

On the other hand, for the case of the instantaneous SFR of the gas cell, the lowest SFR measurable values are found to be $\rm Log \, SFR ~ (\Ms \, \rm yr^{-1}) \simeq -4$ and $\rm Log \, SFR ~ (\Ms \, \rm yr^{-1}) \simeq -3$ for TNG100 and TNG300, respectively.
Galaxies with $\rm Log \, SFR < Log \,  SFR_{min}$, have unresolved SFR values, corresponding to $\rm SFR=0$ in the simulations and will be assigned to the range $10^{-5} - 10^{-4} \, \Ms \rm \, yr^{-1}$.

The effect of the mass resolutions on the averaging timescales is well detectable by-eye in Fig.~\ref{fig:SFR-MSTAR_timescales}, where galaxies are not found below a certain value of SFR. Indeed, individual star particles have such high masses that low-SFR galaxies will very rarely be able to spawn a new star particle, thus in general there will be few young stars with age less than e.g. 10 Myr. This also corresponds to no young stars being able to generate blue light, which is why it is impossible to get blue enough galaxies at poor resolution \citep{2018Nelson,2016Trayford}.

Mass resolution effects are much more pronounced for shorter timescales: this could potentially artificially bias high the main sequence's locus. This is the case, for example, for the shortest timescale (10 Myr) at the low-mass end ($\MS < 10^{9.5} \, \Ms$): see top left panel of Fig.~\ref{fig:systematics}. For this reason, we have not enunciated statements in the paper that are based on such short timescales at low galaxy masses.

We have analyzed the shape of the SFR-$\MS$ planes and of the star-forming MS also for the lower resolution runs of TNG100, namely TNG100-2 and TNG100-3, that have 8 and 64 times worse mass resolution than TNG100, respectively (TNG100-2 being comparable in resolution to TNG300). We find that the worst mass resolution of TNG100-3 leads to a considerably overestimation of the MS, in terms of both slope and normalization, unless the longest timescales -- 1000 Myr -- are considered. On the other hand, the intermediate numerical resolutions of TNG100-2 and TNG300 return results on the star-forming MS that are converged with those of TNG100 if the SFRs are averaged over timescales larger than 50 Myr at masses above $10^{10} \, \Ms$.

These considerations are taken into account when the results of the paper are discussed and motivate the usage of the 200-Myr SFR estimates as fiducial choice for both TNG100 and TNG300. 

\end{document}